\documentclass[journal=jpcbfk,manuscript=article]{achemso}

\usepackage[version=3]{mhchem} % Formula subscripts using \ce{}
\usepackage{longtable}
\SectionNumbersOn

\usepackage{amsmath}
\usepackage{amssymb}
\usepackage{color}
\usepackage{CJK}
\usepackage{framed}
\usepackage{tikz}           % vectorgraphics, used for name tags
\usepackage{hyperref}
\usepackage{algorithm}
\usepackage{algpseudocode}
\algdef{SE}[DOWHILE]{Do}{doWhile}{\algorithmicdo}[1]{\algorithmicwhile\ #1}

\usepackage[scaled=0.85]{beramono}
\usepackage[T1]{fontenc}
\usepackage{changepage}
\usepackage{multirow}
\usepackage{xspace}
\usepackage{tikz}
\usepackage{subfig}
\usepackage{graphicx}
\usepackage{makecell}

\usepackage[draft]{todonotes}   % notes shown
\usepackage{graphicx}
\usepackage{booktabs}
\usepackage{pdflscape}
\usepackage{longtable}
\usepackage{ulem} % for strikeouts

\newcommand{\AAtom}{{\mathrm{AA}}}
\newcommand{\CG}{{\mathrm{CG}}}
\newcommand{\VOX}{{\mathrm{VOX}}}

\author{Kirill Shmilovich}
\affiliation{%
  Pritzker School of Molecular Engineering, University of Chicago, Chicago, IL, United States
}
\email{kirills@uchicago.edu}

\author{Marc Stieffenhofer}
\affiliation{
    Max Planck Institute for Polymer Research, Mainz, Germany
}

\author{Nicholas E. Charron}
\affiliation{
    Weiss School of Natural Sciences, Department of Physics and Astronomy, Rice University, Houston, TX, United States
}
\altaffiliation{
    Department of Physics, Freie Universität Berlin, Berlin, Germany
}

\author{Moritz Hoffmann}
\affiliation{
    Fachbereich Mathematik und Informatik, Freie Universität Berlin, Berlin, Germany
}

\title[]{Temporally coherent backmapping of molecular trajectories from coarse-grained to atomistic resolution}

\begin{document}
%%%%%%%%%%%%%%%%%%%%%%%%%%%%%%%%%%%%%%%%%%%%%%%%%%%%%%%%%%%%%%%%%%%%%
%% The manuscript does not need to include \maketitle, which is
%% executed automatically.  The document should begin with an
%% abstract, if appropriate.  If one is given and should not be, the
%% contents will be gobbled.
%%%%%%%%%%%%%%%%%%%%%%%%%%%%%%%%%%%%%%%%%%%%%%%%%%%%%%%%%%%%%%%%%%%%%

\begin{abstract}
\noindent Coarse-graining offers a means to extend the achievable time and length scales of molecular dynamics simulations beyond what is practically possible in the atomistic regime. Sampling molecular configurations of interest can be done efficiently using coarse-grained simulations, from which meaningful physicochemical information can be inferred if the corresponding all-atom configurations are reconstructed. However, this procedure of backmapping to reintroduce the lost atomistic detail into coarse-grain structures has proven a challenging task due to the many feasible atomistic configurations that can be associated with one coarse-grain structure. Existing backmapping methods are strictly frame-based, relying on either heuristics to replace coarse-grain particles with atomic fragments and subsequent relaxation, or parameterized models to propose atomic coordinates separately and independently for each coarse-grain structure. These approaches neglect information from previous trajectory frames that is critical to ensuring temporal coherence of the backmapped trajectory, while also offering information potentially helpful to produce higher-fidelity atomic reconstructions. In this work we present a deep learning-enabled data-driven approach for temporally coherent backmapping that explicitly incorporates information from preceding trajectory structures. Our method trains a conditional variational autoencoder to non-deterministically reconstruct atomistic detail conditioned on both the target coarse-grain configuration and the previously reconstructed atomistic configuration. We demonstrate our backmapping approach on two exemplar biomolecular systems: alanine dipeptide and the miniprotein chignolin. We show that our backmapped trajectories accurately recover the structural, thermodynamic, and kinetic properties of the atomistic trajectory data.            
\end{abstract}

\clearpage
\newpage


\section{\label{sec:intro}Introduction}

A central limitation of modelling soft-matter systems with molecular dynamics~(MD) simulations are the long characteristic timescales of interesting processes, such as protein folding, compared to the relatively short integration time steps required to accurately propagate the system forward in time. A plethora of strategies strive to overcome this time scale barrier, such as enhanced sampling techniques,~\cite{mitsutake2013enhanced,bernardi2015enhanced,okamoto2004generalized,sidky2020machine,noe2019boltzmann} modern/specialized hardware,~\cite{shaw2009millisecond,hirst2014molecular,shaw2014anton} and hierarchical multiscale modelling~\cite{kremer2002multiscale, horstemeyer2009multiscale, peter2009multiscale,ingolfsson2014power,Praprotnik_Matysiak_Site_Kremer_Clementi_2008,Praprotnik_DelleSite_Kremer_2005}. Coarse-grained (CG) simulations are one such multiscale approach that enables access to spatiotemporal scales entirely out of reach of conventional atomistic molecular dynamics simulations. The process of coarse-graining typically aggregates groups of atoms into `beads' or `superatoms' intended to preserve important properties of the original atomistic system~\cite{voth2008coarse, noid2013perspective, brini2013systematic}. As such, CG simulations require monitoring fewer particles which allows for the study of larger and more complex systems typically untenable in the atomistic regime at comparable computational cost. A typical consequence of this reduction in resolution is an effective `smoothing' of the underlying free energy surface, which helps expedite large-scale and slowly evolving conformational motions that might otherwise be frustrated or kinetically trapped in the more rugged atomistic landscape~\cite{depa2005speed,depa2011coarse,ingolfsson2014power,marrink2007martini,marrink2013perspective} These advantages have led to the growing popularization and use of CG models particularly for simulations of proteins, polymers, molecular self-assembly, membranes, and high-throughput screening~\cite{nielsen2004coarse, baaden2013coarse, wu2011coarse, zhang2019hierarchical, srinivas2004self, shmilovich2020discovery,mohr2022data,sadeghi2020large,bradley2013coarse,wang2018coarse}.

The primary concession of coarse-graining is the sacrifice of fine-grained, atomistic detail. Restoring this lost detail by converting a CG representation into a corresponding atomistic representation is commonly dubbed ``backmapping'' and is important for analyses requiring atomistic resolution, for example electronic structure calculations for determining NMR spectra or dipole moments~\cite{helgaker2014molecular, mcquarrie1997physical}. A major difficulty of backmapping is the inherent ambiguity of many atomistic structures that are feasible candidates to a singular CG configuration. Traditional backmapping strategies rely on geometric heuristics to replace beads with their associated atomic fragments. These approaches typically produce quite poor initial structures that must be subsequently subjected to refinement using energy minimization and/or (restrained) molecular dynamics to equilibrate each backmapped frame~\cite{peter2009multiscale, hess2006long, rzepiela2010reconstruction, wassenaar2014going, Heath_Kavraki_Clementi_2007}. However, significant computational cost is incurred with the required frame-by-frame intervention in these approaches which hinders the applicability of backmapping larger systems and/or longer trajectories. More recently, data-driven backmapping techniques have been proposed which deploy machine learning (ML) models that learn to reconstruct atomistic details from training examples~\cite{stieffenhofer2020adversarial, stieffenhofer2021adversarial, wang2019coarse, li2020backmapping, an2020machine, wang2022generative}. These approaches offer more scalability with higher throughput as they are typically trained to produce well-equilibrated structures that do not require frame-by-frame energy minimization or relaxation. Coarse-graining is an inherently many-to-one operation, with multiple atomic structures corresponding to each CG representation. A favorable feature of any backmapping procedure is the capacity to recapitulate the conformational diversity of atomic structures corresponding to a particular CG representation. A subset of these data-driven methods ~\cite{stieffenhofer2020adversarial,stieffenhofer2021adversarial, wang2022generative, li2020backmapping} that possess this conformational expressibility are therefore capable of non-deterministic backmapping, where a variety of feasible and novel atomistic structures can be generated when backmapping any individual CG configuration. 

A commonality between all existing approaches is backmapping each frame individually and separately from one-another. However, leveraging temporal information can improve reconstruction quality and enable the recovery of dynamic properties. In particular, some important dynamic properties rely on time correlations of local atomistic details. For example, the calculation of diffusion constants is related to the integral of the velocity auto-correlation~\cite{frenkel2001understanding}, infrared absorption spectra is related to the auto-correlation function of the total dipole moment~\cite{bergsma1984electronic, guillot1991molecular}, and scattering functions that are related to Fourier transforms of the van Hove correlation function~\cite{PhysRevE.53.2382, moe1999calculation, chen2008comparison, arbe2012neutron}. Existing backmapping schemes are not temporally aware and correlations between consecutive frames are only maintained via large-scale characteristics. As a consequence, the reintroduced degrees of freedom between consecutive frames might decorrelate locally, and time correlations based on local, atomistic descriptors are typically not reliable for such backmapped trajectories. Therefore, presently absent from this suite of backmapping methods is a data-driven approach for generatively backmapping CG trajectories that also incorporates temporal information.

We present in this work a new method to perform temporally coherent backmapping of molecular simulation trajectories via a deep learning-based solution for all-atom reconstruction of CG simulation trajectories that aims at both generating well-equilibrated molecular structures for each frame and achieves temporal coherence between frames. To this end, we explicitly incorporate configurational information from previous simulation frames when generating reconstructions. This task is accomplished by training a conditional variational autoencoder (cVAE) that learns to up-scale CG configurations into full atomistic resolution by conditioning on the current coarse- and previous fine-grained structures. Our cVAE learns to model a variety of feasible atomistic structures associated with each CG configuration, which allows us to generatively produce novel backmapped trajectories that are not a simple carbon copy of the training data. We show for two exemplar biomolecular systems alanine dipeptide (ADP) and the miniprotein chignolin (CLN) that our approach generates atomic reconstructions which recover atomistic structural, thermodynamic, and kinetic properties. Furthermore, we show our backmapping model performs well on held out in distribution data and generalizes to CG data originating from unseen and approximate CG force fields.

\section{\label{sec:methods}Methods}
\begin{figure}[H]
    \centering
    \includegraphics[width=\linewidth]{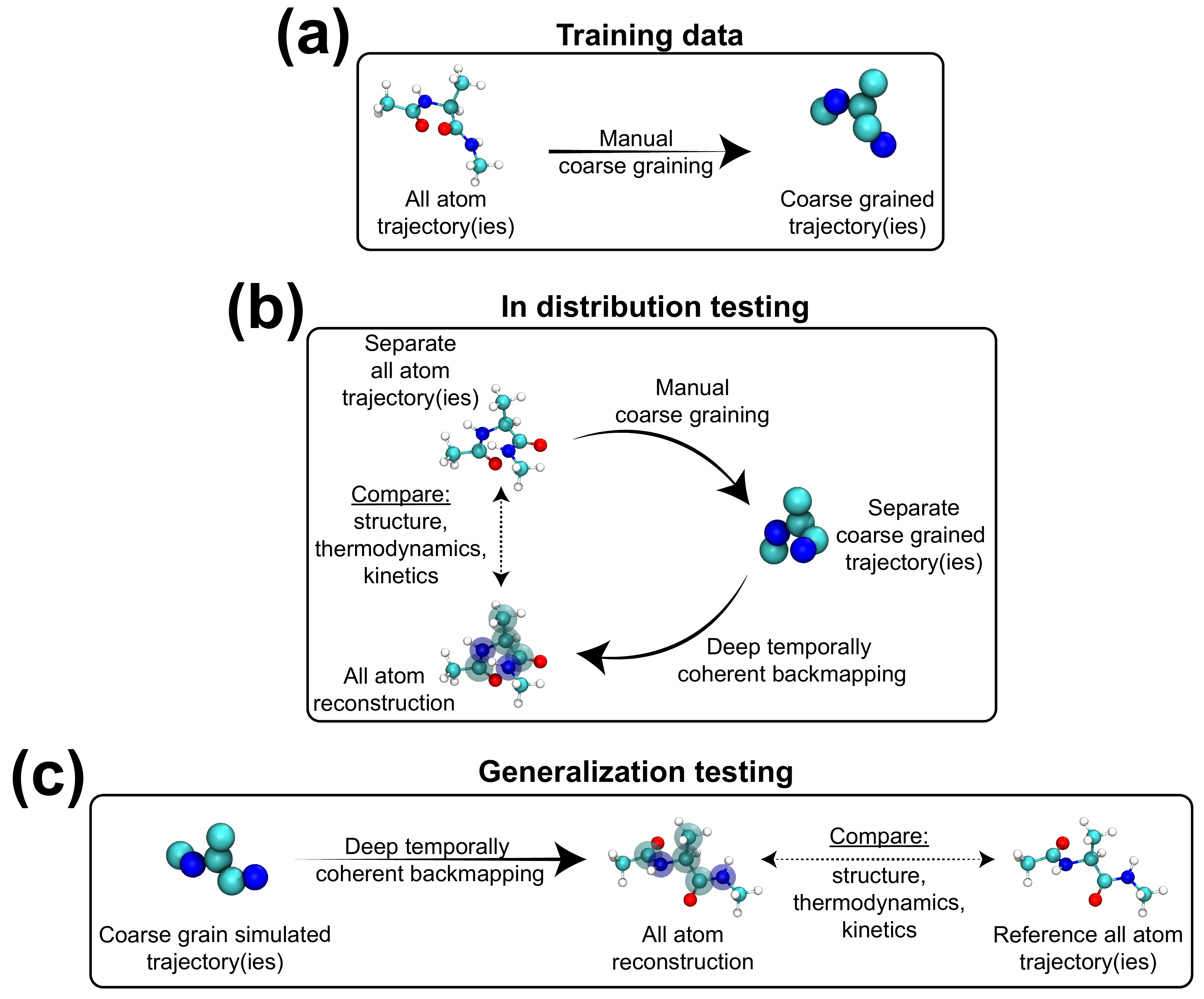}
    \caption{Illustration of training and testing setups.}
    \label{fig:setup}
\end{figure}

Our approach utilizes a reference atomistic trajectory (or set of atomistic trajectories) for a molecule we intend to backmap. The trajectory contains $N$ atoms and is composed of a collection of $T$ frames $\mathbf{ AA } = \{\AAtom^0, \AAtom^1, \ldots, \AAtom^{T-1}\}$, where $\AAtom^t \in \mathbb{R}^{N \times 3}$ are the atomistic coordinates for each frame. We assume there exists a coarse-graining function $f_{cg}$ that maps all-atom coordinates to CG coordinates, such that $f_{cg}(\AAtom^t) = \CG^t \in \mathbb{R}^{n \times 3}$, where $n$ is the number CG beads such that $n < N$. This function is then applied frame-by-frame to the atomistic frames $\mathbf{AA}$ yielding the corresponding CG trajectory $\mathbf{CG} = \{\CG^0, \CG^1, \ldots, \CG^{T-1}\}$. This pair of atomistic and CG trajectories $(\mathbf{AA}, \mathbf{CG})$ compose the data used to train our super-resolution model (Fig.~\ref{fig:setup}a). From this data our model attempts to learn the conditional distribution $P(\AAtom^t |\CG^{t}, \AAtom^{t-1})$ implicitly, such that we can reconstruct an atomisitic configuration given the CG structure $\CG^t$ and a previous atomistic structure $\AAtom^{t-1}$. Each training sample therefore consists of a sequential pair of atomistic configurations and the current CG configuration $(\AAtom^t, \CG^t, \AAtom^{t-1})$. While increasing the number of previous frames to incorporate $\AAtom^{t-2},\AAtom^{t-3},\ldots$ is straight-forward, our experiments show that this simpler Markovian posture already yields accurate temporally coherent reconstructions that reproduce structural, thermodynamic and kinetic properties with remarkable accuracy. A complete PyTorch~\cite{paszke2019pytorch} implementation of our model with all associated analyses is publically available at DOI:10.18126/tf0h-w0jz~\cite{MDF}.  

\subsection{Representing molecules as spatially voxelized particle densities}
\label{sec:vox}

Learning complex and high-order dependencies is a hallmark of computer vision. One of the most successful generative models are convolutional neural networks (CNNs), which have led to groundbreaking successes in image processing~\cite{fukushima1983neocognitron, lecun1989backpropagation, lecun1998gradient, rawat2017deep, voulodimos2018deep}.
In order to take advantage of CNNs for our backmapping task we choose to represent our data as a set of 3D featurized images. To this end, a smooth density representation discretized on a 3D grid is used to encode the positions of atoms and beads. Each particle is placed in a separate tensor channel to avoid overlap of densities that would be difficult to disentangle. More concretely, an atomistic configuration $\AAtom^t \in \mathbb{R}^{N \times 3}$ is represented as a 4D tensor $\VOX^{\AAtom^t} \in \mathbb{R}^{d \times d \times d \times N}$, where the first three dimensions discretize the location of each particle in space and the final dimension represents a channel for each particle:
\begin{equation}
    \VOX^{\AAtom_t}_{i}(\mathbf{x}) = \exp\left(-\frac{(\mathbf{x} - \AAtom^t_i)^2}{2 \sigma ^2}\right).
\end{equation}
Here, $i$ is the particle index and $\mathbf{x} \in \mathbb{R}^3$ is the Cartesian location of each grid point taken from a regular Cartesian grid of width $r_{grid}$ such that the maximal and minimal values of each spatial dimension are $-\frac{r_{grid}}{2}$ and $\frac{r_{grid}}{2}$. The density of each particle is therefore a 3D Gaussian centered about the position $\AAtom_i^t \in \mathbb{R}^3$ with the width $\sigma$. It is critical to ensure $r_{grid}$ is large enough such that each particle density within the trajectory is fully enclosed by the voxelized grid. The parameter $d$ then controls the resolution of our spatial discretization, such that a larger $d$ leads to more spatial resolution at the cost of higher memory and processing requirements. Furthermore, $\sigma$, which controls the effective size of each particle density, will also impact the mass assigned to each voxel. 

From a density profile $\VOX^{\AAtom^t}_i$ we can also recover a set of particle coordinates by performing a weighted average over the voxelized grid,
\begin{equation}
    \Tilde{\AAtom^t_i} = \frac{1}{z}\sum_{j=1}^d \sum_{k=1}^d \sum_{l=1}^d x_{j,k,l}\VOX^{\AAtom^t}_{i}(x_{j,k,l})
\end{equation}
where $z = \sum_{j=1}^d \sum_{k=1}^d \sum_{l=1}^d \VOX^{\AAtom^t}_{i}(x_{j,k,l})$ is a normalization constant, $x_{j,k,l}$ is a particular coordinate value within the three-dimensional spatially voxelized grid and the summations are carried over the three spatial dimensions. Given a predicted density profile $\VOX^{\hat{\AAtom}^t}$ we can perform this weighted average for each $i$ particle channel $\VOX^{\hat{\AAtom}^t}_i$ to recover the complete set of atomistic coordinates $\hat{\AAtom^t}$. We note that transforming these densities into and from Cartesian coordinates constitutes a sequence of differentiable operations and therefore enables us to readily incorporate this particle averaging and voxelization within the computational graph of our model. CG configurations $\CG^t$ are treated in the same way as their atomistic counter-parts, yielding corresponding voxelized representations $\VOX^{\CG^t} \in \mathbb{R}^{d \times d \times d \times n}$ where $n$ is the number of beads associated with each CG configuration.    

\subsection{Conditional Variational Autoencoder}
\begin{figure}
    \centering
    \includegraphics[width=\linewidth]{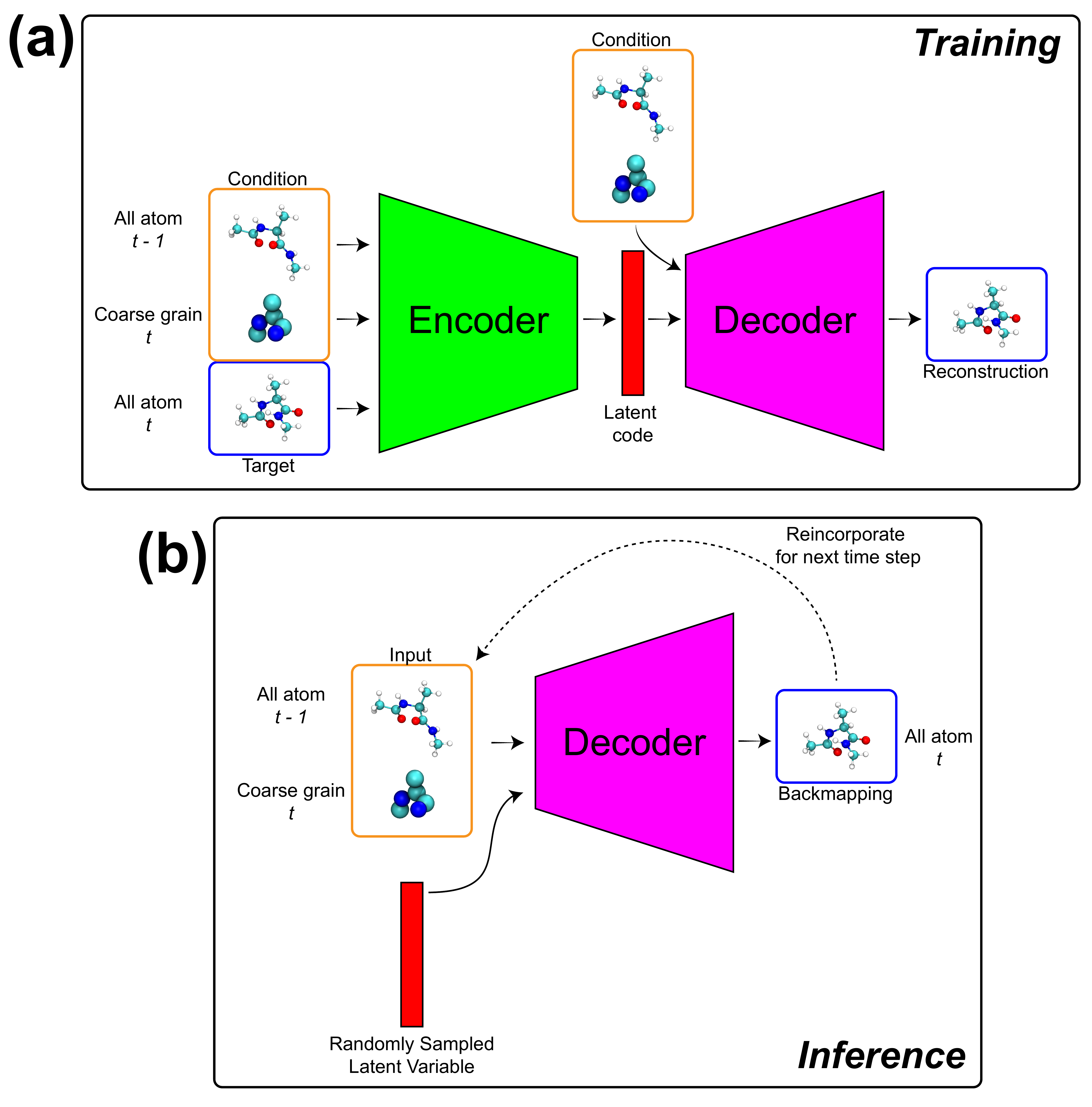}
    \caption{Schematic illustration of our model training and inference setups.}
    \label{fig:model}
\end{figure}

To learn a temporally coherent probabilistic mapping from CG to atomistic configurations we train a conditional variational autoencoder (cVAE)~\cite{sohn2015learning}. A typical VAE~\cite{kingma2013auto} consists of an encoder that compresses high-dimensional inputs into a lower dimensional latent space that captures salient information characterizing the input data. This latent, compressed representation is then inputted to a decoder that aims to reconstruct the original high-dimensional input. The cVAE operates under a similar premise but the decoder is also provided with some partial information about the input along with the latent code when producing reconstructions. Fig.~\ref{fig:model}a shows a schematic illustration of our cVAE architecture. In our application, the cVAE encoder takes as input the triplet of configurations $(\AAtom^t, \CG^t, \AAtom^{t-1})$, where the subset $(\CG^t, \AAtom^{t-1})$ is interpreted as the conditional variable and $\AAtom^t$ the intended reconstruction target. The decoder is a function of both the latent code and the conditional variable $(\CG^t, \AAtom^{t-1})$ and learns to reconstruct atomistic configurations $\hat{\AAtom^t}$ to closely match the data $\AAtom^t$. For a fixed condition $(\CG^t, \AAtom^{t-1})$ we can learn to meaningfully encode information about the target configuration $\AAtom^t$ into a low dimensional latent code, such that the decoder when presented with $(\CG^t, \AAtom^{t-1})$ yields slightly different, yet valid, reconstructions $\hat{\AA^t}$ for different instantiations of the latent code. This data driven approach enables our model to generatively produce reconstructions $\AAtom^t$ in a temporally aware way by conditioning on both the coarse grained configuration and the previous atomistic configuration $(\CG^t, \AAtom^{t-1})$ (Fig.~\ref{fig:model}b).

\subsubsection{The Encoder}
The purpose of the encoder in our cVAE is to distill information from the input configurations $(\AAtom^t, \CG^t, \AAtom^{t-1})$ into a low-dimensional latent space vector that is used in conjunction with the conditional variables $(\CG^t, \AAtom^{t-1})$ for the decoder to generate an atomistic reconstruction $\hat{\AAtom}^t$. These configurations are passed to the encoder as voxelized grids of particle densities $(\VOX^{\AAtom^t}, \VOX^{\CG^t}, \VOX^{\AAtom^{t-1}})$ which are constructed as a single monolithic tensor concatenated along the particle dimension $\mathbf{x} = [\VOX^{\AAtom^t}||\VOX^{\CG^t}||\VOX^{\AAtom^{t-1}}] \in \mathbb{R}^{d \times d \times d \times (2N + n)}$, where $N$ and $n$ are the number of atoms/beads in each atomistic and CG configuration, respectively. The encoder is composed of 3D residual CNNs with a terminal dense layer to extract a fixed dimensional latent vector. The locality of the CNN kernel provides a strong inductive bias by focusing on proximate spatial profiles from multiple high and low resolution scales and multiple time steps. Processing the input $\mathbf{x}$ using many consecutive CNN modules enables us to hierarchically incorporate progressively more distant features, and ultimately yielding a multiscale, spatially and temporally aware latent representation $\mathbf{z} \in d_{\text{latent}}$ for that slice of the trajectory. This latent representation $\mathbf{z}$ is then passed to the decoder in conjunction with the conditional variable $\mathbf{c} = [\VOX^{\CG^t}||\VOX^{\AAtom^{t-1}}]$ to predict the all atom configuration for the subsequent time step $\text{Decoder}(\mathbf{c} , \mathbf{z}) = \hat{\AAtom^{t}}$. 

\subsubsection{The Decoder}
The purpose of the decoder is to use information provided in the conditional variable $\mathbf{y} = [\VOX^{\CG^t}||\VOX^{\AAtom^{t-1}}]$ jointly with the latent code $\mathbf{z}$ produced by the encoder to reconstruct of the current atomistic configuration $\AAtom^{t}$. Upon completing training, we can eschew the encoder and use the decoder as our generative model for backmapping. Using a randomly sampled latent variable $\mathbf{z}$ as a source of noise while specifying the conditional variable as the current CG configuration $\CG^t$ and the previous atomistic configuration $\AAtom^{t-1}$ we can backmap the atomistic reconstruction $\hat{\AAtom^t}$. Similar to the encoder, our decoder is primarily composed of residual 3D CNNs. Unlike the encoder, the neural network backbone of the decoder needs to only output a 4D tensor of equivalent dimensionality to the voxelized atomistic representation $\VOX^{\AAtom^t} \in \mathbb{R}^{d\times d \times d \times N}$. As the voxelized representation $\VOX^{\AAtom^t}$ contains separate channels for each atom, we can simply perform an average over of the spatial density profiles within each particle channel to independently localize each atom coordinate (c.f. Sec.~\ref{sec:vox}). We control the behavior of our model by defining a training loss that captures relevant aspects we ultimately want reflected in our reconstructions. 

\subsection{Training routines} \label{sec:train}

Our model is trained end-to-end using the ADAM optimizer~\cite{kingma2014adam}. For a single sample, the complete loss $\mathcal{L}$ is given by
\begin{align}
    \mathcal{L} &= \mathcal{L}_{\VOX} + \mathcal{L}_{\AAtom} + \mathcal{L}_{\text{CG}} + \mathcal{L}_{\text{EDM}} + \lambda  \mathcal{L}_{\text{ENERGY}} + \beta \mathcal{L}_{\text{KLD}} \label{eqn:loss} \\
    \mathcal{L}_{\VOX} &= \frac{1}{d^3N}\ || \VOX^{\AAtom^t} - \VOX^{\hat{\AAtom^t}} ||_F^2  \notag \\
    \mathcal{L}_{\AAtom} &= \frac{1}{3N}\ || \AAtom^t - \hat{\AAtom^t} ||_F^2 \notag \\
    \mathcal{L}_{\text{CG}} &= \frac{1}{3n}\ || \CG^t - \hat{\CG^t} ||_F^2 \notag \\
    \mathcal{L}_{\text{EDM}} &= \frac{1}{2N^2}\ || EDM^{\AAtom^t} - EDM^{\hat{\AAtom^t}} ||_F^2 \notag \\
    \mathcal{L}_{\text{ENERGY}} &= (U^{\AAtom^t} - U^{\hat{\AAtom^t}})^2 \notag \\
    \mathcal{L}_{\text{KLD}} &= \mathcal{D}_{KL}(\mathbf{z}||\mathcal{N}(0,\mathbf{I})). \notag 
\end{align}
The first five terms in Eqn.~\ref{eqn:loss} are components of an effective reconstruction loss and $\mathcal{L}_{\text{KLD}}$ is the standard Kullback–Leibler divergence loss between the latent code $\mathbf{z}$ and a normal distribution typically used in VAE training~\cite{kingma2013auto}. While we ultimately strive to recover all atom coordinates $\AAtom^t$, the model primarily operates on spatially voxelized particle densities representations $\VOX^{\AAtom^t}$. A critical component of learning will therefore involve reconstructing these density profiles to enable accurate and sharp atomic coordinate generation. The $\mathcal{L}_{\VOX}$ term is a mean squared error (MSE) between the target $\VOX^{\AAtom^t}$ and reconstructed $\VOX^{\hat{\AAtom^t}}$ atomistic voxels, while $\mathcal{L}_{\AAtom}$ is an MSE between the ground truth $\AAtom^t$ and reconstructed $\hat{\AAtom^t}$ atomic coordinates. $\mathcal{L}_{\VOX}$ helps the model learn to reproduce the atomic densities of the intermediate voxelized representations, while $\mathcal{L}_{\AAtom}$ helps ensure sharper coordinate reconstruction when the voxels $\VOX^{\AAtom^t}$ are ultimately collapsed back into atomic coordinates $\AAtom^t$. We also use the coarse graining function $f_{cg}$ to determine the CG representation of an atomistic reconstruction $\hat{\CG^t} = f_{cg}(\hat{\AAtom^t})$, which is used in $\mathcal{L}_{\text{CG}}$ to calculate an MSE with respect to the input CG coordinates $\CG^t$. The motivation to include $\mathcal{L}_{\text{CG}}$ is that a coarse graining of the atomistic backmapping $\hat{\AAtom^t}$ should ultimately match the original CG structure $\CG^t$ from which the reconstruction is derived. In the $\mathcal{L}_{\text{EDM}}$ term we calculate an MSE between the $N \times N$-dimensional Euclidean Distance Matrix (EDM) of the target $EDM^{\AAtom^t}$ and the reconstructed $EDM^{\hat{\AAtom^t}}$ atomic coordinates to help our backmapping better preserve bond lengths and other interatomic distances. Lastly, in the $\mathcal{L}_{\text{ENERGY}}$ term we calculate an MSE between the scalar valued total potential energy of the target $U^{\AAtom^t}$ and reconstruction $U^{\hat{\AAtom^t}}$. $\mathcal{L}_{\text{ENERGY}}$ serves as a regularizer to improve the quality of backmapped structures by penalizing reconstructed configurations that may have suitable geometric contributions but are otherwise energetically unfavorable. As such, it accelerates convergence and helps more precisely match the reconstructed energetics to the ground truth trajectory. During training, before configurations are voxelized, each training configuration is mean centered and we learn covariance with respect to rigid rotations by applying a random Euler rotation $\mathbf{R} \in \mathbb{R}^{3\times3}$ separately augmenting samples in each forward pass $(\AAtom^t, \CG^t, \AAtom^{t-1}) \rightarrow (\AAtom^t\mathbf{R}, \CG^t\mathbf{R}, \AAtom^{t-1}\mathbf{R})$. 

A challenge of incorporating $\mathcal{L}_{\text{ENERGY}}$ within the loss function is that the potential energy function $U$ is sensitive to small perturbations of the atomic coordinates, which is most severe for the bonded and non-bonded Lennard-Jones interaction. As such, it can become dominatingly large during the early stages of training before the model learns to stably localize atomic coordinates. To alleviate this issue, we incorporate the prefactor $\lambda$ in Eqn.~\ref{eqn:loss} which we set to $\lambda=0$ for a fixed number of initial training steps after which point $\lambda$ is slowly annealed up to $\lambda=1$ using an exponential annealing schedule. We also include the $\beta$ prefactor alongside $\mathcal{L}_{\text{KLD}}$ for more flexibility in balancing the impact of the KL regularization against the reconstruction losses. For the ADP model we employ a cyclic annealing schedule for $\beta$ to mitigate KL vanishing~\cite{fu2019cyclical}, while for the CLN model we simply maintain $\beta=1$ throughout training. Complete training settings and hyperparameter details are presented in the Supporting Information.     

\subsection{Inference}

At inference time, we omit the encoder and use the decoder as the primary tool for generatively backmapping a CG trajectory. The decoder thereby reconstructs the atomistic frames in an autoregressive manner, i.e., the previous reconstructed atomistic frame $\hat{\AAtom^{t-1}}$ serves as the input for the reconstruction of the next frame $\hat{\AAtom^t}$. More specifically, the input for the decoder consists of a fixed dimensional latent vector $\mathbf{z}$ and the conditional variable $\mathbf{c} = [\VOX^{\CG^t}||\VOX^{\AAtom^{t-1}}]$ (Fig.~\ref{fig:model}b).
However, for the first trajectory frame $t=0$ there is no preceding frame for us to ascertain $\AAtom^{-1}$. In this case, an atomistic configuration from the training dataset is chosen as this initial seed configuration $\AAtom^{-1}$. We select this configuration by first determining the trajectory frame $t^{*}$ in the training dataset $\AAtom_{\text{train}}$ that minimizes the RMSD with respect to the first test set CG frame $t^{*} = \underset{i}{\text{min}}\ \text{RMSD}(f_{cg}(\AAtom^{i}_{\text{train}}), \CG^{0})$, and then simply set the initial seed configuration as the immediately preceding training set trajectory frame $\AAtom^{-1} = \AAtom^{t^{*}-1}_{\text{train}}$ and lastly apply a rigid rotation to align $\AAtom^{-1}$ with $\CG^{0}$. 

At inference time the decoder uses a fixed dimensional latent vector $\mathbf{z}$ to provide a source of variance when generating atomic reconstructions. For a fixed condition $\mathbf{c}$ our model learns to effectively produce valid yet slightly different atomic reconstructions $\hat{\AAtom^t} = \text{Decoder}(\mathbf{c}, \mathbf{z})$ for different instantiations of the latent code $\mathbf{z}$. Typically, inference is performed for VAEs or cVAEs by simply sampling $\mathbf{z}$ for each inference pass from the prior distribution over $\mathbf{z}$ which is normally taken as an isotropic Gaussian $p(\mathbf{z}) \sim \mathcal{N}(\mathbf{0}, \mathbf{I})$~\cite{kingma2013auto}. At training time, however, the decoder is exposed to latent codes produced by the encoder from only the training set $\mathbf{Z} = \{\mathbf{z} = \text{Encoder}(\mathbf{x})\ : \ \mathbf{x} \in \mathbf{X}_{\text{train}}\}$. Therefore, any mismatch between the true aggregated posterior $\mathbf{Z}$ and the assumed prior $p(\mathbf{z})$ can lead to poor generative performance by the decoder when samples are selected from $p(\mathbf{z})$ because this may lead to operating within regions of latent space previously unseen by the decoder during training. We remedy this issue, following Ref.~\cite{ghosh2019variational}, by performing $\textit{ex-post}$ density estimation fitting a 10-component Gaussian Mixture Model (GMM) over the true posterior $\mathbf{Z}$. We then at inference time randomly sample $\mathbf{z}$ from our fit GMM, instead of from the assumed prior $p(\mathbf{z}) \sim \mathcal{N}(\mathbf{0}, \mathbf{I})$. This process ensures our decoder operates within densely sampled latent space regions, which leads to higher fidelity, and less error-prone, atomic reconstructions.

\begin{figure}[H]
  \centering
  \includegraphics[width=\linewidth]{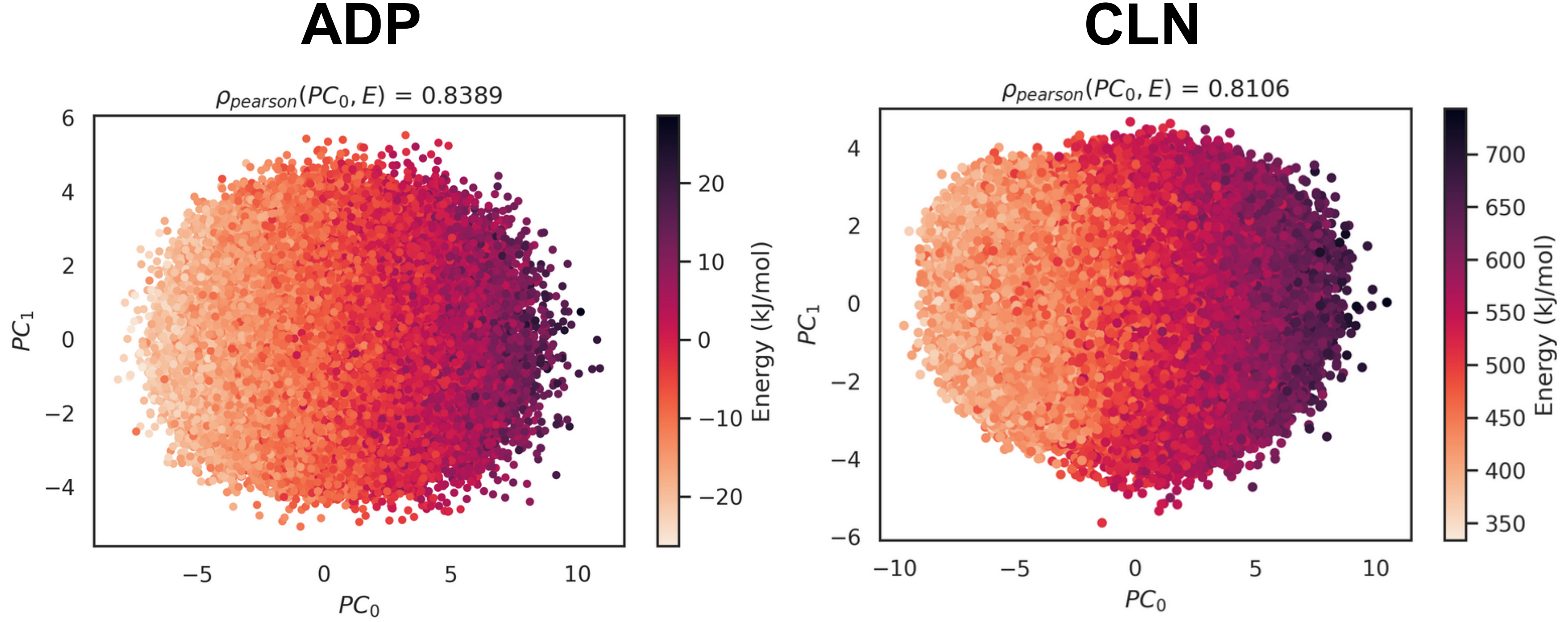}
  \caption{Projection of the full dimensional latent space embedding of all our \textbf{(a)} ADP and \textbf{(b)} CLN training data into the two leading principal components color-coded by the target atomistic configuration internal potential energy.}
  \label{fig:latent_energy}
\end{figure}

After completing training for both our ADP and CLN models, we can visualize our latent space posterior $\mathbf{Z}$ by projecting our full-dimensional latent space into the first two Principal Components (PCs) of $\mathbf{Z}$ (Fig.~\ref{fig:latent_energy} in the Supporting Information). Color-coding latent codes by the potential energy of the corresponding target configuration $U_{target}$ in each sample reveals a strong correlation between the leading PC and internal energy for both of our test systems ADP ($\rho_{\text{pearson}}(PC_0^{ADP}, U_{target}) \approx 0.84$) and CLN ($\rho_{\text{pearson}}(PC_0^{CLN}, U_{target}) \approx 0.81$). These strong correlation suggests our encoder effectively extracts features from the input data to encode physically meaningful features, reflected in the internal energy, of the target atomistic configuration within our latent space embedding. We can also interrogate the generative capabilities of our model by using our decoder to produce reconstructions of a fixed CG input under many different latent code instantiations (Fig.~\ref{fig:latent_generation} in the Supporting Information). The diversity of valid generated structures which still adhere to the CG input suggest that the decoder is attentive to subtle variations in the latent code when reconstructing atomistic configurations. In all, these properties of our cVAE model and latent space embedding validate expected behavior: our encoder effectively distills information to the latent code from the input data, while our decoder purposes the latent code together with the condition to non-deterministically and generatively reconstruct atomistic structures.

\subsection{Data curation}
Our model learns to backmap a CG trajectory of a molecule by training on a reference atomistic trajectory of that molecule that we coarse-grain after the fact to yield exemplar pairs of atomistic and coarse grained frames (Fig.~\ref{fig:setup}a). When backmapping to predict an atomistic structure $\AAtom^t$ we consider both the current CG structure $\CG^t$ along with the previous atomistic structure $\AAtom^{t-1}$. An important consideration when obtaining training data is the temporal spacing between consecutive frames $\AAtom^{t-1}, \AAtom^{t}$, as it may be impossible to accurately recover molecular motions that occur faster than this time. We also find that it is critical to ensure the reference atomistic trajectory is sufficiently sampled and captures the relevant atomistic state transitions that are expected to be reflected in the backmapped atomisitic trajectory.

We train separate models on reference atomistic trajectories of alanine dipeptide (ADP) and the mini protein chignolin (CLN). Separate held-out trajectories for ADP and CLN are used as test sets for evaluating the performance of our model on in distribution data (Fig.~\ref{fig:setup}b). As a more challenging generalization test of our method, we also backmap CG trajectories generated from a bespoke CG force-field, CGSchNet (Fig.~\ref{fig:setup}c). We then evaluate structural, thermodynamic and kinetic statistics of our atomistic reconstructions against reference atomistic trajectories to evaluate performance of our method. In the following sections, we describe the details of the simulations methods used to generate the ADP and CLN trajectories we use to train and test our model.      

\subsubsection{Alanine dipeptide}
\textbf{Atomistic data}

Atomistic trajectories of alanine dipeptide (ADP) used for training are collected by performing molecular dynamics simulations in explicit solvent using OpenMM~\cite{eastman2017openmm}. We closely mimic the simulation procedures outlined Ref.~\cite{nuske2017markov}. Langevin dynamics simulations are performed with a 2 fs time-step in the NVE ensemble using the AMBER ff-99SB-ILDN force field~\cite{lindorff2010improved} within a cubic box containing 651 TIP3P water molecules randomly placed within a volume of (2.7273~nm)$^3$. Electrostatics are treated using the particle-mesh Ewald (PME) method~\cite{darden1993particle} using a 1.0 nm cutoff for the direct space interactions. The length of all bonds involving hydrogen atoms are constrained. Steepest descent energy minimization is used to relax the initial system configuration to within an energy of 10 kJ/mol. We then assign initial velocities to the energy minimized configuration by sampling from a Maxwell-Boltzmann distribution at 300 K. A short 100 ps equilibration run is then performed followed by a 500 ns production run. This 500 ns production run comprises our training data. A separate 250 ns production run is performed to generate the in distribution data used for testing. In each case trajectory snapshots are saved every 1 ps, yielding 500,000 simulation frames for the training data and 250,000 simulation frames for the test data.

\noindent \textbf{Coarse-grained data}

For the CG representation of ADP, we choose to remove all solvent and represent the molecule using the five backbone carbon and nitrogen atoms (C, N, CA, C, N) and the carbon beta (CB) of the alanine residue, resulting in a total of six coarse grain atoms. A CG trajectory is generated from the above-mentioned all atom trajectory by slicing the coordinates to retain only the specified coarse grain atoms. The same coarse grain mapping is also applied to the all-atom forces to produce an associated set of instantaneous coarse grain forces. Using both the coarse grain coordinates and forces, bespoke CG force fields are recovered using CGSchNet~\cite{husic2020coarse} neural network models. These ADP CG models are trained using the same data set from Refs.\cite{wang2019machine,husic2020coarse}, and are then used to generate out-of-sample data in the form of CG trajectories as a generalization test for our backmapping method to illustrate performance on real, noisy data. The trajectories used for backmapping consist of 100 separate simulations initalized at random configurations from the reference atomistic data set and containing a total of 4,000 frames each. Sequential CG simulation frames are temporally separated by the same 1 ps spacing as used in the training data trajectory. The training routines and hyperparameters of these CG force field models, as well as the simulation parameters for the out-of-sample CG simulations, are described in the Supporting Information. 

\subsubsection{Chignolin}
\textbf{Atomistic data}

We use reference Chignolin (CLN) trajectories generated from atomistic simulations performed in Ref.~\cite{wang2019machine}. Briefly recapping the protocols, these simulations are performed using the ACEMD program~\cite{harvey2009acemd} on GPUgrid~\cite{buch2010high} at 350 K mimicking the setup originally used on the Anton supercomputer simulation~\cite{lindorff2011fast}. To sufficiently sample folding/unfolding transitions in CLN the data is produced through an MSM-based adaptive sampling strategy~\cite{doerr2014fly} consisting of an aggregated $\sim$187 $\mu$s of molecular dynamics simulation split into 3,744 short trajectories. Simulation snapshots are spaced by 100 ps culminating in a total of 1,868,861 frames. As each of the 3,744 trajectories are independent we simply split-off 3,650 for training and the remaining 94 for testing to comprise our in distribution test set.

\noindent \textbf{Coarse-grained data}

For a CG representation of CLN, we choose to remove all solvent and represent the molecule using just the 10 sequential $\alpha$-carbons (CA) atoms along the molecular backbone. Following the same procedure described above for ADP, a set of CG trajectories and associated forces are generated. The same CG coordinates and forces mapped from the atomistic data described in the preceding section are then used to train CGSchNet~\cite{husic2020coarse} neural network force fields, which in turn are used to generate out-of-sample data for generalization tests of our backmapping model. We produce 1,000 separate trajectories containing 4,000 frames each and, similar to ADP, are initialized at random configurations from the reference atomistic dataset. In this case for CLN each frame is temporally separated by 1 ps, which differs from the 100 ps frame spacing in the training data. While we appreciate this frame spacing represents a different regime than our model is trained to operate in, the inherently accelerated nature of CG dynamics makes direct comparison between CG and atomistic time steps difficult. We appeal to a smaller frame spacing in this work to account for this inherent acceleration of CG dynamics and enable better sampling of short-lived, transient states within these trajectories. As with ADP, the training routines, model hyperparameters, and CG simulation parameters for the CLN coarse grain force fields are detailed in the Supporting Information.

\section{\label{sec:results}Results}

We present a data-driven and temporally coherent approach for backmapping coarse grained trajectories into full atomistic resolution. Our approach is based on training a conditional variational autoencoder (cVAE) to generate atomistic coordinates of a coarse grain (CG) configuration while also incorporating information from the previous atomistic configuration within the trajectory. The proposed method is applied to two biomolecular systems: alanine dipeptide (ADP) and the miniprotein Chignolin (CLN). The performance of our model is evaluated by measuring its ability to generate backmapped trajectories that preserve atomistic structural, thermodynamic and kinetic properties. Our model is trained on data consisting of pairs of atomistic and corresponding CG trajectories, that are obtained by mapping the atomistic data to CG resolution (Fig.~\ref{fig:setup}a). For one evaluation test we apply our method to backmap CG trajectories generated from mapping separate held-out atomistic trajectories to CG resolution, which will be referred to as the \textit{in-distribution} test set, or simply referred to as the \textit{test set} (Fig.~\ref{fig:setup}b). For a more challenging test of the generalization capabilities of our model, we apply our method to CG trajectories generated using a bespoke CG force-field CGSchNet~\cite{husic2020coarse}, which will be referred to as \textit{out-of distribution} or \textit{generalization} set (Fig.~\ref{fig:setup}c). For both molecules ADP and CLN the in-distribution and generalization tests show excellent agreement in measures of structural, thermodynamic and kinetic similarity between our backmapped and atomistic trajectory data. 

\subsection{Alanine dipeptide}
\subsubsection{Energetics}

\begin{figure}
    \centering
    \includegraphics[width=\linewidth]{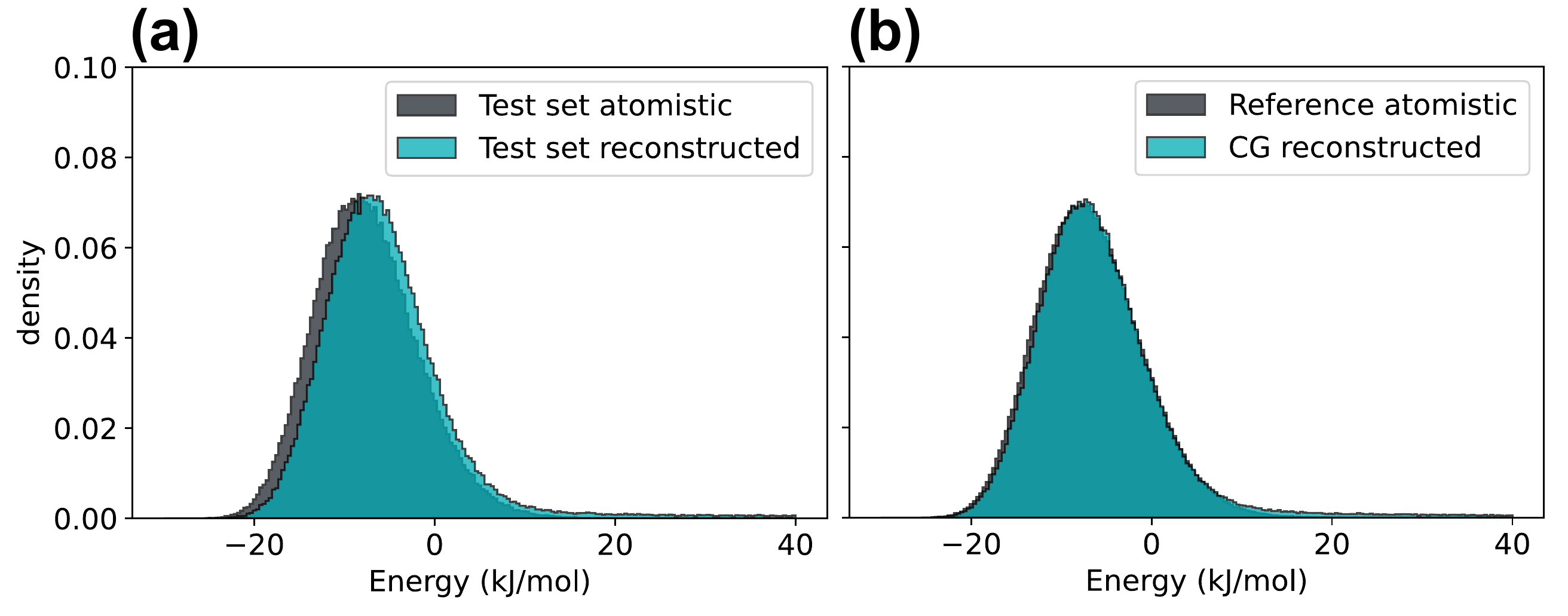}
    \caption{Comparison of the internal potential energy distributions for atomistic and backmapped ADP trajectories. \textbf{(a)} Internal energies for the held out test set trajectory and the atomistic reconstructed trajectory generated by backmapping a manual coarse graining of the original atomistic trajectory. \textbf{(b)} Generalization test comparing reference atomistic trajectories obtained from Ref.~\cite{wehmeyer2018time} to our model-generated backmappings of a real CG simulation conducted using CGSchNet~\cite{husic2020coarse}.}
    \label{fig:adp_energy}
\end{figure}

\begin{figure}[H]
  \centering
  \includegraphics[width=\linewidth]{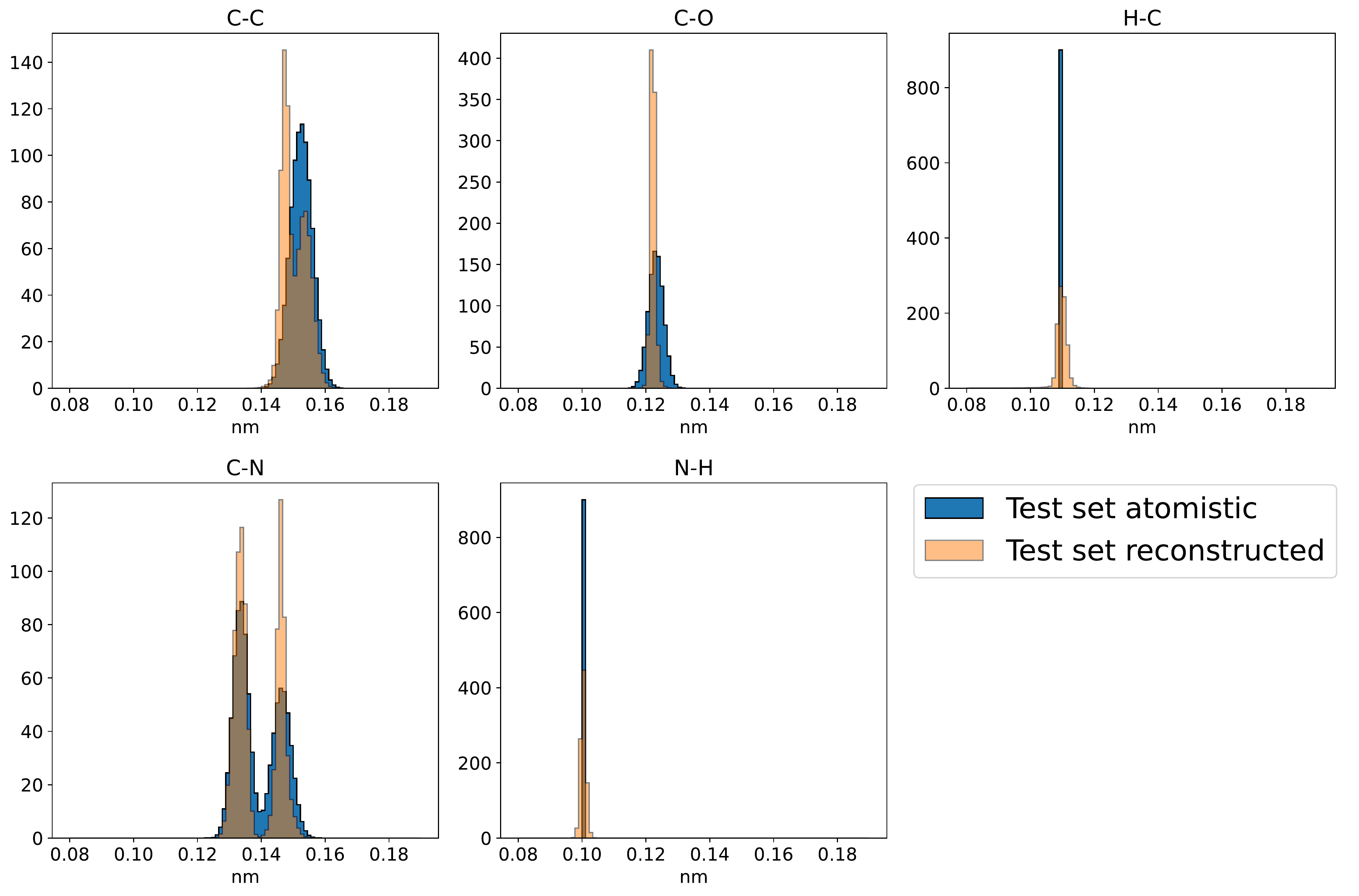}
  \caption{Bond length distributions delineated by element participation for the in distribution test set atomistic and backmapped trajectories for ADP.}
  \label{fig:adp_bonds_testset}
\end{figure}

\begin{figure}[H]
  \centering
  \includegraphics[width=\linewidth]{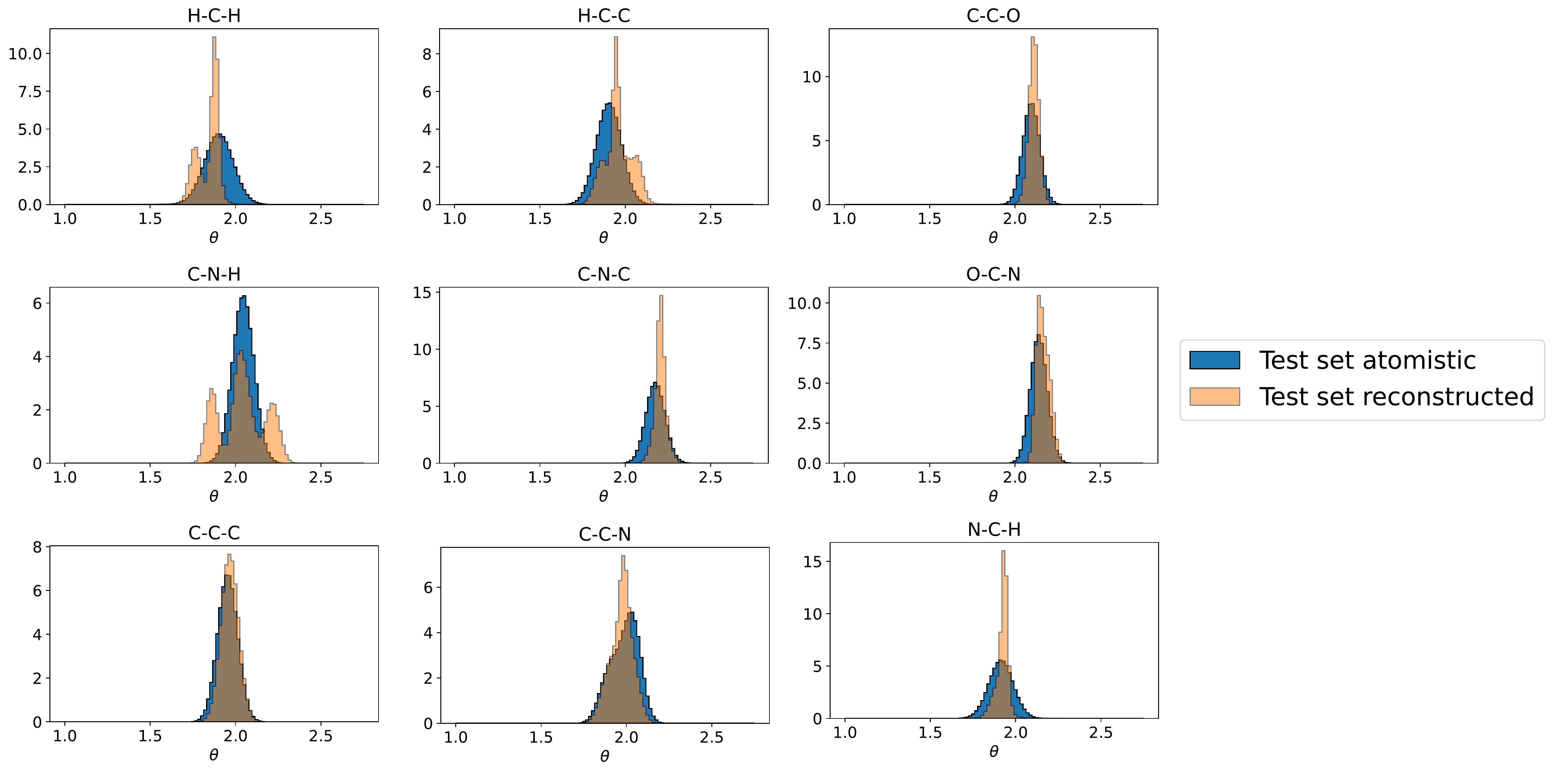}
  \caption{Atomic angle distributions delineated by element participation for the in distribution test set atomistic and backmapped trajectories for ADP.}
  \label{fig:adp_angles_testset}
\end{figure}

\begin{figure}[H]
  \centering
  \includegraphics[width=\linewidth]{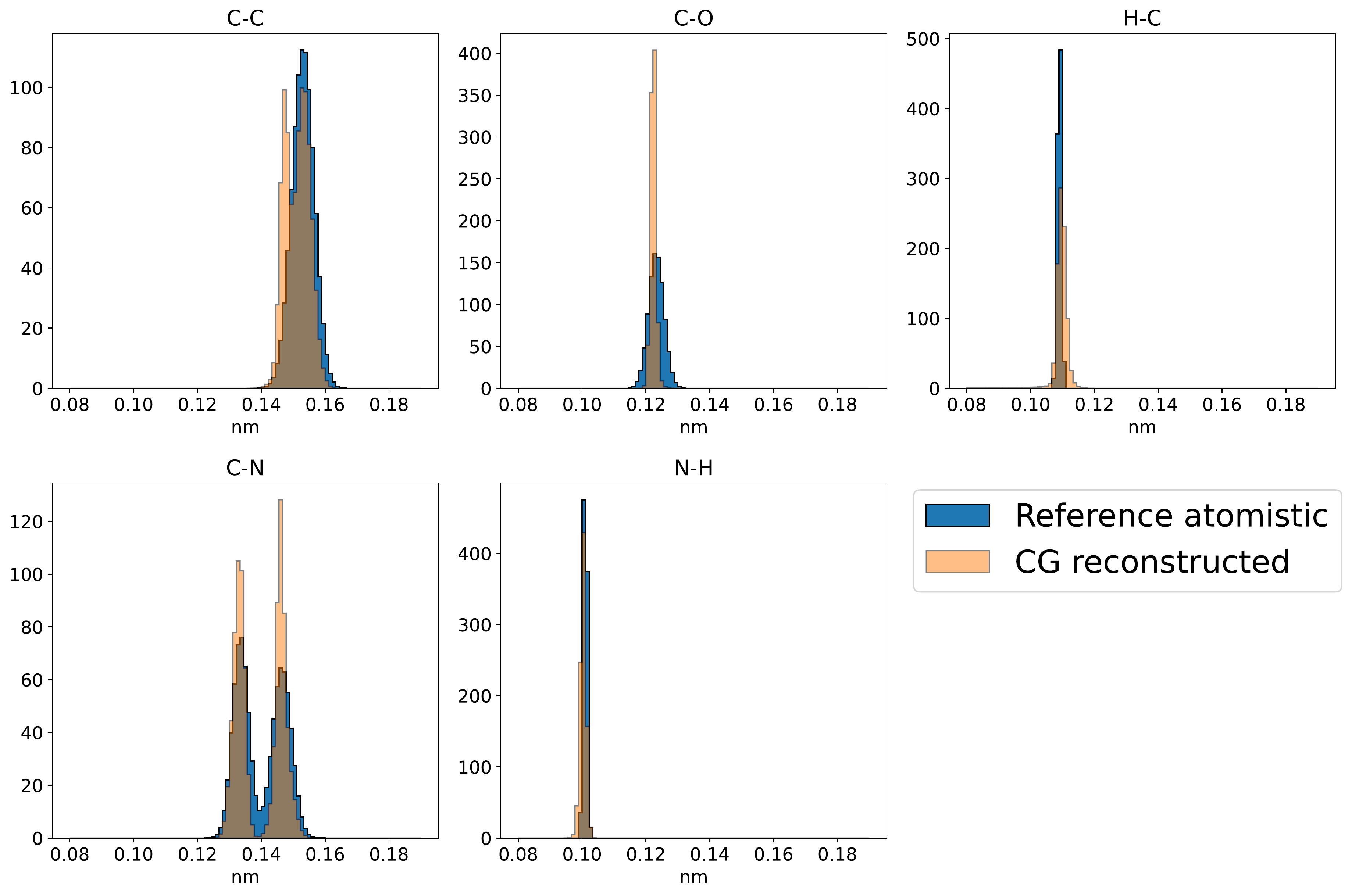}
  \caption{Bond length distributions delineated by element participation for the generalization test reference atomistic and backmapped CGSchNet trajectories for ADP.} 
  \label{fig:adp_bonds_cgset}
\end{figure}

\begin{figure}[H]
  \centering
  \includegraphics[width=\linewidth]{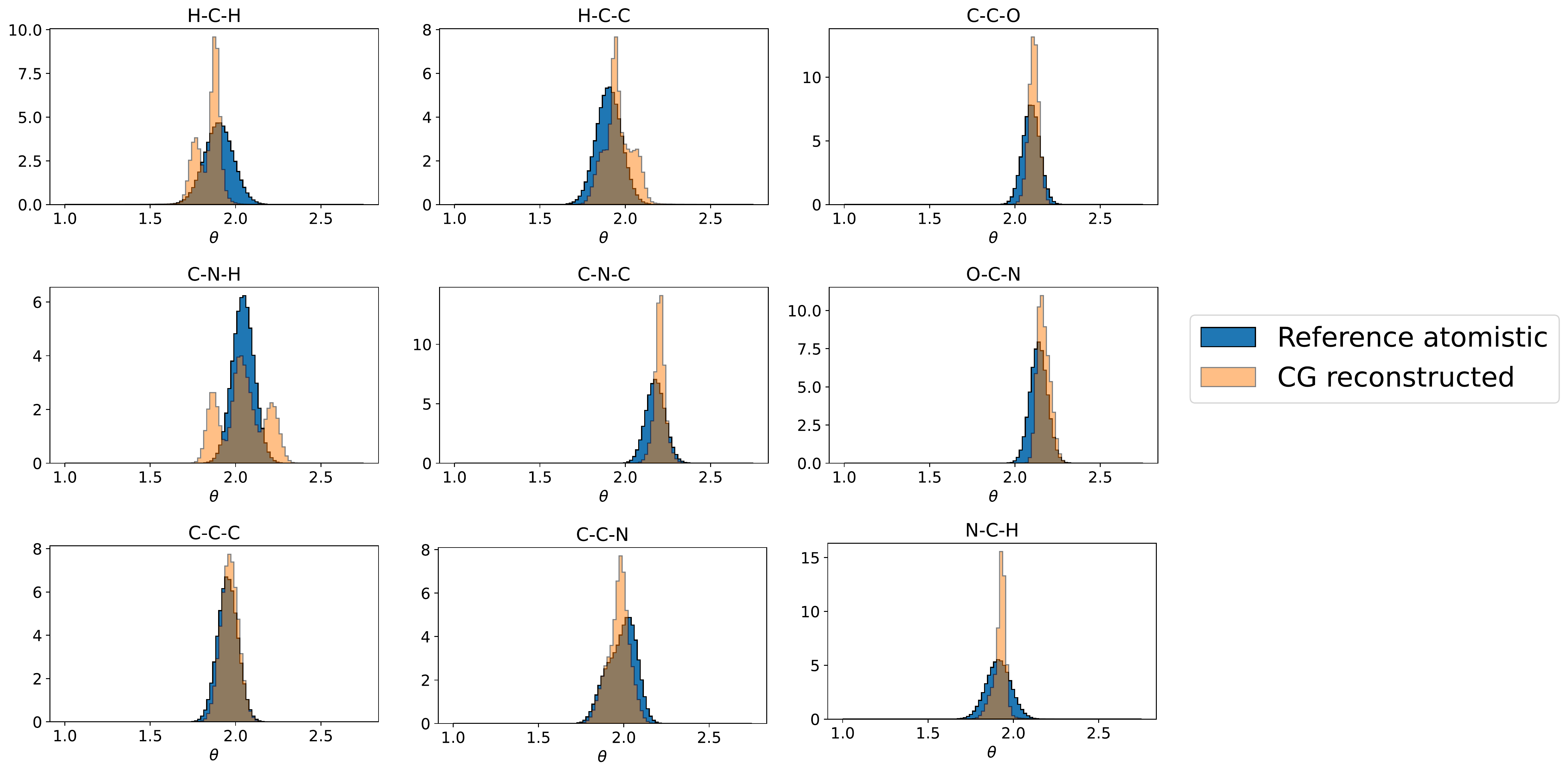}
  \caption{Atomic angle distributions delineated by element participation for the generalization test reference atomistic and backmapped CGSchNet trajectories for ADP.} 
  \label{fig:adp_angles_cgset}
\end{figure}

As a first example, we apply our backmapping method to the small molecule alanine dipeptide (ADP) composed of 22 atoms from which we consider a coarse-graining into six CG beads along the peptide backbone. We evaluate structural similarity between atomistic and our backmapped trajectories by comparing distributions of the internal potential energy. The internal energy aggregates contributions from bonded and non-bonded interactions. As such, agreement between atomistic and reconstructed energy distributions serves as a good indicator of overall structural similarity. The energy distribution for the in distribution test set shown in Fig.~\ref{fig:adp_energy}a reveals that our model nearly identically reproduces the energy distribution of the original atomistic data. Similarly, the energy distribution for the generalization test set obtained by backmapping a trajectory generated with CGSchNet~\cite{husic2020coarse} is displayed in Fig.~\ref{fig:adp_energy}b. This out-of distribution test represents a more difficult exercise for our model, as it must generalize to real CG simulated data generated by a different, approximate force field than our model was not exposed to during training. Our model yields backmapped structures that nearly identically match the energetics of the atomistic reference data. The only noticeable deviation for both test sets are small high-energy tails of the backmapped reconstructions (Fig.~\ref{fig:adp_energy}). While these rare high-energy configurations can be attributed to minor deviations in a few bond length and angle distributions, we find overall excellent agreement between distributions of local intramolecular features between backmapped and reference structures (Fig.~\ref{fig:adp_bonds_testset}-\ref{fig:adp_angles_cgset} in the Supporting Information). Our backmapping method in general reconstructs atomistic ADP trajectories with high structural and energetic similarity to reference atomistic data while also generalizing well to unseen and real CG force fields, and as a result generates visually identical backmaped structures (Fig.~\ref{fig:adp_fes}e and Fig.~\ref{fig:latent_generation} in the Supporting Information). 

\newpage
\subsubsection{Thermodynamics}
\begin{figure}[H]
    \centering
    \includegraphics[width=\linewidth]{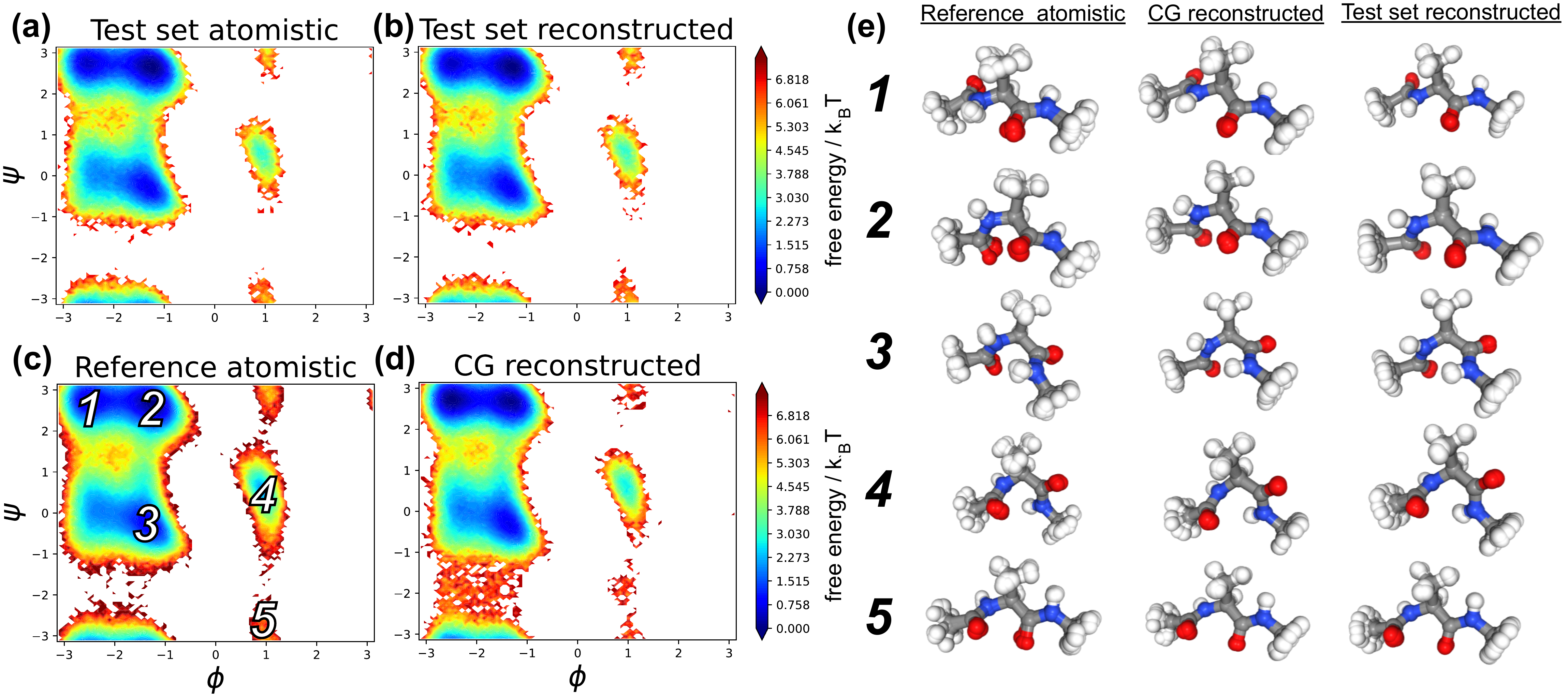}
    \caption{Comparison of atomistic and backmapped MSM-reweighted Free Energy Surfaces (FES) for ADP. Ramanchandran plots for the in distribution test set that includes \textbf{(a)} a held-out atomistic trajectory and \textbf{(b)} our model generated backmapping of the manually coarse grained atomistic trajectory. For a more challenging test of generalizability we compare the Ramachandran plots for \textbf{(c)} a reference atomistic trajectory taken from Ref.~\cite{wehmeyer2018time} and \textbf{(d)} a backmapping of a CG simulation performed using CGSchNet~\cite{husic2020coarse}. Labeled in \textbf{(c)} are phase space locations for the five meta-stable states of ADP. \textbf{(e)} Seven superimposed configurations near each of these five major meta-stable states from \textbf{(c)} the reference atomistic, \textbf{(d)} the CG reconstructed and \textbf{(b)} the test set reconstructed trajectories.}
    \label{fig:adp_fes}
\end{figure}

\begin{figure}[H]
  \centering
  \includegraphics[width=\linewidth]{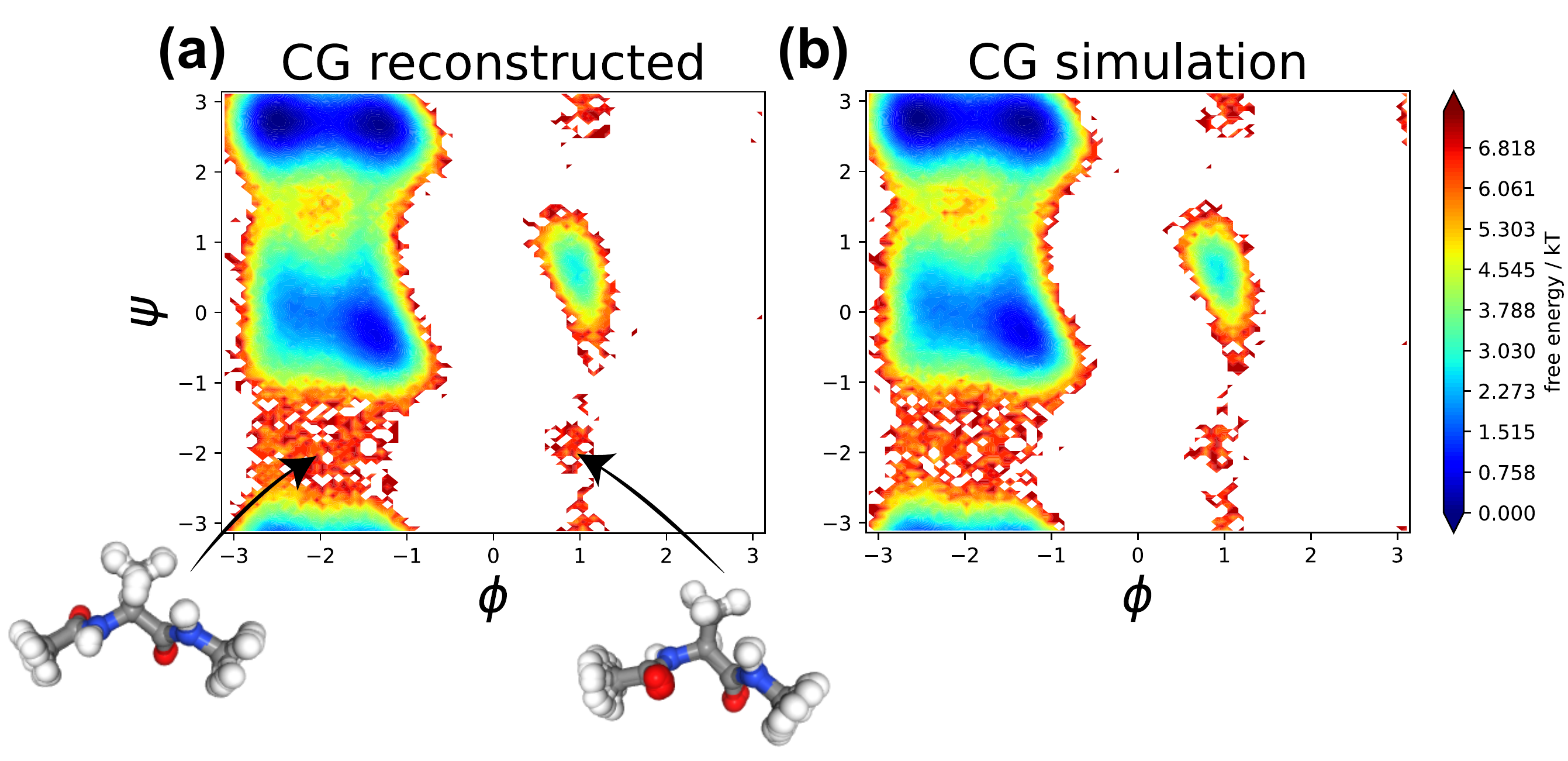}
  \caption{Comparison of MSM-reweighed ADP FES between the \textbf{(a)} the backmapped CG reconstructed trajectory and \textbf{(b)} the original CG simulation performed with CGSchNet. Insets show a superposition of seven configurations within the transition paths between select meta-stable states. Although these configurations are rarely seen in the atomistic training data, our model effectively generalizes to reliably reconstruct these high-energy configurations from the data simulated with CG force fields.} 
  \label{fig:adp_cg_recon_fes}
\end{figure}

We evaluate thermodynamic similarity by comparing free energy surfaces~(FES) of our reconstructed trajectories to reference atomistic data. We construct our FES in the space of the backbone dihedral angles $(\phi,\psi)$ as they have proven to be good collective variables for characterizing the conformational states of ADP~\cite{nuske2014variational,vitalini2015dynamic}. In Fig.~\ref{fig:adp_fes}a we present the FES in $(\phi,\psi)$ for the in distribution test set atomistic trajectory compared to the CG backmapped trajectory FES in Fig.~\ref{fig:adp_fes}b. Our atomistic reconstruction here reproduces a nearly identical FES to the ground truth atomisitic data demonstrating that for in distribution data our model accurately captures the ground truth thermodynamics. For the generalization test we also present a FES generated from a reference atomistic trajectory (Fig.~\ref{fig:adp_fes}c) compared to the FES of a CG backmapped trajectory generated from a CG simulation performed with CGSchNet~\cite{husic2020coarse} (Fig.~\ref{fig:adp_fes}d). Once again our reconstruction is in excellent agreement with the atomistic reference, importantly correctly identifying the five major meta-stable states of ADP. In Fig.~\ref{fig:adp_fes}e we show a superimposed collection of configurations for each of these meta-stable states from the reference, test set reconstructed and CG reconstructed trajectories. For both in distribution and out of distribution data our model reconstructs visually identical configurations with remarkable similarity to the atomistic reference data. Note that the CG model yields configurations throughout the transition paths between meta-stable states, for example ($\phi\approx-2$,$\psi\approx-2$). While those configurations are underrepresented in the atomistic trajectory due to high-energy barriers, the smoothed energy landscape of the CG force field enables broader and more frequent exploration of these regions of phase space. We find our model generalizes well to those sparsely sampled areas and accordingly reconstructs these high-energy configurations (Fig.~\ref{fig:adp_cg_recon_fes} in the Supporting Information). Overall, our backmapping scheme reproduces the FES for ADP that are in excellent thermodynamic agreement with reference atomistic data for both our in distribution and generalization tests. 

\subsubsection{Kinetics} \label{sec:adp_kinetics}

An important aspect of the proposed method is the incorporation of the previous trajectory configuration as a conditional input for our ML model. This provides temporal information required to achieve temporal coherence between consecutive frames, which is typically omitted in traditional backmapping strategies. Here, we test the temporal coherence of our backmapped trajectories by analyzing kinetics in terms of implied process timescales and velocity distributions.

\begin{figure}
    \centering
    \includegraphics[width=\linewidth]{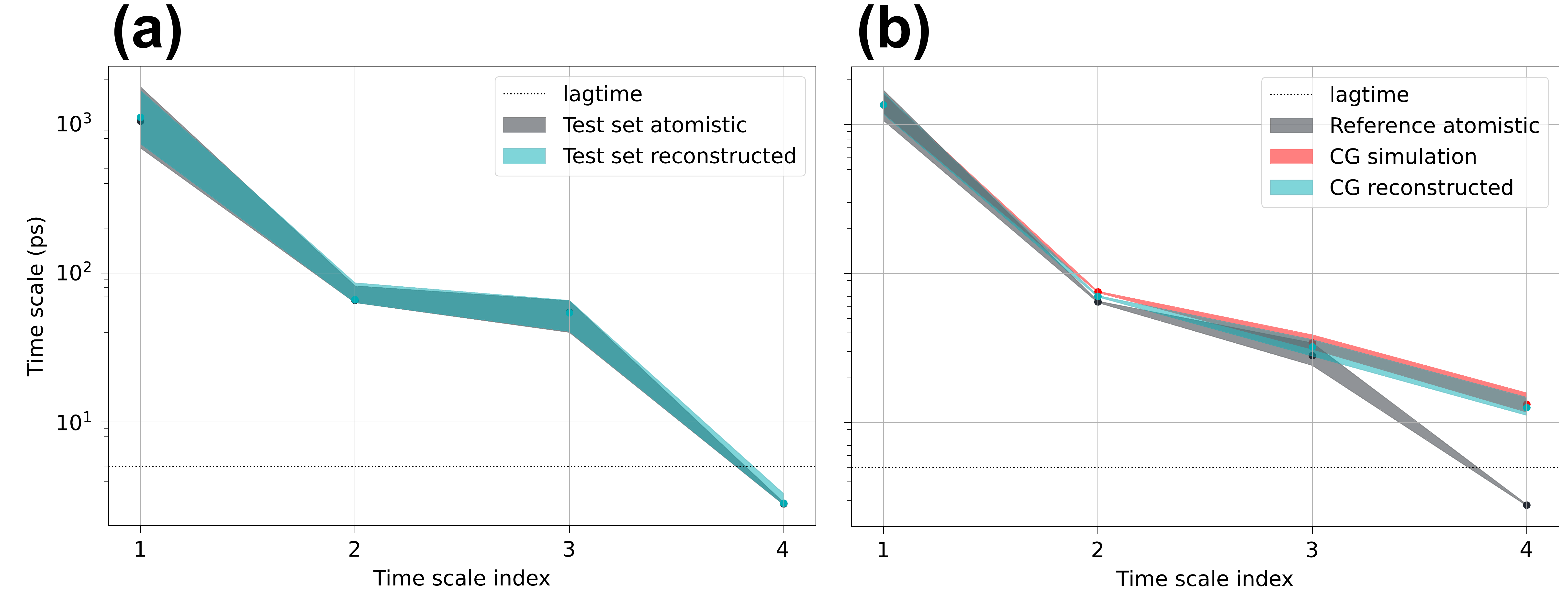}
    \caption{Implied timescales of atomistic and backmapped CG trajectories for ADP. \textbf{(a)} Comparisonon of timescales recovered for the test set atomistic and test set reconstructed trajectories. \textbf{(b)} Implied timescales calculated for the reference atomistic data taken from Ref.~\cite{wehmeyer2018time} compared to the backmapped CG simulation performed with CGSchNet~\cite{husic2020coarse} along with the original CG simulation. Timescales for the backampped and CG simulated data are normalized such that the dominant processes is the same as the reference atomistic trajectory. Errors are estimated with Bayesian sampling and represent a $95\%$ confidence interval of the timescales of the sampled MSMs.}
    \label{fig:adp_kinetics}
\end{figure}

\begin{figure}[H]
  \centering
  \includegraphics[width=\linewidth]{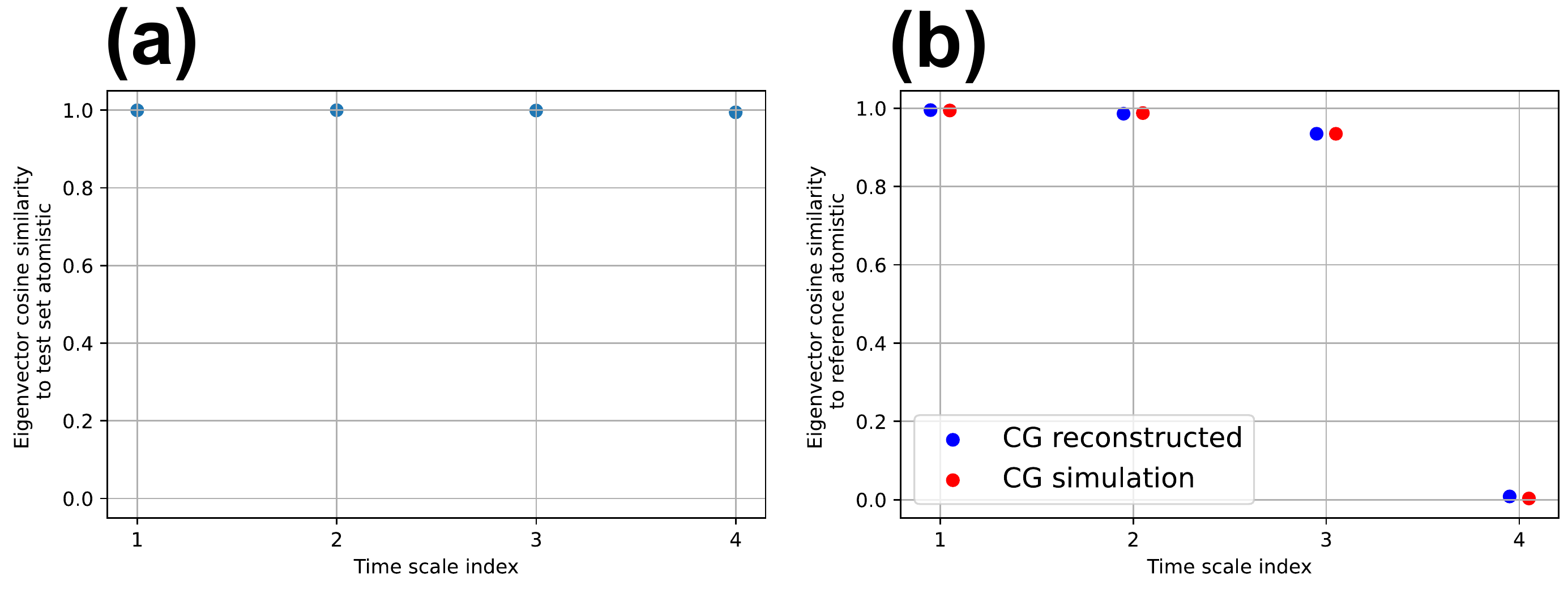}
  \caption{Cosine similarity between recovered MSM eigenvectors of atomistic and backmapped ADP trajectories. \textbf{(a)} The in distribution test set cosine similarity between atomistic and backmapped trajectories. \textbf{(b)} Cosine similarity between MSM eigenvectors of the backmapped data and the original CG data with respect to the reference atomistic data.} 
  \label{fig:adp_cosine_sim}
\end{figure}

We compare kinetic agreement between ground truth atomistic data and our backmapped reconstructions by building Markov State Models (MSMs)~\cite{prinz2011markov,bowman2013introduction,trendelkamp2015estimation,scherer2015pyemma,hoffmann2021deeptime} to characterize and compare the recovered physical processes and timescales. Construction and estimation of MSMs are carried out using the Deeptime~\cite{hoffmann2021deeptime} software library. We construct MSMs in $\phi$,$\psi$ space for ADP and perform state assignment by discretizing the phase space with 100 k-means centroids fit onto the atomistic data. These same 100 centroids are then also used to generate state assignments for the backmapped trajectory to which we compare. Complete validation of the MSM with associated clustering plots and implied timescale analysis for lag time selection is provided in the Supporting Information. Shown in Fig.~\ref{fig:adp_kinetics}a are the recovered timescales for the test set atomistic and test set backmapped data. Our backmapped trajectory reproduces the atomistic timescales within error for all recoverable processes. A result of using the same state assignments when building the original atomistic and backmapped MSMs is that the elements of the recovered eigenvectors characterize eigenfluxes between the same states. We can therefore measure the cosine similarity between these eigenvectors to quantitatively validate that these timescales correspond to the same physical processes between the original atomistic and the backmapped trajectories. Indeed, shown in Fig.~\ref{fig:adp_cosine_sim}a in the Supporting Information we confirm via this cosine similarity measure that the MSM eigenvectors are effectively identically recovered for the backmapped trajectory of in distribution test set.       

Shown in Fig.~\ref{fig:adp_kinetics}b is a comparison of the implied timescales between reference atomistic data, backmapped trajectories generated by our model and the original CGSchNet CG simulation. For both the CG backmapped data and the original CG simulation we normalize the implied timescales such that the dominant (slowest) process matches the reference atomistic data. We perform such a normalization to correct for the fact that CG models typically sample configurational landscapes at an accelerated rate compared to atomistic force fields due to absence of explicit solvent and elimination of atomistic degrees of freedom that can cause the coarse-grained degrees of freedom to be accelerated by different scaling factors~\cite{depa2005speed,depa2011coarse,fritz2011multiscale,marrink2007martini,marrink2013perspective}. This normalization also serves as a visual convenience when comparing deviations of timescales between the accelerated CG dynamics and the atomistic data. An alternative method to this rescaling approach that provides the same information would simply measure timescale ratios instead. We report once again overall excellent agreement of the implied timescales, within error, between our backmapped and reference atomistic trajectories in Fig.~\ref{fig:adp_kinetics}b, with the exception of the fourth, and subsequent, timescales that are faster than our lag time of 5 ps and therefore below the resolution limit of our MSM. We also validate again in Fig.~\ref{fig:adp_cosine_sim}b in the Supporting Information by comparing MSM eigenvector similarity that these first three timescales indeed correspond to the same physical processes between these two data sets. The backmapped CG simulated data ultimately reflects the kinetics produced by the coarse-grained model, evidenced by the nearly identical timescales of the CG simulation and backmapping in Fig.~\ref{fig:adp_kinetics}b. Nevertheless, upon normalization of the timescales our excellent agreement for these generalization results suggesting the ratio of timescales between processes is preserved upon backmapping with our model.    

\begin{figure}[H]
    \centering
    \includegraphics[width=\linewidth]{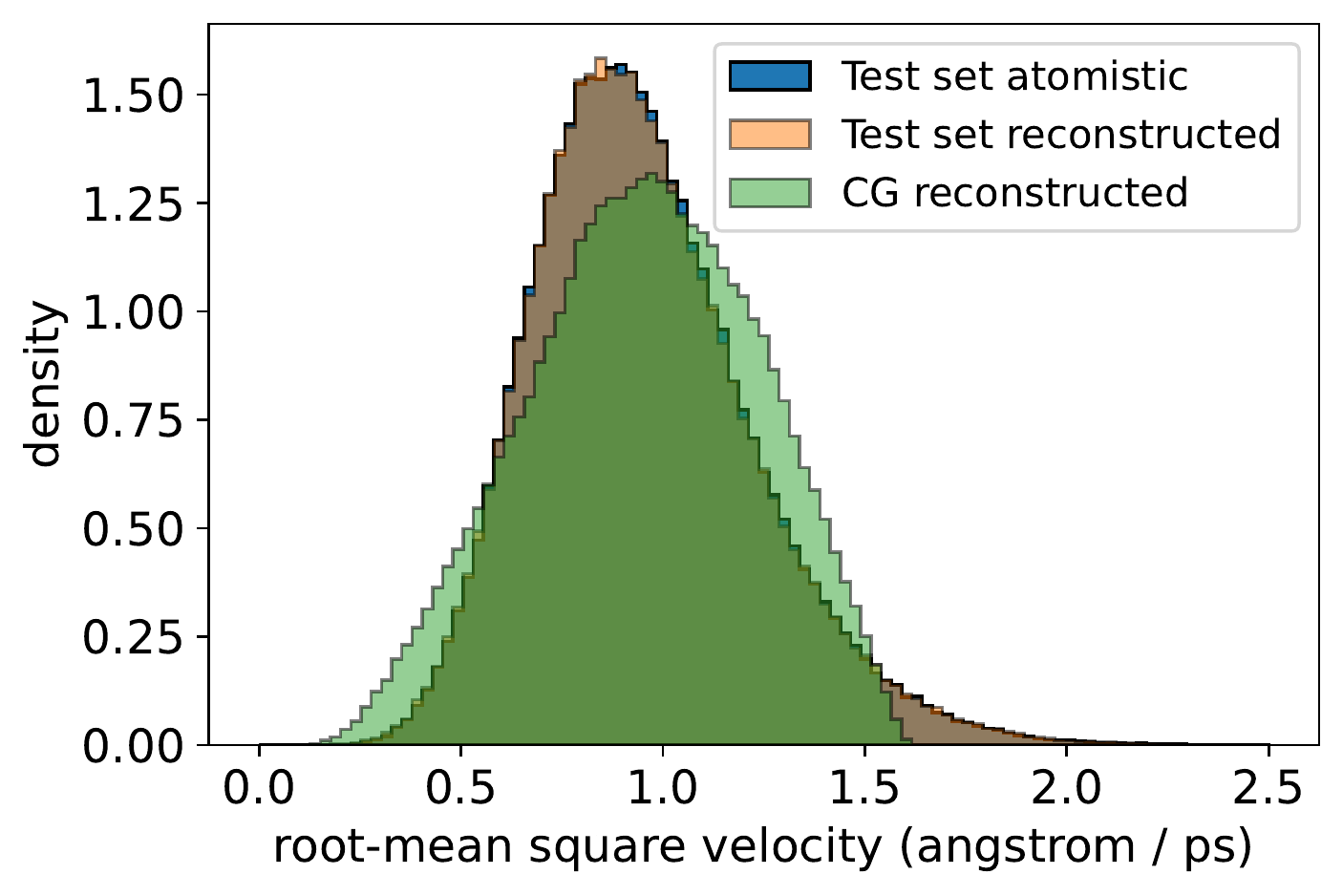}
    \caption{Comparison of atomistic and backmapped root mean square velocity distributions for ADP. Velocities are calculated atom-wise between sequential frames using a finite forward difference method and aggregated into a distribution over all atomic velocities. Velocities for the CG reconstructed data corresponding to the generalization set are rescaled by a constant factor ($\sim$3.25) such that the mean of the CG reconstructed velocities matches the mean of the test set atomistic velocities.} 
    \label{fig:adp_velocities}
\end{figure}

A result of conditioning our backmappings using previous atomistic configurations we find our model is capable of successfully generating backmapped trajectories that reproduce intra-frame root mean square velocity distributions (Fig.~\ref{fig:adp_velocities}). The distribution of velocities is an effective measure of the deviation of atomic coordinates between consecutive frames. Although our model is not explicitly trained to match intra-frame velocities, the temporal coherence built into our network architecture and training procedure enables the trained network to accurately reproduce these velocity distributions. The excellent agreement we observe for the in distribution data is enabled by the known and directly comparable spacing between frames. The temporal spacing between consecutive frames for the in distribution test set is well-defined by the frame spacing of the atomistic reference simulation, i.e., 1 ps. However, the specific temporal spacing is more obscure for the generalization set, as a direct consequence of the CG dynamics. While most CG models target thermodynamic consistency with a yet higher resolution model or experimentally observed properties, kinetic consistency is typically neglected. In general, CG force fields effectively accelerate simulation dynamics as a consequence of smoothing the energy landscape and lowering energy barriers. However, the timescales for transitions between meta-stable states are typically not rescaled uniformly~\cite{depa2005speed,depa2011coarse,fritz2011multiscale,marrink2007martini,marrink2013perspective}. To account for this inherent acceleration due to CG force fields, we rescale the velocity distribution of the backmapped generalization set data by a constant factor such that the mean of the velocity distribution matches the mean of the velocity distribution for the test set atomistic data. Applying this empirical correction we see reasonable agreement in the shape of the velocity distributions between the backmapped generalization set data and the native atomistic data, suggesting for ADP here our backmapping model is capable of reconstructing realistic atomic velocities up to a constant scaling factor for data originating from CG force fields.         

\subsection{Chignolin}

\subsubsection{Energetics}

\begin{figure}
    \centering
    \includegraphics[width=\linewidth]{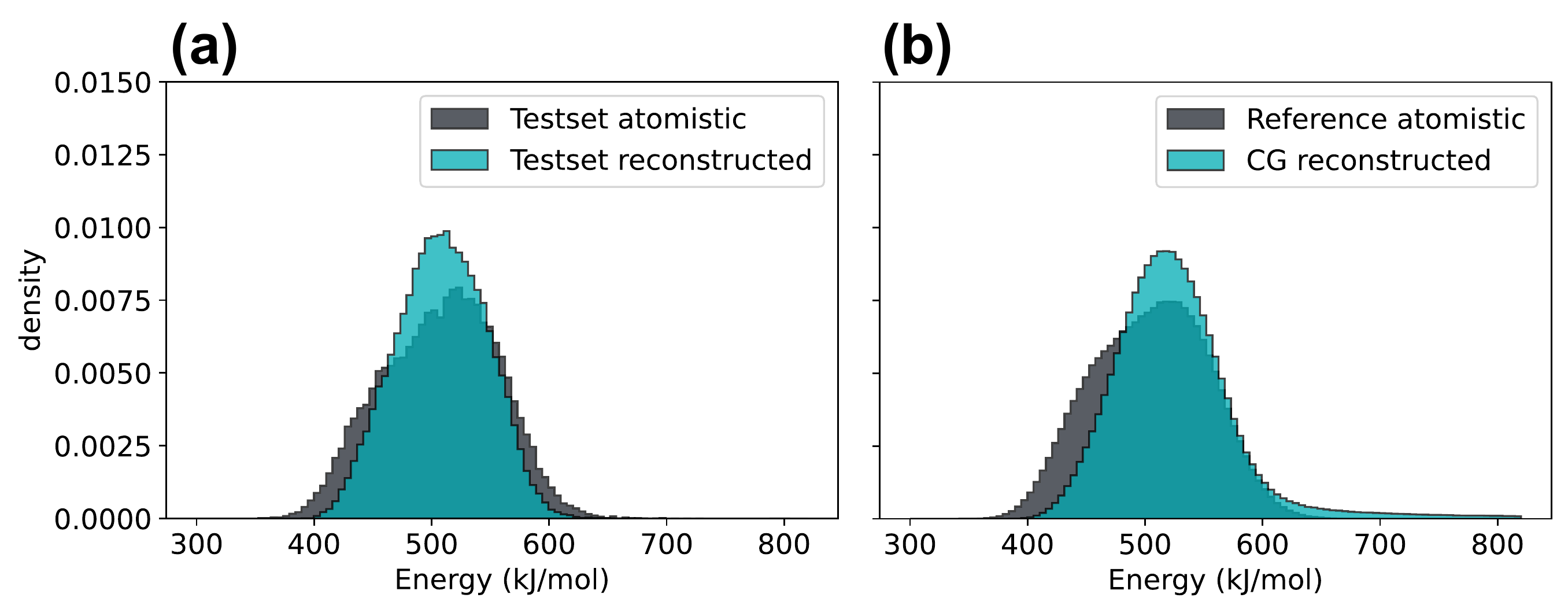}
    \caption{Potential energy distributions of atomistic and backmapped CLN trajectories for \textbf{(a)} the in distribution test set and \textbf{(b)} the generalization set.}
    \label{fig:cln_energy}
\end{figure}

\begin{figure}[H]
  \centering
  \includegraphics[width=0.8\linewidth]{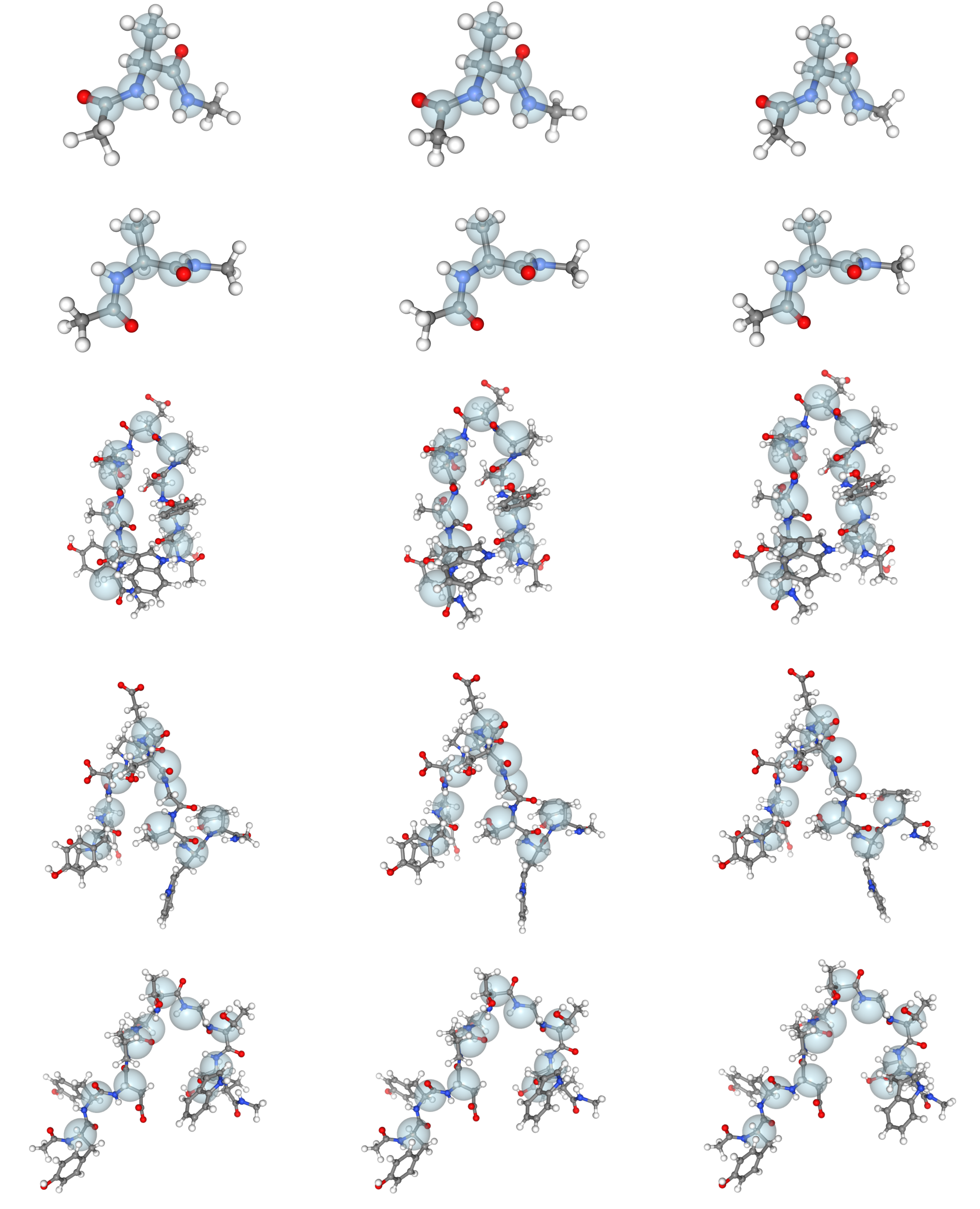}
  \caption{Atomic reconstructions generated from a fixed condition with different latent space instantiations. Each row represents a set of three atomic reconstructions for a fixed CG configuration. Faded blue spheres represent the input CG configuration overlaid on the atomic reconstruction. For ADP along the first three rows, the primary variation in these latent space samples are rotations of the methyl group hydrogens, as the rest of the peptide backbone and side chains are defined by the input CG configuration alone. For CLN along the final three rows, we notice the majority of configurational variation stems from side-chain motions along the C- and N-terminus.}
  \label{fig:latent_generation}
\end{figure}

\begin{figure}[H]
  \centering
  \includegraphics[width=\linewidth]{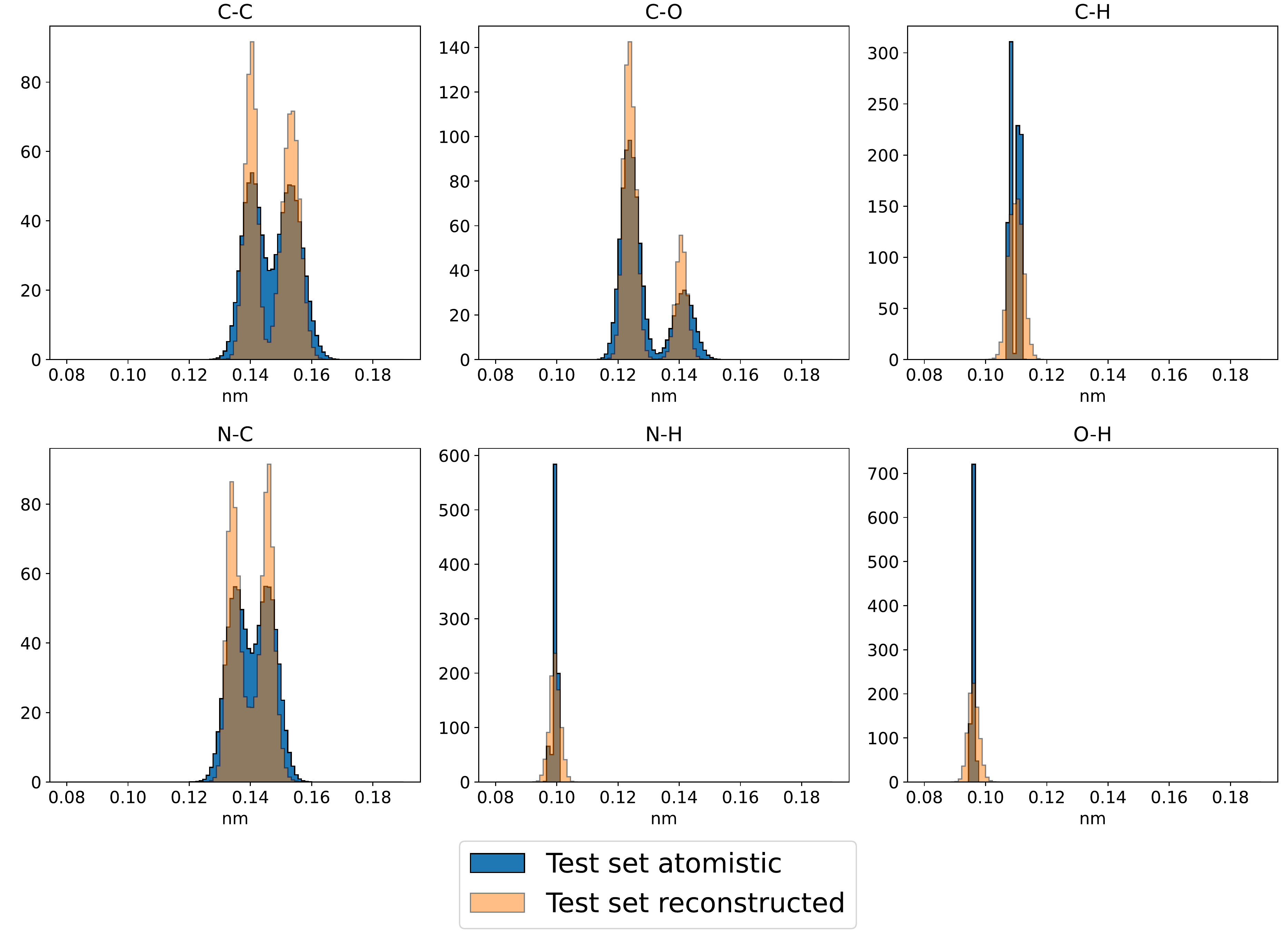}
  \caption{Bond length distributions delineated by element participation for the in distribution test set atomistic and backmapped trajectories for CLN.}
  \label{fig:cln_bonds_testset}
\end{figure}

\begin{figure}[H]
  \centering
  \includegraphics[width=\linewidth]{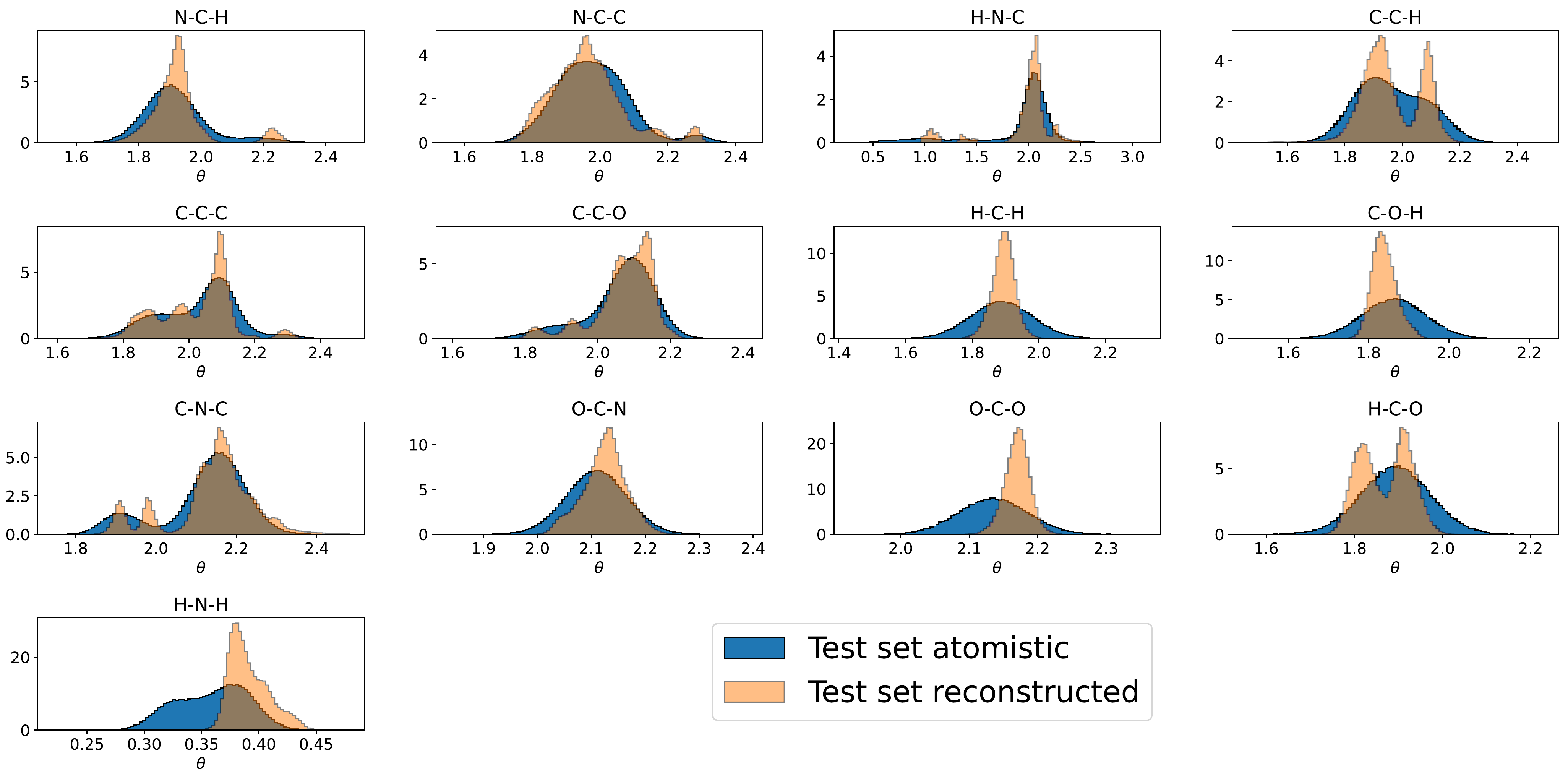}
  \caption{Atomic angle distributions delineated by element participation for the in distribution test set atomistic and backmapped trajectories for CLN.}
  \label{fig:cln_angles_testset}
\end{figure}

\begin{figure}[H]
  \centering
  \includegraphics[width=\linewidth]{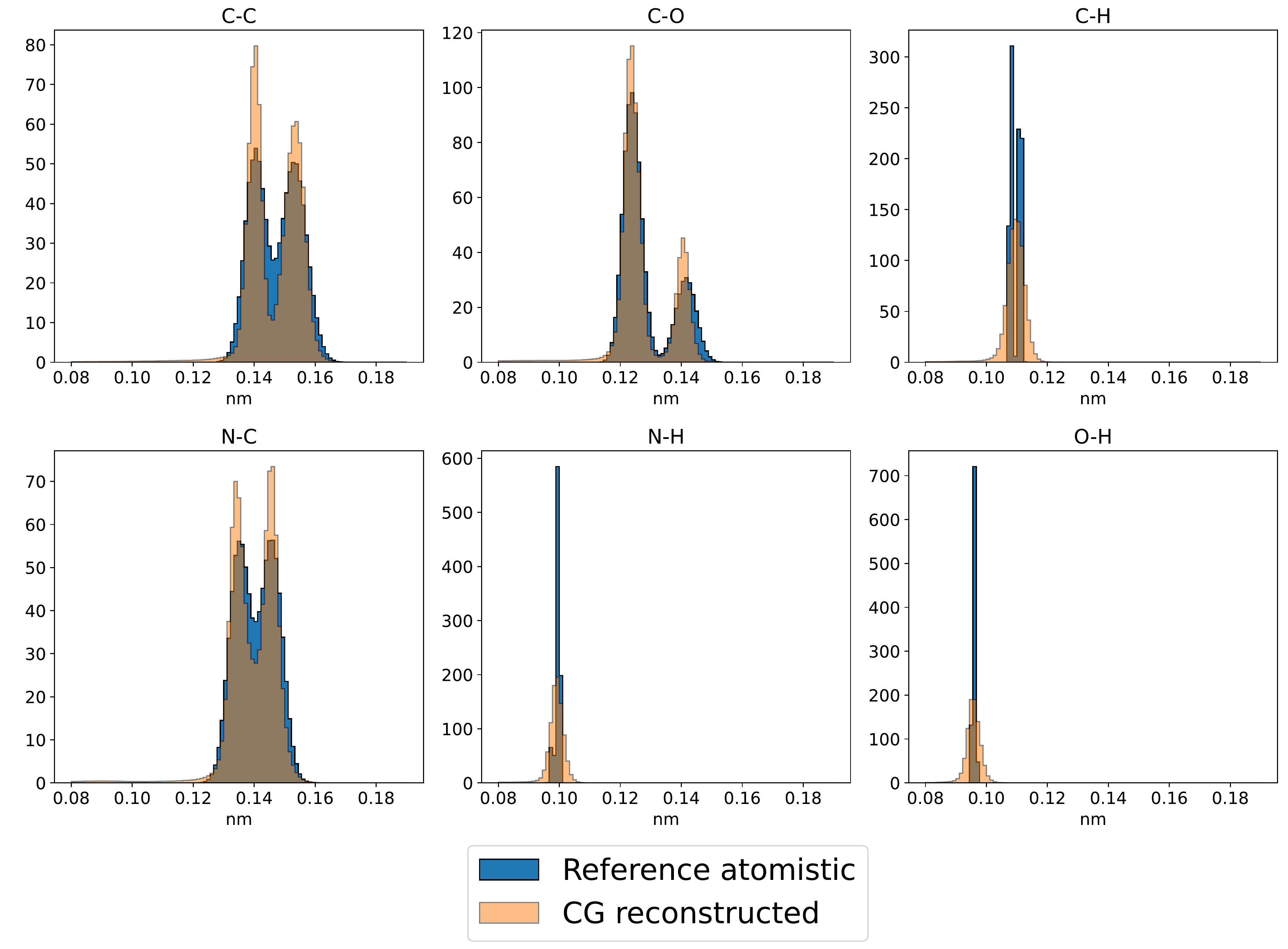}
  \caption{Bond length distributions delineated by element participation for the generalization test reference atomistic and backmapped CGSchNet trajectories for CLN.} 
  \label{fig:cln_bonds_cgset}
\end{figure}

\begin{figure}[H]
  \centering
  \includegraphics[width=\linewidth]{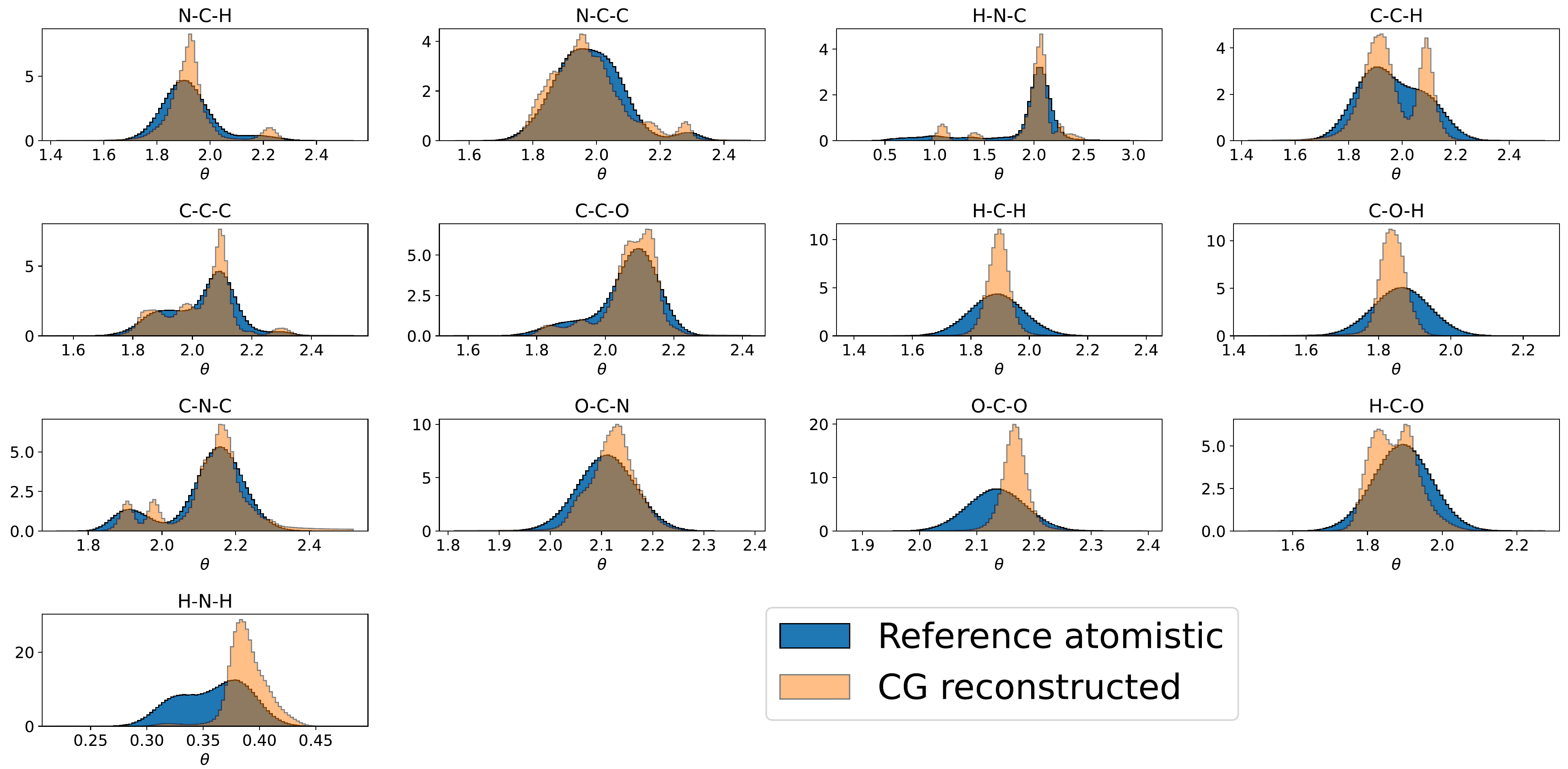}
  \caption{Atomic angle distributions delineated by element participation for the generalization test reference atomistic and backmapped CGSchNet trajectories for CLN.} 
  \label{fig:cln_angles_cgset}
\end{figure}

The second molecular system we apply our backmapping method to is the miniprotin Chignolin (CLN), which is composed of 10 residues with 175 atoms from which we consider a coarse graining into the 10 $\alpha$-carbons along the peptide backbone. Compared to ADP, the scale and complexity of CLN presents a more challenging test case for both our backmapping and accurate CG force field construction. We once again compare distributions of the internal energy as an indicator for structural similarity between reference atomistic and our backmapped data. Shown in Fig.~\ref{fig:cln_energy}a is a comparison of the internal energies for the in distribution test set atomistic and backmapped trajectories, while a comparison for the generalization set data is presented in Fig.~\ref{fig:cln_energy}b. These results show excellent energetic overlap between the reference atomistic and our backmapped data for both the in distribution and generalization sets. The internal energies recovered from the backmapped CGSchNet simulation in Fig.~\ref{fig:cln_energy}b from the generalization test show longer high-energy tails compared to the in distribution test in Fig.~\ref{fig:cln_energy}a. The origins of these higher-energy reconstructed configurations can be explained by an examination of the bond lengths and angles revealing the presence of very slightly contracted bond length distributions in the generalization set compared to the in distribution test set (Figs.~\ref{fig:cln_bonds_testset}-\ref{fig:cln_angles_cgset} in the Supporting Information). Our backmapped trajectories nevertheless show overall excellent agreement in these local intermolecular features and energetics, and therefore also produce visually convincing atomistic reconstructions (Fig.~\ref{fig:cln_fes}e and Fig.~\ref{fig:latent_generation} in the Supporting Information).  

\subsubsection{Thermodynamics}

We construct Free Energy Surfaces (FES) using the basis recovered from Time-lagged Independent Component Analysis (TICA)~\cite{perez2013identification, nuske2014variational, wu2020variational, noe2015kinetic} to compare thermodynamic similarity between reference atomistic data and our backmapped trajectories. Unlike ADP, there are no simple intuitive variables capable of compactly representing the CLN FES, so we instead use TICA as a dimensionality reduction technique to extract a low-dimensional representation for the CLN phase space. Using our entire reference atomistic dataset we first learn a TICA embedding into the first two non-trivial Independent Components (ICs) using all 45 pairwise $\alpha$-carbon distances as features~\cite{wang2019machine, husic2020coarse,chen2021machine,kohler2022computational}. This learned TICA model provides us with a fixed basis set we use for constructing an FES in these two leading ICs for the atomistic and backmapped data from both our in distribution and generalization data sets. 

\begin{figure}[H]
    \centering
    \includegraphics[width=\linewidth]{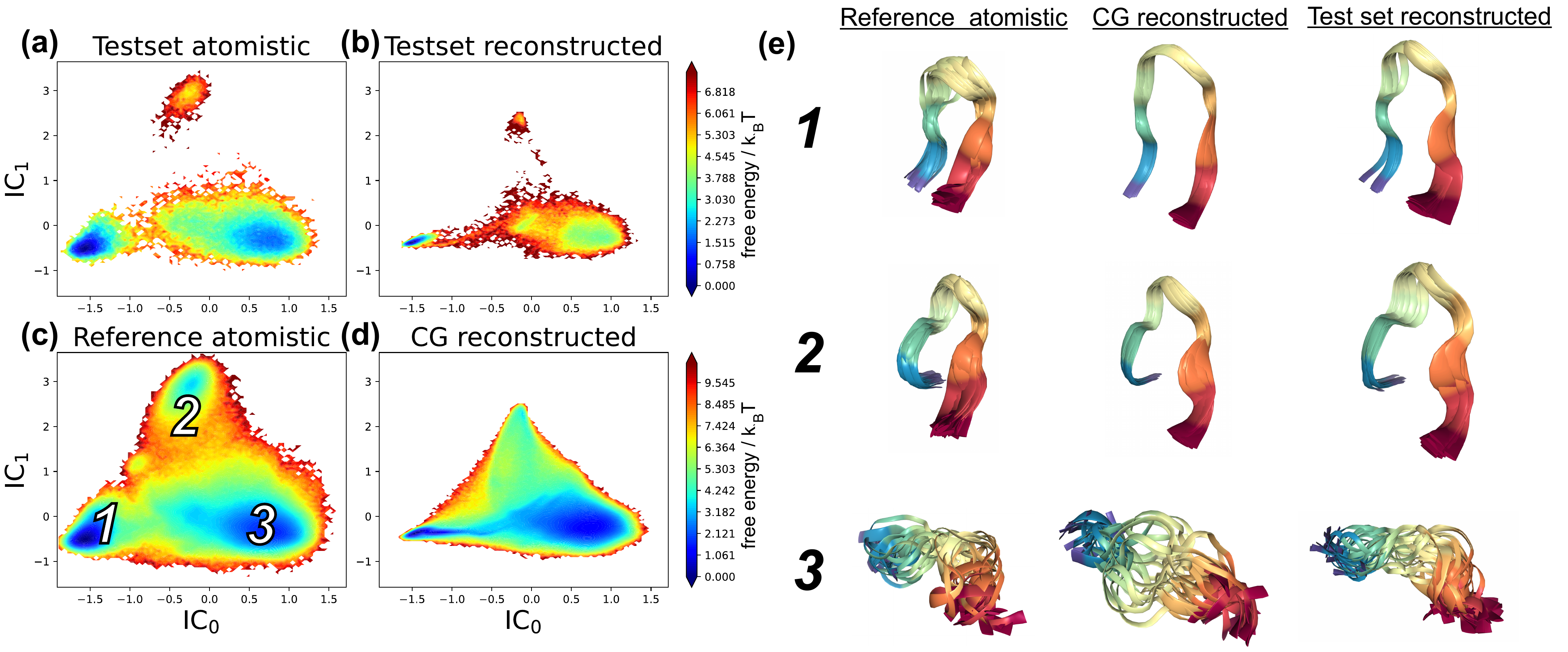}
    \caption{Comparison of the atomistic and backmapped MSM-reweighted Free Energy Surface (FES) for CLN. The FES for CLN is represented in the basis of the two leading non-trivial Independent Components (ICs) of a Time-lagged Independent Component Analysis (TICA) model fit to the reference atomic dataset. Shown are FES of the in distribution test set atomistic \textbf{(a)} and backmapped \textbf{(b)} trajectories, alongside the reference atomistic data \textbf{(c)} and the corresponding backmapped CGSchNet simulation \textbf{(d)} from the generalization set. \textbf{(e)} Visualization for a collection of 25 superposed structures from the three meta-stable states taken from the reference atomistic data \textbf{(c)}, the backmapped CG reconstructed data \textbf{(d)} and backmapped in distribution test set trajectory \textbf{(b)}.}
    \label{fig:cln_fes}
\end{figure}

\begin{figure}[H]
  \centering
  \includegraphics[width=\linewidth]{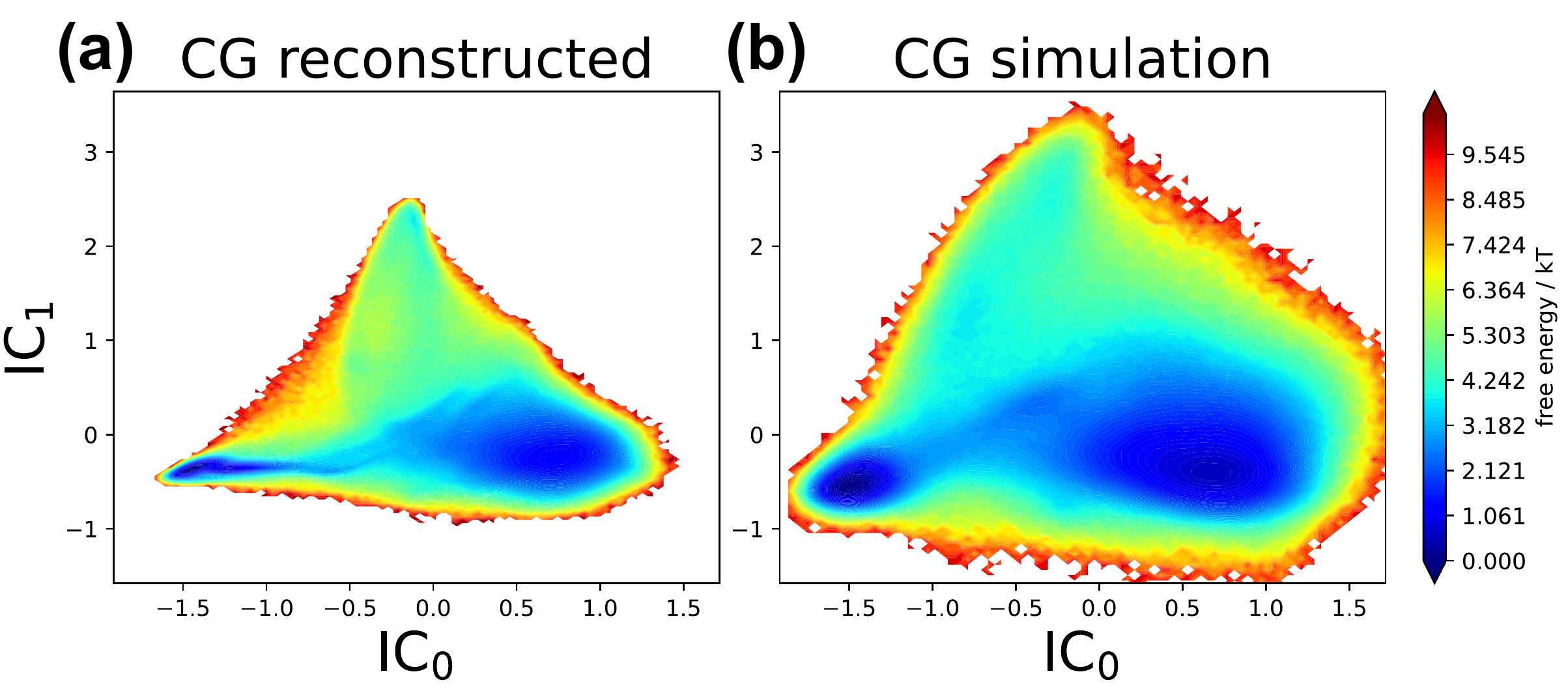}
  \caption{Comparison of MSM-reweighed FES for CLLN along the first two Independent Components (ICs) of a TICA model fit to the reference atomistic dataset between the \textbf{(a)} the backmapped CG reconstructed trajectory and \textbf{(b)} the original CG simulation performed with CGSchNet.} 
  \label{fig:cln_cg_recon_fes}
\end{figure}

Shown in Fig.~\ref{fig:cln_fes} is a comparison of these CLN FESs for in distribution test set atomistic (Fig.~\ref{fig:cln_fes}a) and backmapped (Fig.~\ref{fig:cln_fes}b) trajectories, alongside the generalization set reference atomistic FES (Fig.~\ref{fig:cln_fes}c) and the FES obtained from backmapping a CG simulation performed with CGSchNet~\cite{husic2020coarse} (Fig.~\ref{fig:cln_fes}d). In each case the backmapped FES recovers the presence of the three primary meta-stable states in CLN and produces visually identical atomic reconstructions compared to the atomistic reference (Fig.~\ref{fig:cln_fes}e). We notice our backmapped FES are contracted near regions corresponding to the folded state (labeled \textit{1}) and the misfolded state (labeled \textit{2}) compared to the unfolded ensemble (labeled \textit{3}). Correspondingly, backmapped structures extracted from the folded and misfolded states display slightly less configurational variability compared to atomistic reference than structures visualized from the unfolded ensemble (Fig.~\ref{fig:cln_fes}e). We suspect this is due to a loss-conserving strategy by the network: comparatively less loss is sacrificed by predicting a (nearly) fixed configuration for CG structures corresponding to folded and misfolded states compared to the benefit of capturing the diversity of possible structures in the unfolded ensemble. This results in more comprehensive coverage of the unfolded ensemble in our backmapped data compared to the folded and misfolded states. We also notice that the misfolded state is noticeably less stable in the generalization set backmapping (Fig.~\ref{fig:cln_fes}d) compared to the reference atomistic data (Fig.~\ref{fig:cln_fes}c). Indeed, we find this property to be a reflection of the misfolded state being inherently less stable in the original CG simulation as well (Fig.~\ref{fig:cln_cg_recon_fes} in the Supporting Information). While our backmapping method does a reasonable job of reproducing the atomistic FES, we find that are ultimately limited by the underlying accuracy of the CG simulation that we backmap.

\subsubsection{Kinetics} \label{sec:cln_kinetics}

We evaluate kinetic similarity between reference atomistic data and our backmapped trajectories by comparing similarity of MSM-recovered implied timescales and processes. We perform state space discretization for our MSMs within the same two-dimensional TICA basis used to construct our FES. Data from the reference atomistic trajectories in this two-dimensional TICA space is fit to determine 150 k-means centroids identifying the state space decomposition. We then use these same 150 centroids to generate state assignments when building separate MSMs for other atomistic and backmapped data within the in distribution and generalization data sets. Using the same TICA projection and set of cluster centers between MSMs ensures direct comparability of the recovered eigenvectors and allows us to determine similarity of the recovered processes. We use the same methodology as with ADP here for CLN for quantifying similarity of processes by measuring the cosine similarity of the MSM eigenvectors. In the cases where not all 150 states are utilized, we construct MSMs using only occupied states and compute cosine similarity using only this subset of mutually occupied states between the two MSMs being compared. Using this approach we can quantitatively ensure the that implied timescales in fact correspond to the same physical processes between the two data sets. As with ADP, complete details on MSM construction and validation are provided in the Supporting Information.             

\begin{figure}
    \centering
    \includegraphics[width=\linewidth]{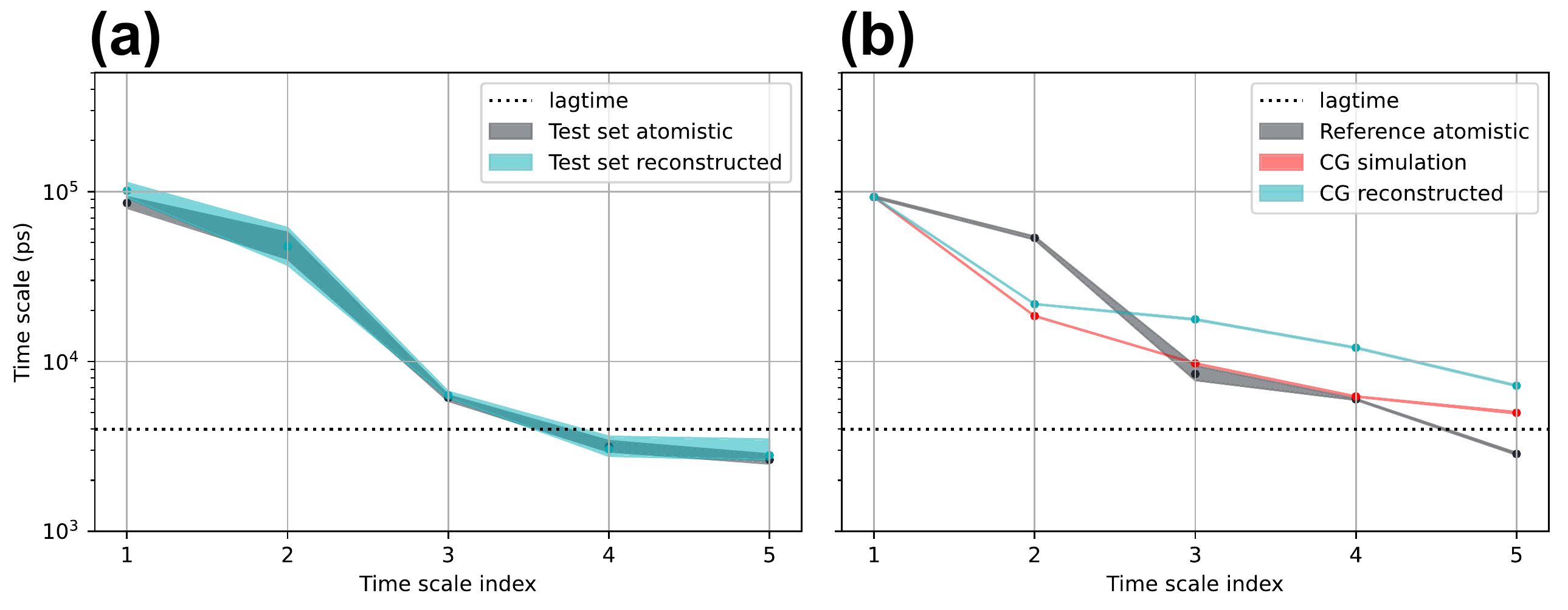}
    \caption{Implied timescales of atomistic and backmapped CG trajectories for CLN. \textbf{(a)} Comparison of implied timescales from atomistic and backmapped trajectories for the in distribution test set. \textbf{(b)} Implied timescales for the reference atomistic data, the original CGSchNet simulation and the backmapped CGSchNet simulation. Timescales for the original CG simulation and backmapped data are normalized such that the dominant (slowest) processes matches the slowest reference atomistic timescale.}
    \label{fig:cln_kinetics}
\end{figure}

\begin{figure}[H]
  \centering
  \includegraphics[width=\linewidth]{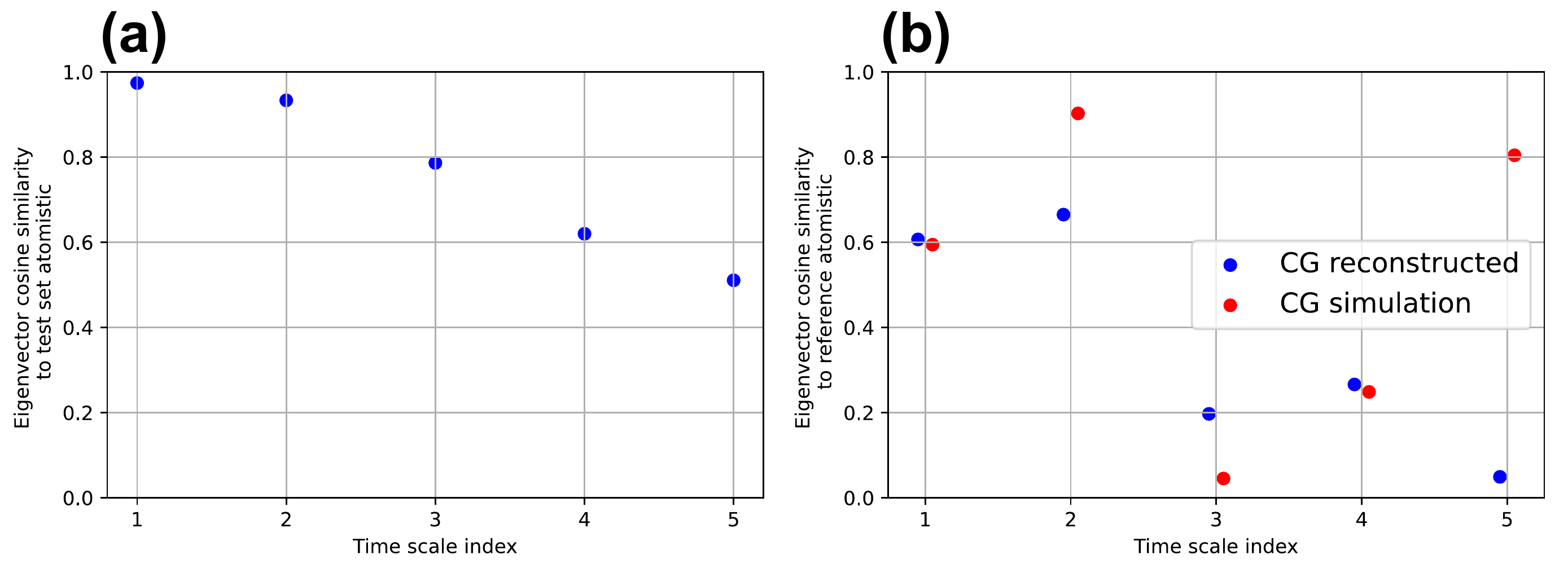}
  \caption{Cosine similarity between recovered MSM eigenvectors of atomistic and backmapped CLN trajectories. \textbf{(a)} The in distribution test set cosine similarity between atomistic and backmapped trajectories. \textbf{(b)} Cosine similarity between MSM eigenvectors of the backmapped data and the original CG data with respect to the reference atomistic data.} 
  \label{fig:cln_cosine_sim}
\end{figure}

Presented in Fig.~\ref{fig:cln_kinetics}a is a comparison of implied timescales between atomistic and backmapped trajectories for the in distribution test set. For the in distribution test set our backmapped reconstruction precisely reproduces within error all implied timescales. The large gap between the 2$^{\text{nd}}$ and 3$^{\text{rd}}$ timescales suggests the majority kinetic variance is captured in the first two processes, which we recover with >90\% cosine similarity, and for the final remaining processes greater than our lag time -- and therefore the only other resolvable process by our MSM -- we also recover with $\sim$80\% cosine similarity (Fig.~\ref{fig:cln_cosine_sim}a in the Supporting Information). Comparison of implied timescales for the generalization set between the reference atomistic data, backmapped CGSchNet~\cite{husic2020coarse} simulation and the original CG simulation is shown in Fig.~\ref{fig:cln_kinetics}b. We follow the same approach here as with ADP, where we normalize the implied timescales for data generated from CG simulations by a constant factor such that their slowest timescale matches the slowest timescale from the reference atomistic data. This normalization accounts for the inherently accelerated dynamics of CG simulations and enables us to effectively compare the ratio of implied timescales~\cite{depa2005speed,depa2011coarse,fritz2011multiscale,marrink2007martini,marrink2013perspective}. The ratio between the first two timescales is poorly recovered for both the backmapped data and the original CG simulation compared to the reference atomistic data. Comparison of the cosine similarity between these processes reveals the 1$^{\text{st}}$ processes is recovered with $\sim$60\% similarity for both the backmapped data and the CG simulation, while the 2$^{\text{nd}}$ processes is recovered with also $\sim$66\% similarity for the backmapped data and $\sim$90\% similarity for the original CG simulation (Fig.~\ref{fig:cln_cosine_sim}b in the Supporting Information). While the ratio in the timescales for the faster 3$^{\text{rd}}$ and 4$^{\text{th}}$ processes seems to be better conserved in the original CG simulation, our backmapped data actually recovers these processes with slightly better similarity than the original CG simulation (Fig.~\ref{fig:cln_cosine_sim}b in the Supporting Information). The fact that our method precisely recovers kinetics for the in distribution data but can only approximately recovers kinetics when backmapping real CG simulated data confirms that our method produces backmapped trajectory data that is largely a reflection of the kinetics expressed in the underlying CG data.    

\begin{figure}[H]
    \centering
    \includegraphics[width=\linewidth]{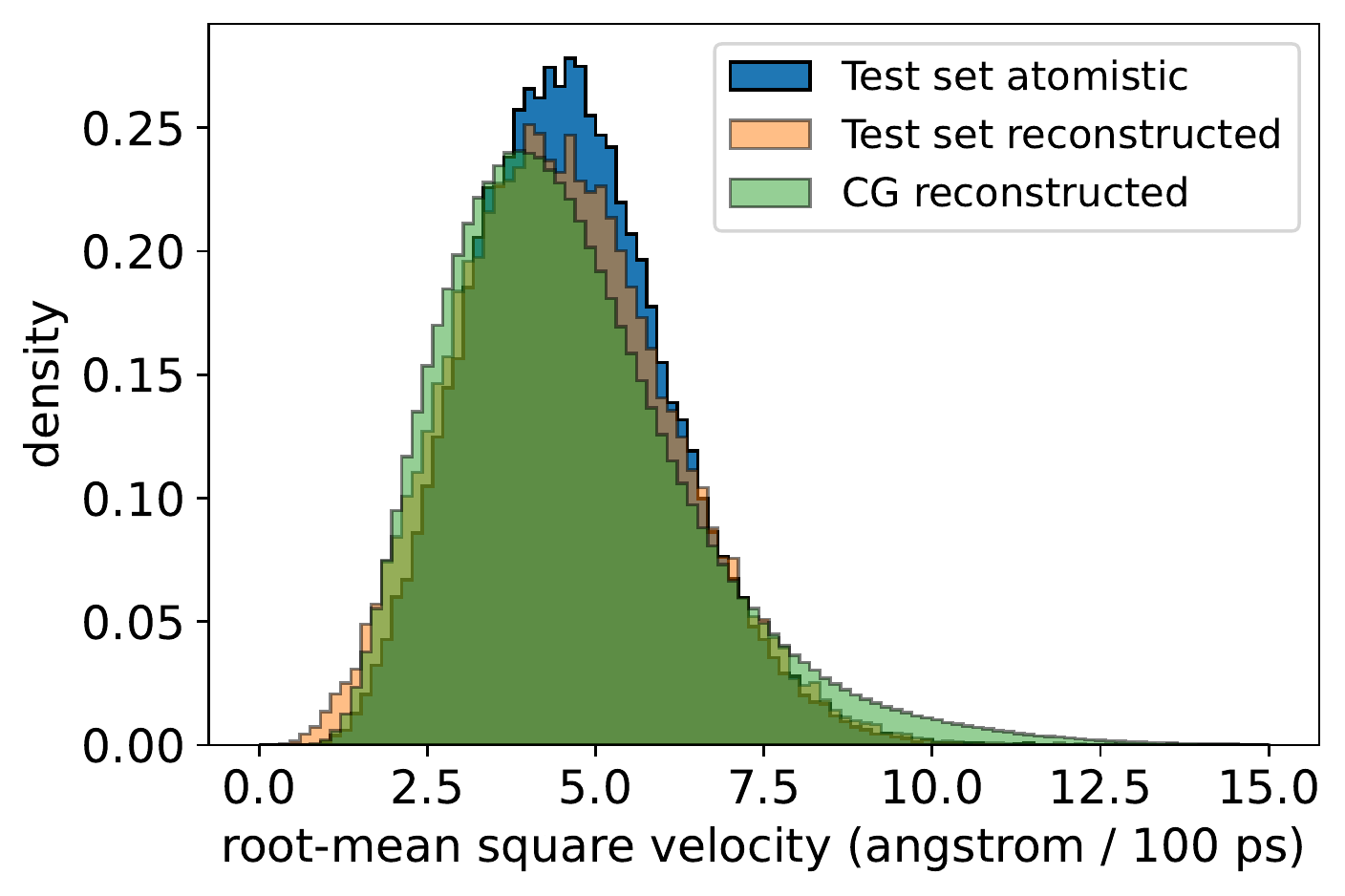}
    \caption{Comparison of atomistic and backmapped root mean square velocity distributions for CLN. Velocities are calculated frame-by-frame using a finite forward difference method and represented in angstrom / 100 ps, as 100 ps is the native temporal spacing between frames in the reference atomistic data~\cite{wang2019machine}. The backmapped generalization set velocities are rescaled by a constant factor ($\sim$17.81) such that the mean of the CG reconstructed velocities equals the mean of the test set atomistic velocities.}
    \label{fig:cln_velocities}
\end{figure}

Lastly, we compare velocity distributions as an indicator of temporal coherence between sequential frames in our backmapped reconstructions. Shown in Fig.~\ref{fig:cln_velocities} is a comparison of frame-by-frame root mean square velocity distributions for the reference and backmapped in distribution test set data alongside the backmapped CGSchNet simulation data. For the in distribution test set the velocities are an excellent match between the reference and backmapped data. Mimicking our approach with ADP, to account for the inherently accelerated dynamics of the CGSchNet force field we use a constant factor to rescale the atomic velocities such that the mean of the backmapped generalization set velocities matches the mean of the test set atomistic velocities. After applying this constant scaling we notice excellent agreement in the shape of the velocity distributions between the backmapped generalization set data and the original atomistic data. We note that the scaling factor used to correct the CLN velocities ($\sim$17.81) is much larger than the scaling factor used for the ADP velocities ($\sim$3.25), suggesting a more substantial acceleration of backmapped CLN dynamics due to the CG force field compared to ADP. This relative speed-up of the backmapped CLN CG simulation could be attributed to the more severe coarse-graining of CLN from 175 atoms to 10 beads (17.5$\times$ reduction) compared to the ADP model from 22 atoms to 6 beads ($\sim$3.67$\times$ reduction). This more dramatic reduction in the degrees of freedom upon coarse-graining for the CLN model could result in the CLN CG simulation operating in a comparatively 'smoother' free energy surface than ADP, and therefore leading to the relatively faster atomic motions identified by the velocity distributions.     

\section{Conclusions} \label{sec:concl} 

We present in this work a data-driven and temporally coherent scheme for backmapping CG trajectories into atomistic resolution. Our approach trains a conditional variational autoencoder (cVAE) to reconstruct atomistic detail given the target CG configuration and the previous atomistic structure. Our method is showcased here to backmap two popular biomolecular systems: alanine dipeptide (ADP) and the miniprotein chignolin (CLN). We train our model using a reference atomistic trajectory which we coarse-grain \textit{post hoc} to produce exemplar pairs of atomistic and CG configurations (Fig.~\ref{fig:setup}a). We test that our backmapping method on both in distribution data generated from backmapping a CG trajectory produced by coarse-graining held-out atomistic data (Fig.~\ref{fig:setup}b), and out of distribution data generated from a real CG simulation performed using CGSchNet~\cite{husic2020coarse}. We evaluate the performance of our model in terms of capability to reproduce structural, thermodynamic and kinetic properties of reference atomistic systems. To this end, structural similarity is probed by comparing distributions of potential energies and local structural features, such as bond lengths and angles. Thermodynamic similarity is tested by analyzing free energy surfaces that are constructed in terms of collective variables. Kinetic agreement is tested by comparing implied timescales of processes identified by MSMs, while temporal coherence between consecutive frames is analyzed in terms of intra-frame velocity distributions. Our model yields backmapped trajectories for the in distribution test set that are in good agreement with atomistic data. Moreover, our model generalizes well producing convincing atomic reconstructions for out of distribution data obtained from CGSchNet simulations. While we notice slightly better generalizability to CGSchNet simulations of ADP compared to the more complex CLN molecule, we find our method generates backmapped trajectories that largely maintain thermodynamic and kinetic properties reflected in the original CG simulation while also reconstructing atomistic velocities up to a constant scaling factor. Errors in our backmapped reconstructions on the out of distribution generalization data can be attributed to both the performance and expressivity of our backmapping cVAE model compounded with biases and inaccuracies of the CGSchNet force field used to generate the CG trajectories. In the case of ADP, the highly accurate CGSchNet model leads to overall good backmapping generalization performance. For CLN, the CGSchNet model is comparatively lower-fidelity which we find results in slightly poorer generalization by our backmapping model in comparison to ADP. Ultimately, the backmapping scheme presented in this work can serve as a tool to analyze thermodynamic and kinetic properties of a CG system at atomistic resolution while also generating temporally coherent backmapped trajectories.

Future work will strive to improve upon the data efficiency and training/inference routines of our method. The backbone of our model primarily uses convolutional neutral networks (CNNs) operating on voxelized representations that are converted to and from Cartesian coordinates. While CNNs offer excellent expressibility via naturally hierarchical processing, they are not rotationally covariant and therefore require that we train our network with random rigid rotations to implicitly learn sensitivity to arbitrary alignments. Using explicitly covariant network architectures~\cite{miller2020relevance,thomas2018tensor,batzner2021se}, such as those employed in the backmapping scheme by Wang et al.~\cite{wang2022generative}, can lead to superior data efficiencies without the need to train with random rotations. Training routines could also be augmented to incorporate more inductive biases that may benefit backmapping, such as: (i) better emphasizing sparsely populated regions of configurational space, which could be accomplished by accompanying training samples with thermodynamic or dynamical path weights; (ii) an autoregressive training protocol could be employed to improve the temporal coherence by using a recurrent approach to predict multiple consecutive frames each forward pass; (iii) further encourage the model to utilize knowledge of preceding trajectory frames by augmenting the training loss to incorporate information that is explicitly based on velocities or higher order time derivatives.

\section*{Data Availability}
A complete PyTorch implementation of our model with associated training routines, data, and pretrained model weights for ADP and CLN along with Jupyter Notebooks demonstrating the energetic, thermodynamic, and kinetic analyses performed in this work is provided via the Materials Data Facility (MDF)~\cite{blaiszik2016materials,blaiszik2019data} at DOI:10.18126/tf0h-w0jz~\cite{MDF}. 

\section*{Acknowledgments}
K.S. was supported by the National Science Foundation’s Graduate Research Fellowship (Grant No. DGE-1746045). We gratefully acknowledge computing time on the University of Chicago high-performance GPU-based cyberinfrastructure supported by the National Science Foundation under Grant No. DMR-1828629. Part of this research was performed while the authors were visiting the Institute for Pure and Applied Mathematics (IPAM), which is supported by the National Science Foundation (Grant No. DMS-1440415). M.S. was supported in part by the Collaborative Research Center “Multiscale Simulation Methods for Soft Matter” of Deutsche Forschungsgemeinschaft under Grant No. SFB-TRR146, as well as the Max Planck Graduate Center. M.H. acknowledges financial support from Deutsche Forschungsgemeinschaft DFG (SFB 1114, Projects A04 and C03). N.C. was supported by the Einstein Foundation Berlin, the Deutsche Forschungsgemeinschaft (DFG, German Research Foundation) GRK 2433/1/Project number 384950143, the NLM Training Program in Biomedical
Informatics and Data Science (Grant No. 5T15LM007093-27), and the Welch Foundation (Grant No. C-1570). The authors would like to thank Andrew White, Andrew L. Ferguson, Cecilia Clementi, and Frank Noe for critical reading and constructive criticism of the manuscript.  

\section*{Supporting Information}

\subsection{cVAE architectural details and training settings}

We discuss here details and hyperparameters in our cVAE architecture and training setup, with all code and datasets publicly available at DOI:10.18126/tf0h-w0jz~\cite{MDF}. Typical VAEs are composed of an encoder-decoder architecture where the encoder process an input into a fixed dimensional latent code that the decoder purposes to generate reconstructions of the input. Conditional VAEs (cVAEs) differ from this paradigm by defining an auxiliary conditional variable alongside the reconstruction target, which is passed to the decoder alongside the latent code when generating reconstructions. This cVAE framework enables the decoder alone to act as a generative model provided a conditional variable as input and a randomly sampled latent code as a source of generative noise. 

In our application, we strive to produce temporally coherent backmappings by incorporating as input previous atomistic configurations $\AAtom^{t-1}$ alongside the current coarse grain $\CG^{t}$ configuration when generating reconstructions $\hat{\AAtom^t}$. This temporal coherence is achieved in our cVAE framework by defining the reconstruction target $\mathbf{y}=\AAtom^t$ and the conditional variable as the tuple of configurations $\mathbf{c} = (\CG^t, \AAtom^{t-1})$.

\begin{figure}[H]
  \centering
  \includegraphics[width=\linewidth]{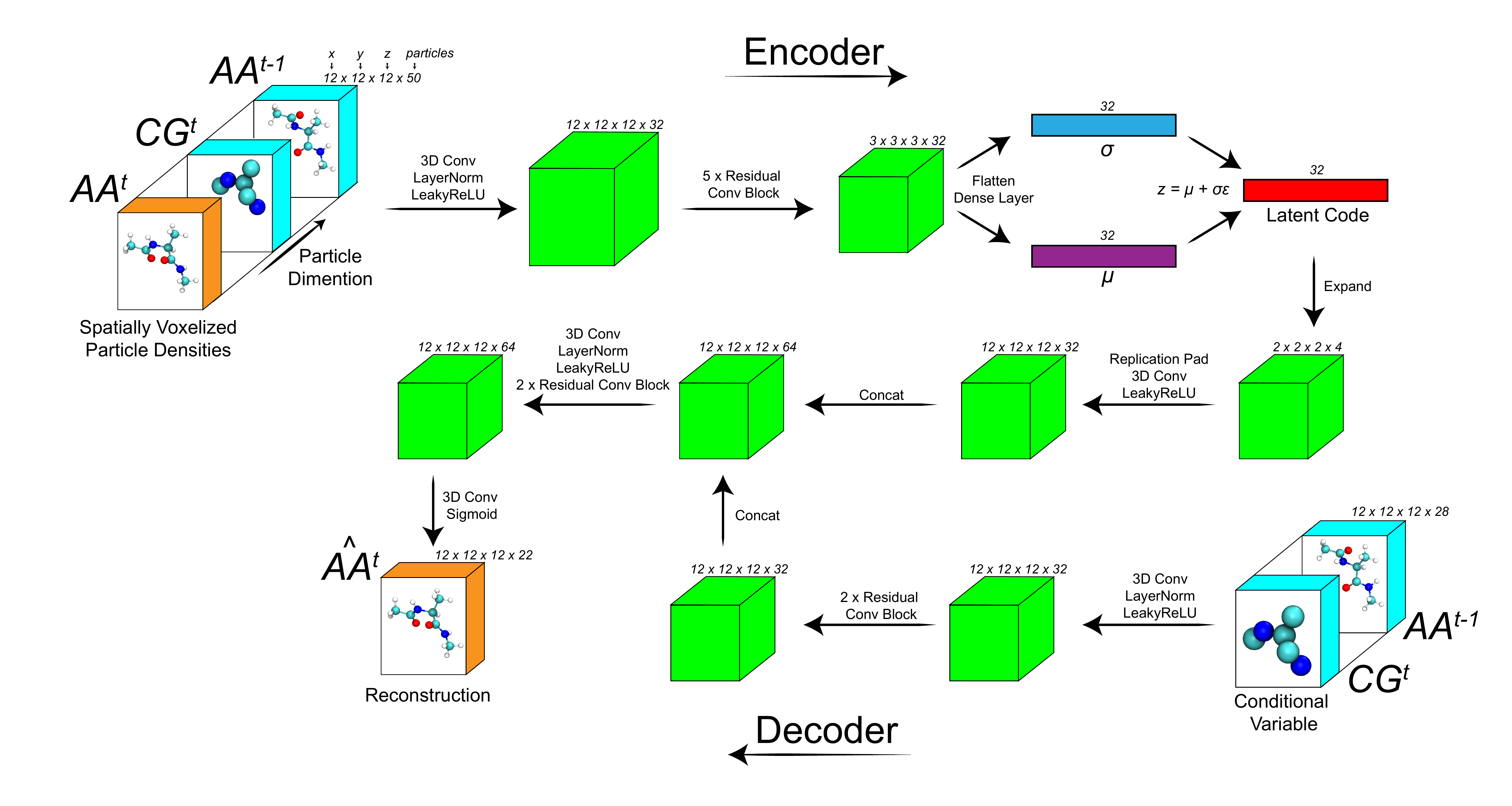}
  \caption{Schematic illustration of our cVAE architecture and data flow exemplified with particle dimensions above each tensor corresponding to our model trained on ADP.  }
  \label{fig:arch}
\end{figure}

Our model operates on 3D spatially voxelized particle densities delineated for each particle which we can readily transform to/from Cartesian coordinates (cf. Sec.~\ref{sec:vox} of the main text). Presented in Fig.~\ref{fig:arch} is an illustration of our cVAE architecture and data flow exemplified for the alanine dipeptide (ADP) molecule. Beginning in the top left of Fig.~\ref{fig:arch}, the input to our encoder $\mathbf{x}$ are the three configurations ($\AAtom^{t-1}$, $\CG^{t}$, $\AAtom^{t-1}$) voxelized and concatenated along their particle dimension $\mathbf{x} = [\VOX^{\AAtom^{t-1}}|| \VOX^{\CG^{t}}|| \VOX^{\AAtom^{t-1}}] \in \mathbb{R}^{d \times d \times d \times (2N + n)}$, where $d$ is the number of grid points used for the spatial discretization, $N$ is the number of atoms within each atomistic configuration and $n$ is the number of beads within each CG configuration. For this exemplar case of ADP shown here we use a spatial discretization of $d=12$ with $N=22$ and $n=6$, yielding an input tensor with dimensionality $\mathbf{x} \in \mathbb{R}^{12 \times 12 \times 12 \times 50}$. The different components of the input tensor are color-coded to designate the configurations associated with the conditional variable ($\CG^{t}$, $\AAtom^{t-1}$) (light blue) and the reconstruction target $\AAtom^t$ (orange).

\begin{figure}[H]
  \centering
  \includegraphics[width=\linewidth]{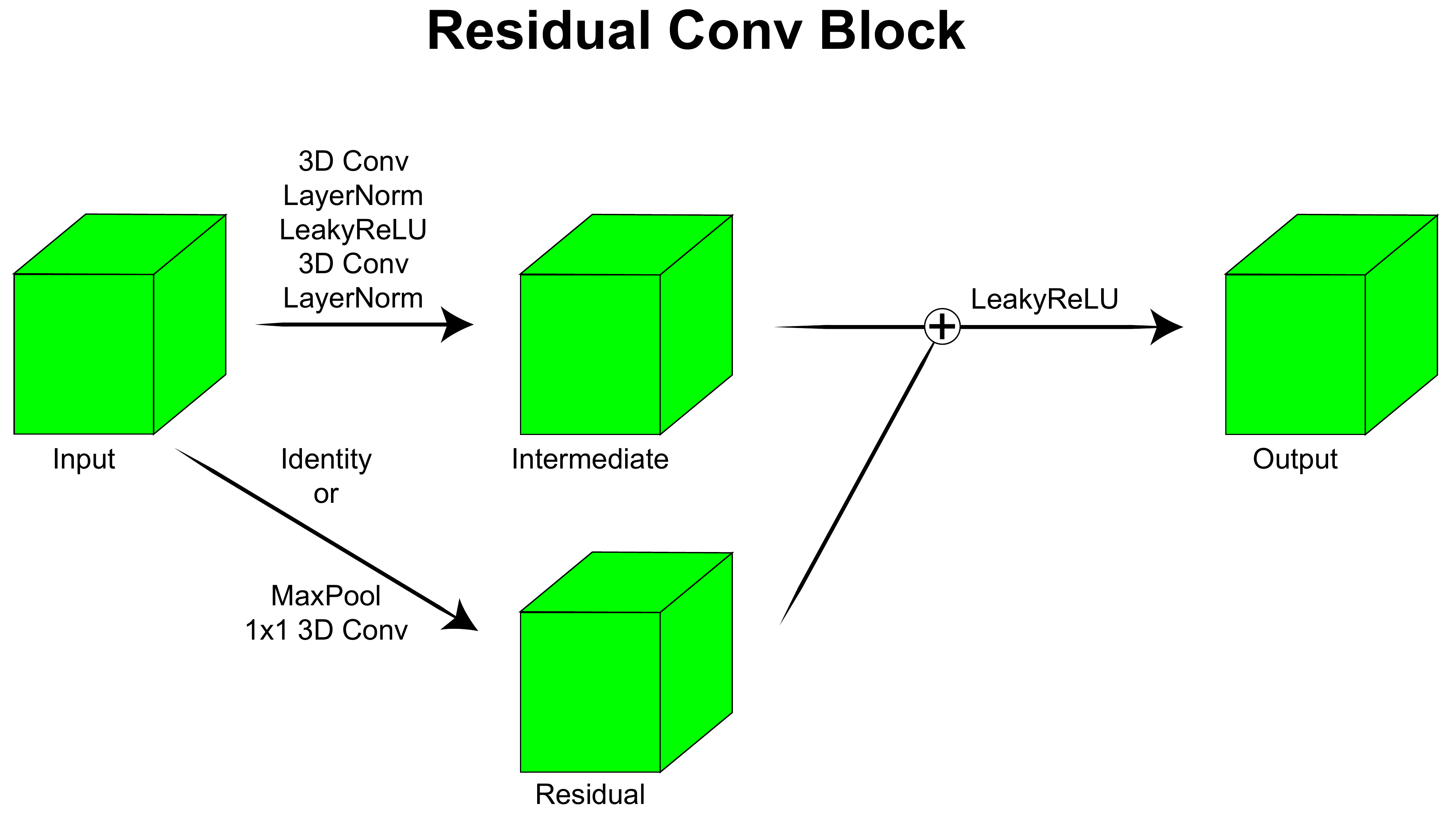}
  \caption{Schematic illustration of residual convolutional block used within our cVAE architecture.} 
  \label{fig:convblock}
\end{figure}

The encoder processes the input $\mathbf{x}$ using a series of (residual) CNN modules. Each arrow in Fig.~\ref{fig:arch} represents a series of operations performed on a representation with the resultant tensor dimensionality shown above the subsequent block. For example, the first arrow in the top left of Fig.~\ref{fig:arch} indicates a series of computations on the input $\mathbf{x}$ involving a 3D convolution, followed by a LayerNorm and lastly a LeakyReLU activation yielding an intermediate representation with dimension $12 \times 12 \times 12 \times 32$. Here, $d_{hidden}=32$ represents the hidden channel dimensionality of the network. The next step within the encoder involves sequential application of a 5$\times$ series of residual convolutional blocks. In some cases we use different hidden channel dimensions for the encoder $d_{hidden}^{Encoder}$ and the decoder  $d_{hidden}^{Decoder}$. A schematic illustration of our residual convolution block is shown in Fig.~\ref{fig:convblock}. In the case where the dimensionality of the input changes within the residual convolutional block, instead of a simple identity operation the residual is accommodated to the appropriate dimensionality using max pooling and 1x1 3D convolutions. A variety of strides and kernel sizes are used within the five sequential residual convolutional blocks of the encoder to yield a representation with dimensionality $3 \times 3 \times 3 \times 32$. This representation is then flattened into a $3\times 3 \times 3 \times 32 = 864$-dimension vector which is processed with two separate dense layers to yield two $32$-dimensional representations corresponding to the mean $\boldsymbol{\mu}$ and standard deviation $\boldsymbol{\sigma}$ of the cVAE latent space. At training time, the latent code $\mathbf{z}$ is then sampled from $\boldsymbol{\mu}$ and $\boldsymbol{\sigma}$ using the reparamterization trick $\mathbf{z} = \boldsymbol{\mu} + \boldsymbol{\epsilon} \odot \boldsymbol{\sigma}$ where $\boldsymbol{\epsilon}$ is sampled from an isotropic Gaussian $\boldsymbol{\epsilon} \sim \mathcal{N}(\mathbf{0}, \mathbf{I})$. Here, the dimension of the latent space is $d_{latent}=32$ such that $\mathbf{z} \in \mathbb{R}^{d_{latent}=32}$.        

Following extraction of the latent code $\mathbf{z}$ from the encoder, the decoder is passed the latent code $\mathbf{z}$ together with the conditional variable $\mathbf{c}=[\VOX^{\CG^t}||\VOX^{\AAtom^t}]$ to generate the voxelized reconstruction $\VOX^{\hat{\AAtom}^{t}}$. We can ultimately transform this output voxelized representation $\VOX^{\hat{\AAtom}^{t}}$ into Cartesian coordinates $\hat{\AAtom}^t$ for downstream application using the methods outlined in Sec.~\ref{sec:vox} of the main text. The decoder begins with separately processing $\mathbf{z}$ and $\mathbf{c}$, which are later concatenated and further processed to yield the target reconstruction. First, displayed in the bottom right of Fig.~\ref{fig:arch} the conditional variable $\mathbf{c}$ of dimension $12 \times 12 \times 12 \times 28$ (there are 28=22+6 particle channels for $\mathbf{c}$ due to the 22 atoms within the atomistic structure the 6 beads of the coarse grain structure) is processed using (residual) 3D convolutions into an intermediate $12 \times 12  \times 12 \times 32$-dimensional representation. Concurrently, the $d_{latent}=32$-dimensional latent code $\mathbf{z}$ is reshaped into a 4D tensor of shape $2 \times 2 \times 2 \times 4$. This representation is then expanded using the PyTorch replication padding and processed with 3D convolutions to generate a compatible $12 \times 12 \times 12 \times 32$-dimensional intermediate representation of the latent code. The processed conditional variable and latent code are then  concatenated along the final dimensional to yield a $12 \times 12 \times 12 \times 64$-dimensional representation, which we further process with (residual) convolutions to yield the $12 \times 12 \times 12 \times 22$-dimensional output. The final particle-dimension of this output tensor $\VOX^{\hat{\AAtom}^t}$ corresponds to the number of particles in the target atomistic configuration. A terminal sigmoid activation scales the voxel values between 0-1 and we interpret each channel as a voxelized particle density, which we can ultimately then collapse back into Cartesian coordinates $\hat{\AAtom}^t$ as the complete atomistic reconstruction.                  

The training loss for our model described in detail in Sec.~\ref{sec:train} of the main text is 
\begin{equation}
    \mathcal{L} = \mathcal{L}_{\VOX} + \mathcal{L}_{\AAtom} + \mathcal{L}_{\text{CG}} + \mathcal{L}_{\text{EDM}} + \lambda  \mathcal{L}_{\text{ENERGY}} + \beta \mathcal{L}_{\text{KLD}} \label{eqn:si_loss}.\\
\end{equation}
Initially during training we find that $\mathcal{L}_{\text{ENERGY}}$ can become dominatingly large, and possibly NaN due to numerical overflow, motivating us to incorporate the prefactor $\lambda$ which we set to $\lambda=0$ for an initial $t_{\lambda=0}$ training steps. After which point once the model learns to roughly, but consistently, localize atomic coordinates we slowly anneal $\lambda$ from a value of $\lambda_{initial}$ to $\lambda_{final}$ over the course of $n_{\lambda}$ training steps using the following exponential annealing schedule
\begin{equation}
    \lambda^{(t)} = 
    \begin{cases}
      0 & t < t_{\lambda=0}\\
      \lambda_{initial} (\frac{\lambda_{final}}{\lambda_{initial}})^{\frac{(t - t_{\lambda=0})}{n_{\lambda}}} & t_{\lambda=0} + n_{\lambda} > t \geq t_{\lambda=0} \\
      \lambda_{final} & t \geq t_{\lambda=0} + n_{\lambda}
    \end{cases}       
\end{equation}
where $\lambda^{(t)}$ is the value of $\lambda$ at training step $t$ (not to be confused with $\AAtom^t$/$\CG^t$ which are the atomistic/CG configurations at frame index $t$ in our dataset). To further help stabilize training once $\lambda > 0$ and the $\mathcal{L}_{\text{ENERGY}}$ takes effect, we use gradient clipping to clip gradient norms to a maximum value of $w_{clip}$, and for the CLN model we also clamp the maximum possible value of the reconstructed energy $U^{\hat{\AAtom^t}}$ to be at most $U_{max}$ effectively limiting the maximum achievable value of $\mathcal{L}_{\text{ENERGY}} = \mathcal{L}_{\text{ENERGY}}(U^{\AAtom^t}, \text{max}(U^{\hat{\AAtom}^t}, U_{max}))$. The internal potential energy is calculated entirely differentiably with OpenMM~\cite{eastman2017openmm} using the AMBER99SB-ILDN~\cite{lindorff2010improved} force field for ADP and the CHARMM~\cite{brooks2009charmm, lee2016charmm} force field generated with the CHARMM-GUI~\cite{jo2008charmm} for CLN. 

The $\beta$ prefactor to $\mathcal{L}_{\text{KLD}}$ controls the weight of the KLD loss relative to the other reconstruction loss terms. To mitigate KL vanishing for the ADP model we employ a cyclic annealing schedule on $\beta$ with a sigmoid increasing function as described in Ref.~\cite{fu2019cyclical}. We define the schedule for determining the value $\beta^{(t)}$ at each training step $t$ as,
\begin{equation} \label{eqn:beta_anneal}
  \beta^{(t)} =
    \begin{cases}
      [1 + \exp(-k(\text{mod}(t, R) - \tau))]^{-1} & t < MR\\
      1 & t \geq MR\\
    \end{cases}       
\end{equation}
In Eqn.~\ref{eqn:beta_anneal} $M$ represents the number of cycles, $R$ is the length in training steps of each cycle, and $k$ and $\tau$ are parameters for width and center of the sigmoid, respectively. For our CLN model we simply maintain $\beta=1$ throughout training as we find KL vanishing is not problematic. A possible explanation for why our model benefits from cyclic annealing for the ADP data set could be related to ADP trajectory frames being spaced by 1 ps compared to the 100 ps trajectory frame spacing for the CLN data. As the previous atomistic configuration $\AAtom^{t-1}$ is provided to the decoder alongside the current CG configuration $\CG^t$ as the condition, a possible failure mode for the model would be for the decoder to simply reproduce a replica the previous atomistic frame $\hat{\AAtom}^t \approx \AAtom^{t-1}$ as a low-loss reconstruction when sequential atomistic configurations are nearly identical. In this situation the decoder is no longer generative failing to effectively purpose the latent $\mathbf{z}$ and resulting in KL vanishing $\mathcal{L}_{\text{KLD}} \rightarrow 0$. For the CLN model we find KL vanishing is not a problem when simply setting $\beta$=1 throughout training, possibly due to atomistic configurations in sequential frames being sufficiently different from one-another as the trajectory frames for the CLN data are separated by 100 ps, instead of the 1 ps spacing in the ADP data.

Our ADP model contains $\sim$1.2M parameters and is trained on a singe NVIDIA RTX 2080Ti GPU, and our CLN model contains $\sim$4.9M parameters and is trained on two NVIDIA RTX 2080Ti GPUs using distributed data parallel as implemented in PyTorch Lightning~\cite{falcon2019pytorch}. We train our model until the $\mathcal{L}_{\text{KLD}}$ loss term has approximately converged, suggesting the posterior latent space has equilibrated. Provided in Table~\ref{tab:hparams} is a complete list of model and training setting hyperparameter values with corresponding descriptions for both our ADP and CLN models.

\begin{center}
    \begin{table}
        \centering
        \resizebox{\columnwidth}{!}{%
        \begin{tabular}{|c|c|c|}
             \hline
             Hyperparameter & ADP & CLN \\
             \hline\hline 
             Mini-batch size & 32 & 32 \\
             \hline
             Learning rate & 1.398 $\times$ 10$^{-4}$ & 4.233 $\times$ 10$^{-4}$ \\
             \hline
             Voxelized grid discretization ($d$) & 12 & 12 \\
             \hline
             Width of Cartesian grid ($r_{grid}$)  & 1.8 & 5.5 \\
             \hline
             Gaussian width of voxelized particle density ($\sigma$) & 0.05398 & 0.3076 \\
             \hline
             Dimension of cVAE latent space ($d_{latent}$) & 32 & 64 \\
             \hline
             $\mathcal{D}_{KL}$ weight schedule ($\beta$) & \makecell{Cyclic annealing (Eqn.~\ref{eqn:beta_anneal}) with \\ $k=0.0025$, $\tau_0=10000$, \\ $M=4$ and $R=25000$}  & None; $\beta$=1 \\
             \hline
             \makecell{Maximum potential energy \\ value (in kJ/mol) during training ($U_{max}$)} & None & 1 $\times$ 10$^5$ \\ 
             \hline
             \makecell{Number of initial training steps \\ with prefactor $\lambda=0$ to $\mathcal{L}_{\text{ENERGY}}$  ($t_{\lambda=0}$)} & 2 $\times$ 10$^{5}$ & 7.5 $\times$ 10$^{5}$ \\
             \hline
             Initial value of $\lambda$ when starting to anneal $\lambda$ $(\lambda_{initial})$ & 1 $\times$ 10$^{-6}$ & 1 $\times$ 10$^{-6}$ \\  
             \hline
             Final value of $\lambda$ after anneal $\lambda$ $(\lambda_{final})$ & 1.0 & 1.0 \\  
             \hline
             Number of steps to anneal $\lambda$ $(n_{\lambda})$ & 1.25 $\times$ 10$^{6}$ & 1.25 $\times$ 10$^{6}$ \\  
             \hline
             \makecell{Gradient clipping value used \\ after $t_{\lambda=0}$ training steps $(w_{clip})$} & 0.1 & 0.1 \\  
             \hline
             Encoder hidden channel dimension $(d_{hidden}^{Encoder})$ & 32 & 32 \\  
             \hline
             Decoder hidden channel dimension $(d_{hidden}^{Decoder})$ & 32 & 64 \\  
             \hline
             Number of training steps & $\sim$14.8M & $\sim$9.4M \\  
             \hline
        \end{tabular}}
        \caption{ADP and CLN model and training setting hyperparameters with associated descriptions.}
        \label{tab:hparams}
    \end{table}
\end{center}

\subsection{Coarse Grain CGSchNet Force Fields}

Coarse grain force field models of ADP are assessed using a 5-fold cross-validation strategy. Model creation and training was done using force matching via TorchMDNet~\cite{doerr2021torchmd} and PyTorch. Each of the model hyperparameters are described in \ref{tab:cgschnet_model_params}:

\begin{center}
    \begin{table}
        \centering
        \begin{tabular}{|c|c|c|}
             \hline
             Hyperparameter & ADP & CLN \\
             \hline\hline 
             Interaction Blocks & 2 & 2 \\
             Hidden Channels & 128 & 128 \\
             Filters & 128 & 128 \\
             Neighbor Cutoff & 5 Angstroms & 30 Angstroms \\
             Filter Network Cutoff Function & None & Cosine Cutoff \\
             Radial Basis Functions & 20 & 300 \\
             Radial Basis Function Span & 0 to 5 Angstroms & 0 to 30 Angstroms \\
             Radial Basis Function Type & Exponential Normal & Gaussian \\
             Terminal Network Layers & 1 & 1 \\
             Terminal Network Width & 128 & 128 \\
             Activation Function & Tanh & Tanh \\
             Embedding Strategy & Atomic Number & Amino Acid Identity\footnote{Terminal TYR residues received a unique terminal embedding.} \\
             Priors & Bonds, Angles & Bonds, Angles, Repulsions \\
             \hline
        \end{tabular}
        \caption{CGSchNet Model Hyperparameters}
        \label{tab:cgschnet_model_params}
    \end{table}
\end{center}

The following prior energy terms were used to construct a baseline force field and to constrain basic molecular structure for each model. These prior terms are summarized in \ref{tab:cgschnet_model_params} and \ref{tab:cgschnet_prior_params}.
\begin{center}
    \begin{table}
        \centering
        \begin{tabular}{|c|c|c|}
             \hline
             Prior Type & Energy \\
             \hline\hline 
             Bonds & $\frac{k}{2}(r-r_0)^2$ \\
             Angles & $\frac{k}{2}(\theta-\theta_0)^2$ \\
             Repulsions & $\left( \frac{\sigma}{r_{ij}}\right)^6$ \\
             \hline
        \end{tabular}
        \caption{CGSchNet Model Hyperparameters}
        \label{tab:cgschnet_prior_params}
    \end{table}
\end{center}
Following previous work\cite{wang2019coarse, husic2020coarse}, the spring constants ($k$) for angles and bonds were fit directly from reference data distributions. For ADP, all physical bonds and sequential triplet angles were used. For CLN, all sequential CA bonds and triplet angles were used. For the CLN repulsions, all beads pairs outside of bonds were assigned to the repulsion set, and the excluded volumes ($\sigma$) were uniformly set to $3.5$ angstroms for all pairs in the repulsion set.

Following the procedure in Chen et al~\cite{chen2021machine}, each model was training on the \emph{delta forces}; that is, each model was trained on the difference between the reference forces and the forces predicted by the prior model:

\begin{equation}
F_{Delta} = F_{Ref} - F_{Prior}
\end{equation}

\begin{figure}[!tbp]
  \centering
  \includegraphics[width=\linewidth]{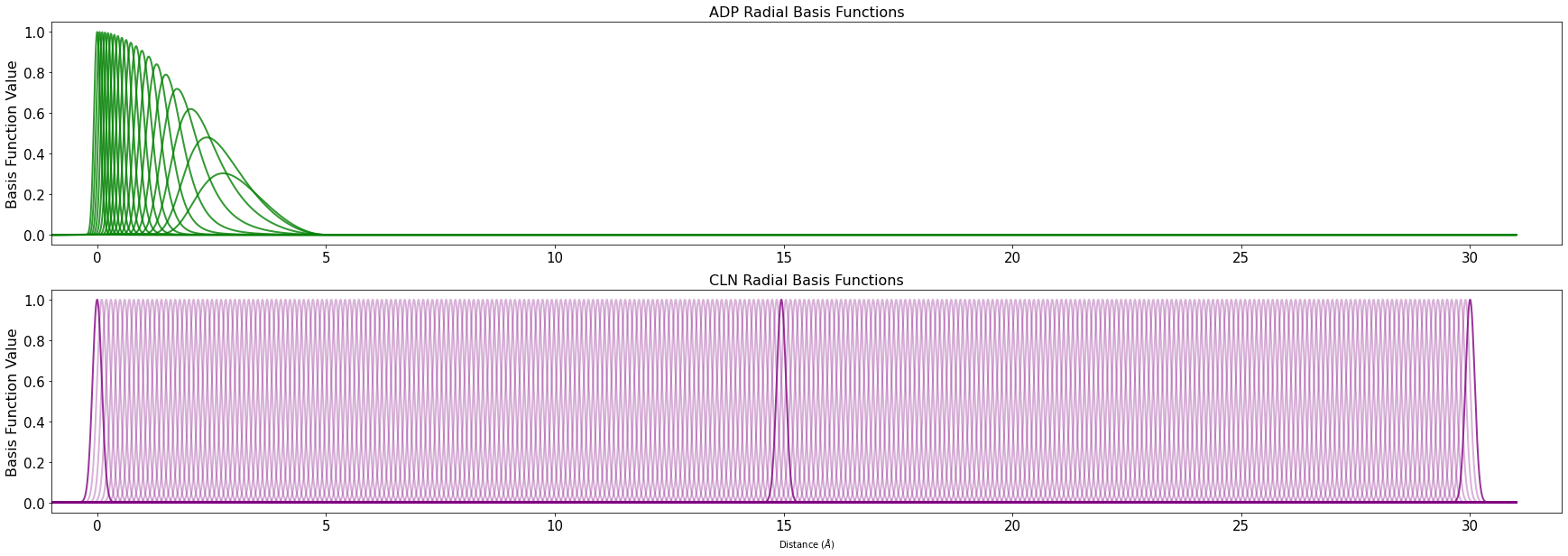}
  \caption{CGSchNet Radial basis function sets for ADP and CLN. For visual clarity, the first, middle, and last basis functions for CLN have been highlighted.}
\end{figure}

Separate training routines were adopted for each molecule, following the procedure described in Husic et al~\cite{husic2020coarse}. They are reproduced in \ref{tab:cgschnet_hparams} for convenience.

\begin{center}
    \begin{table}
        \centering
        \begin{tabular}{|c|c|c|}
             \hline
             Hyperparameter & ADP & CLN \\
             \hline\hline 
             Number of Epochs & 100 & 100 \\
             Batch Size & 512 & 512 \\
             Optimizer & ADAM & ADAM \\
             Weight Initialization & Xavier Uniform & Xavier Uniform \\
             Bias Initialization & Uniform(-$\sqrt{k}$,$\sqrt{k}$) & 0 \\
             Initial LR & 0.0006 & 0.0001 \\
             LR Scheduler & ReduceLROnPlateau & None \\
             Loss Function & MSE Force Matching & MSE Force Matching \\
             Model Selection & Last Epoch & Minimum Average Validation Loss \\
             \hline
        \end{tabular}
        \caption{Model Training Hyperparameters}
        \label{tab:cgschnet_hparams}
    \end{table}
\end{center}

\begin{figure}[!tbp]
  \centering
  \subfloat{\includegraphics[width=0.4\textwidth]{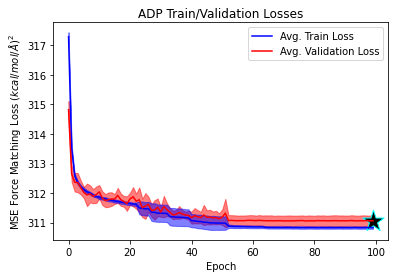}\label{fig:a}}
  \hfill
  \subfloat{\includegraphics[width=0.4\textwidth]{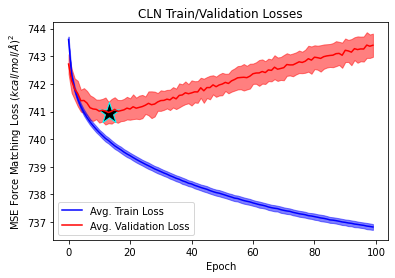}\label{fig:f2}}
  \caption{Epochal training and validation loss curves for the ADP and CLN CGSchNet models over 5 cross validation folds. The solid curves represent the average train/validation loss, while the shaded regions represent a moving standard deviation across the 5 models. The black star represents the epoch at which the final model was chosen.}
\end{figure}

To assess the performance of each CGSchNet Model, implicit solvent Langevin simulations were carried out with the parameters described in \ref{tab:cgschnet_sim_params}.

\begin{center}
    \begin{table}
        \centering
        \begin{tabular}{|c|c|c|}
             \hline
             Parameter & ADP & CLN \\
             \hline\hline 
             Integration Time step & 4 fs & 4 fs \\
             Friction & 1 ps$^{-1}$ & 1 ps$^{-1}$ \\
             Initial Structures & Random Sample From Train Set & Random Sample From Train Set\\
             Number of Trajectories & 100 & 1000 \\
             Temperature & 300K & 350K \\
             \hline
        \end{tabular}
        \caption{CG Simulation Parameters}
        \label{tab:cgschnet_sim_params}
    \end{table}
\end{center}
For both ADP and CLN, the following BAOA(F)B Langevin scheme was used:

\begin{center}
\begin{flushleft}
\hspace{5cm}1. $\bar{F} = -\nabla U(x_t) $\\
\hspace{5cm}2. $v_{t+1} = v_t + dt \frac{\bar{F}}{m}$\\
\hspace{5cm}3. $x_{t+1/2} = x_t + v_t \frac{dt}{2}$ \\
\hspace{5cm}4. $v_{t+1} = v_{t+1/2} \eta_v + dW_t \eta_n$ \\
\hspace{5cm}5. $x_{t+1} = x_{t+1/2} v\frac{dt}{2}$
\end{flushleft}
\end{center}

where $\bar{F}$ is an average force predicted by averaging the force predictions from each model produced from cross validation,

\begin{equation}
\bar{F} = \sum_{i=1}^{\text{num folds}} f_i    
\end{equation}
$\eta_v$ and $\eta_n$ are velocity and noise scales respectively, and $dW$ is a stochastic Wiener process. Forward and backward transitions through all major metastable states are observed in the CG simulations for both molecular systems.

\subsection{Latent space embedding and samples}

\subsection{Alanine dipeptide}
\textbf{Energetics}

We notice overall excellent agreement between our atomistic and backmapped ADP trajectories for both the in distribution and generation test sets. This agreement results in nearly identical distributions of the internal potential energy(Fig.~\ref{fig:adp_energy}). The backmapped data tends to slightly deviate by producing small high-energy tails. These rare high-energy configurations produced by our model are a result of a select few bond length and angle distributions that are moderately deformed compared to the reference atomistic data. The largest discrepancies are atomic angles and bonds that include hydrogen atoms, and particularly terminal methyl group hydrogens. While the MD engine restrains all bond lengths involving hydrogen, our model can struggle to exactly reproduce these bond lengths due to the achievable resolution of the voxelized grid. The terminal methy group of ADP is also entirely unspecified by the CG configuration, and methyl group hydrogen atoms are a significant source of variation rotating about the central carbon atom with relaxation times of $\sim$0.1-1 ps which is faster than the 1 ps spacing between consecutive frames~\cite{yi2013dynamic}.

\textbf{Thermodynamics}

\textbf{Kinetics}

\textbf{ADP MSM construction and validation}

We construct Markov State Models (MSMs) to facilitate the comparison of the kinetics between the reference atomistic data and our backmapped trajectories in Sec.~\ref{sec:adp_kinetics}. 

We construct MSMs for ADP over the phase space of the backbone $\phi$,$\psi$ angles as they are known to be good CVs for this system. When comparing kinetics between backmapped/CG data and atomistic data we perform k-means clustering for state space decomposition first on the atomistic dataset. We use these same k-means clusters to then produce state assignments for the backmapped/original CG data. Independent MSMs are subsequently fit to the atomistic and backmapped/original CG data from these cluster assignments to ensure direct comparability bewteen the recovered timescales and processes. Associated state space clustering plots, implied timescale analysis for lag time selection and Chapman-Kolmogorow (CK) tests for Markovianity for MSMs built on the different ADP datasets in this work are shown below.

For the in distribution test set we first perform k-means clustering using 100 centroids on the reference atomistic data, the same 100 clusters of which we then also use to build the backmapped in distribution test set MSM.

\textbf{ADP in distribution test set atomistic}
\begin{figure}[H]
  \centering
  \includegraphics[width=0.8\linewidth]{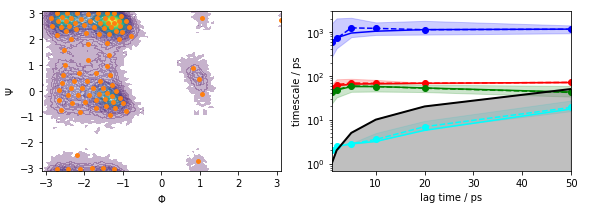}
  \caption{(left) State space clustering using 100 k-means centers fit to the in distribution test set atomistic data shown as orange dots over a contour of the data density. (right) Implied timescale analysis for MSM construction. Solid lines indicate the maximum likelihood MSM estimate, while the dashed lines show the mean implied timescale bounded by a 95\% confidence interval. A lag time of 5 ps was selected.} 
\end{figure}

\begin{figure}[H]
  \centering
  \includegraphics[width=0.6\linewidth]{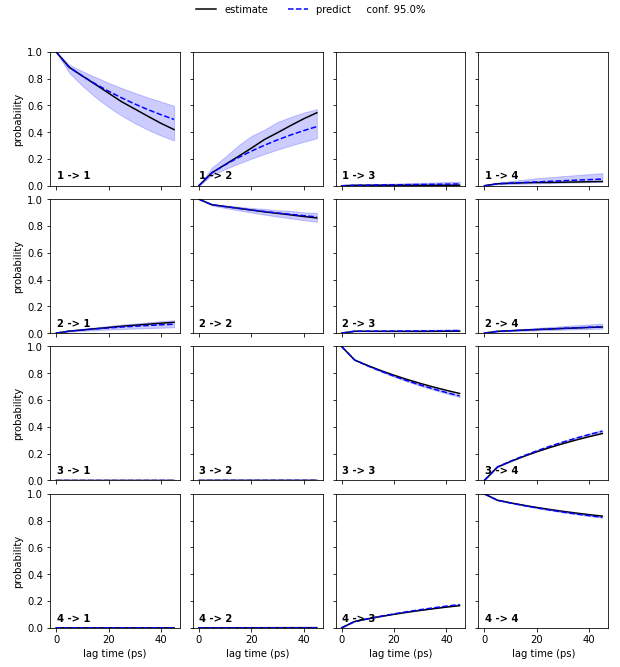}
  \caption{(Chapman-Kolmogorov (CK) test using 4 macrostates. Black lines indicate estimates, while the dashed blue lines are the predictions bounded by a 95\% confidence interval)} 
\end{figure}

\textbf{ADP in distribution test set reconstructed}
\begin{figure}[H]
  \centering
  \includegraphics[width=0.8\linewidth]{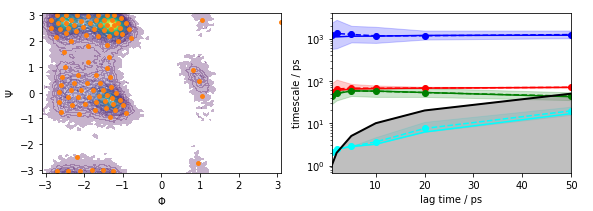}
  \caption{(left) State space clustering of the backmapped in distribution test set using 100 k-means centers taken from the in distribution test set atomistic data shown as orange dots over a contour of the data density. (right) Implied timescale analysis for MSM construction. Solid lines indicate the maximum likelihood MSM estimate, while the dashed lines show the mean implied timescale bounded by a 95\% confidence interval. A lag time of 5 ps was selected.} 
\end{figure}

\begin{figure}[H]
  \centering
  \includegraphics[width=0.6\linewidth]{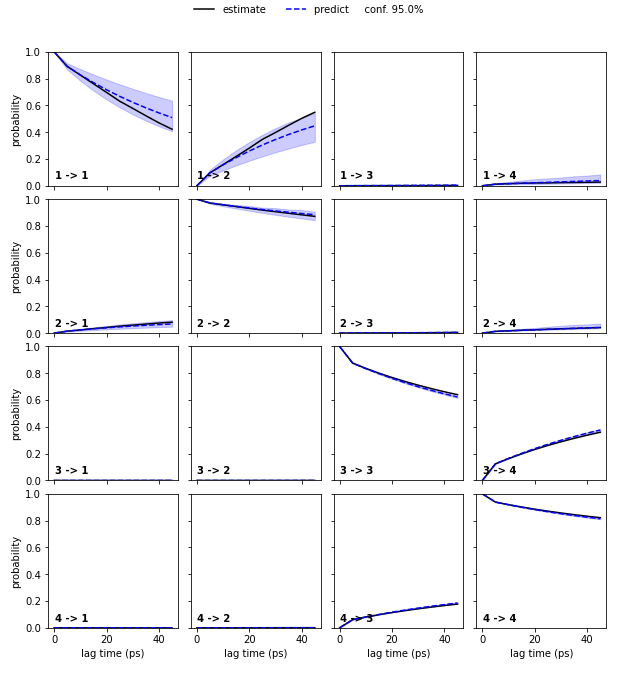}
  \caption{(Chapman-Kolmogorov (CK) test using 4 macrostates. Black lines indicate estimates, while the dashed blue lines are the predictions bounded by a 95\% confidence interval)} 
\end{figure}

\textbf{ADP reference atomistic}

For the generalization set we perform state space clustering using 100 kmeans centers first over the reference atomistic dataset which is taken from Ref.~\cite{wehmeyer2018time}. These same 100 clusters then define state assignments for MSMs built on both the backmapped CG reconstructed data and the original CGSchNet~\cite{husic2020coarse} simulation.

\begin{figure}[H]
  \centering
  \includegraphics[width=0.8\linewidth]{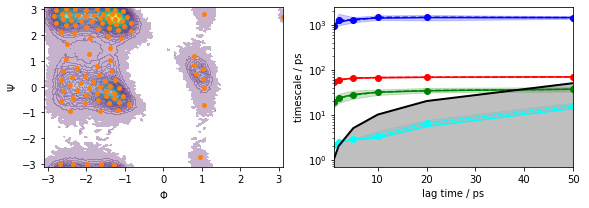}
  \caption{(left) State space clustering using 100 k-means centers fit to the reference atomistic data taken from Ref.~\cite{wehmeyer2018time} shown as orange dots over a contour of the data density. (right) Implied timescale analysis for MSM construction. Solid lines indicate the maximum likelihood MSM estimate, while the dashed lines show the mean implied timescale bounded by a 95\% confidence interval. A lag time of 5 ps was selected.} 
\end{figure}

\begin{figure}[H]
  \centering
  \includegraphics[width=0.6\linewidth]{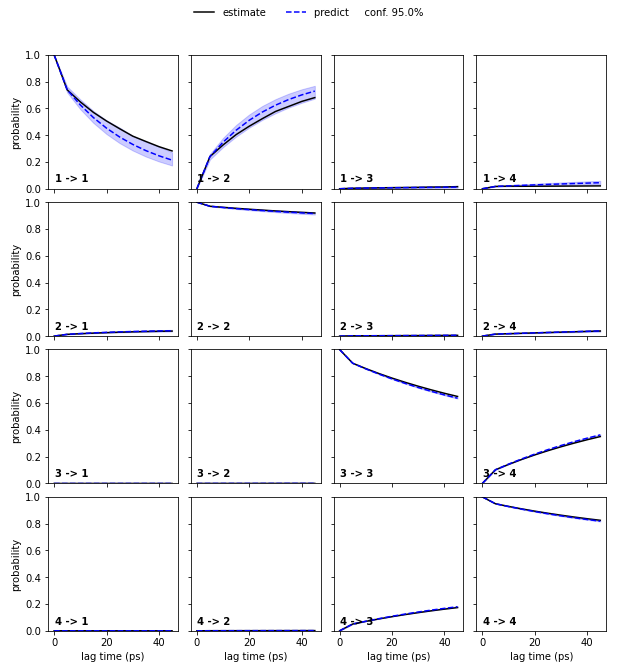}
  \caption{(Chapman-Kolmogorov (CK) test using 4 macrostates. Black lines indicate estimates, while the dashed blue lines are the predictions bounded by a 95\% confidence interval)} 
\end{figure}

\textbf{ADP generalization set reconstructed}
\begin{figure}[H]
  \centering
  \includegraphics[width=0.8\linewidth]{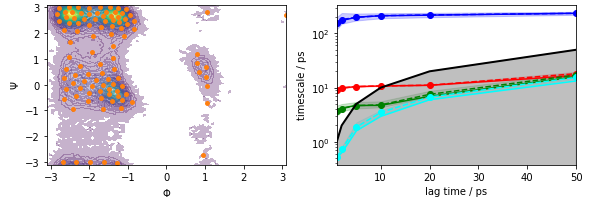}
  \caption{(left) State space clustering of the backmapped CGSchNet data using 100 k-means centers taken from the reference atomistic data shown as orange dots over a contour of the data density. (right) Implied timescale analysis for MSM construction. Solid lines indicate the maximum likelihood MSM estimate, while the dashed lines show the mean implied timescale bounded by a 95\% confidence interval. A lag time of 5 ps was selected.} 
\end{figure}

\begin{figure}[H]
  \centering
  \includegraphics[width=0.6\linewidth]{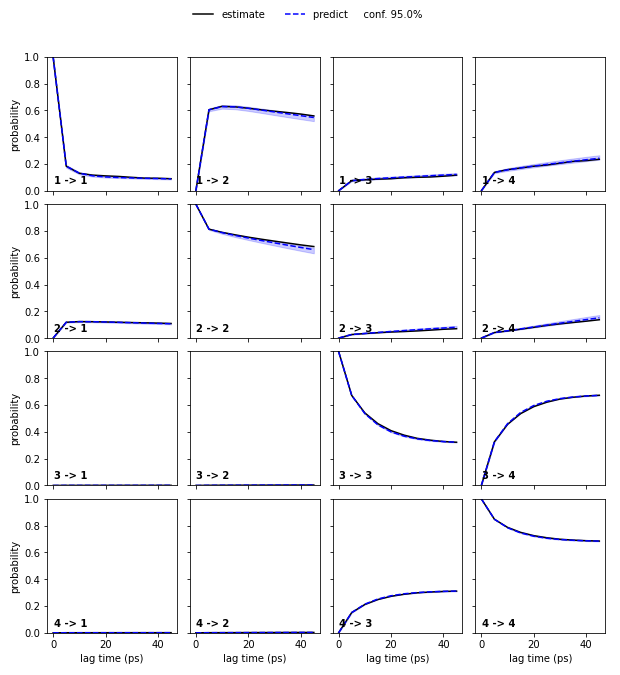}
  \caption{(Chapman-Kolmogorov (CK) test using 4 macrostates. Black lines indicate estimates, while the dashed blue lines are the predictions bounded by a 95\% confidence interval)} 
\end{figure}

\textbf{ADP CGSchNet simulation}
\begin{figure}[H]
  \centering
  \includegraphics[width=0.8\linewidth]{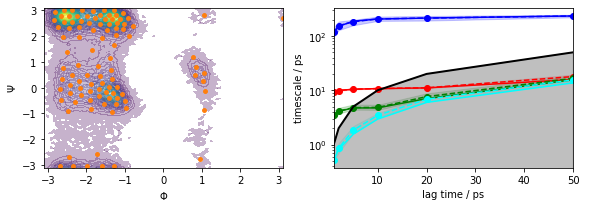}
  \caption{(left) State space clustering of the original coarse grained CGSchNet data using 100 k-means centers taken from the reference atomistic data shown as orange dots over a contour of the data density. (right) Implied timescale analysis for MSM construction. Solid lines indicate the maximum likelihood MSM estimate, while the dashed lines show the mean implied timescale bounded by a 95\% confidence interval. A lag time of 5 ps was selected.} 
\end{figure}

\begin{figure}[H]
  \centering
  \includegraphics[width=0.6\linewidth]{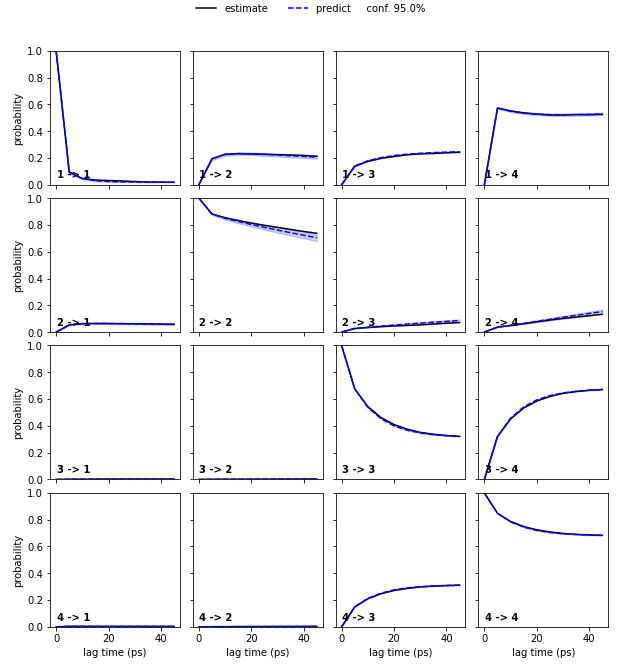}
  \caption{(Chapman-Kolmogorov (CK) test using 4 macrostates. Black lines indicate estimates, while the dashed blue lines are the predictions bounded by a 95\% confidence interval)} 
\end{figure}

\subsection{Chignolin}

\textbf{Energetics}

\textbf{Thermodynamics}

\textbf{Kinetics}

\textbf{CLN MSM construction and validation}

We construct Markov State Models (MSMs) to facilitate the comparison of the kinetics between the reference atomistic data and our backmapped trajectories in Sec.~\ref{sec:cln_kinetics}. 

To construct MSMs for CLN we first featurize the reference atomistic data taken from Ref.~\cite{wang2019machine} using the 45 pairwise $\alpha$-carbon distances between each pair of amino acid residues. We then perform Time-lagged Independent Component Analysis (TICA) on this atomistic data using these pairwise distance features where we retain the first two Independent Components (ICs). We perform state decomposition in this TICA space using k-means clustering with 150 centroids. With the same residue contact featurization for other atomistic, backmapped or CG data we use our learned TICA model to project other data into a common TICA space. Independent MSMs are subsequently built for each separate dataset with state assignments generated in this common TICA space from the same previously identified 150 k-means centers in the reference atomistic dataset. In the case if not all 150 states are occupied MSMs are constructed only using the subset of occupied states for that data set. Operating within this shared TICA and clustering space allows direct comparison between the different MSM recovered timescales and processes. Associated state space clustering plots, implied timescale analysis for lag time selection and Chapman-Kolmogorow (CK) tests for Markovianity for MSMs built on the different CLN datasets in this work are shown below.

\textbf{CLN reference atomistic}
\begin{figure}[H]
  \centering
  \includegraphics[width=0.8\linewidth]{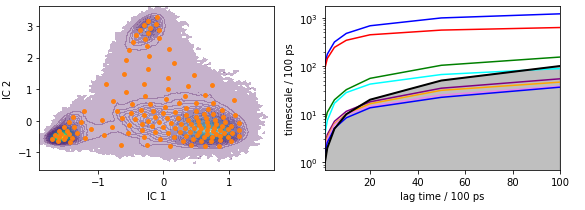}
  \caption{(left) Clustering in TICA space using 150 k-means centers fit to the reference atomistic dataset taken from Ref.~\cite{wang2019machine} shown as orange dots over a contour of the data density. (right) Implied timescale analysis for MSM construction. Solid lines indicate the maximum likelihood MSM estimate. A lag time of 4 ns (40 frames) was selected.} 
\end{figure}

\begin{figure}[H]
  \centering
  \includegraphics[width=0.6\linewidth]{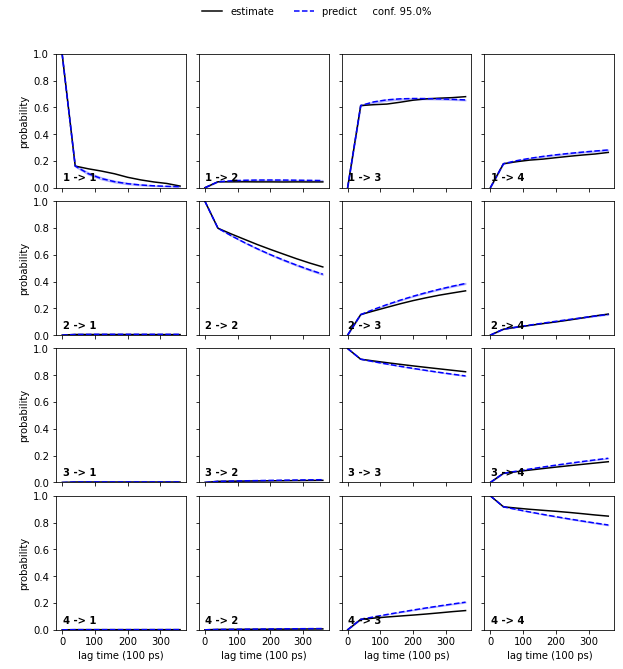}
  \caption{(Chapman-Kolmogorov (CK) test using 4 macrostates. Black lines indicate estimates while the dashed blue lines are the predictions bounded by a 95\% confidence interval)} 
\end{figure}

\textbf{CLN in distribution test set atomistic}
\begin{figure}[H]
  \centering
  \includegraphics[width=0.8\linewidth]{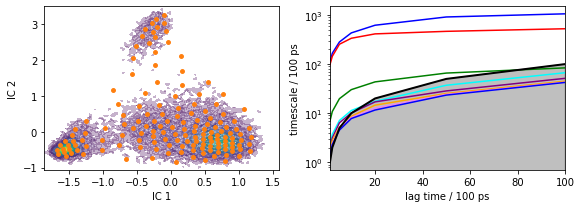}
  \caption{(left) State space clustering for the atomistic in distribution test set data using the same TICA projection and 150 k-means centers from the reference atomistic dataset shown as orange dots over a contour of the data density. (right) Implied timescale analysis for MSM construction. Solid lines indicate the maximum likelihood MSM estimate. A lag time of 4 ns (40 frames) was selected.} 
\end{figure}

\begin{figure}[H]
  \centering
  \includegraphics[width=0.6\linewidth]{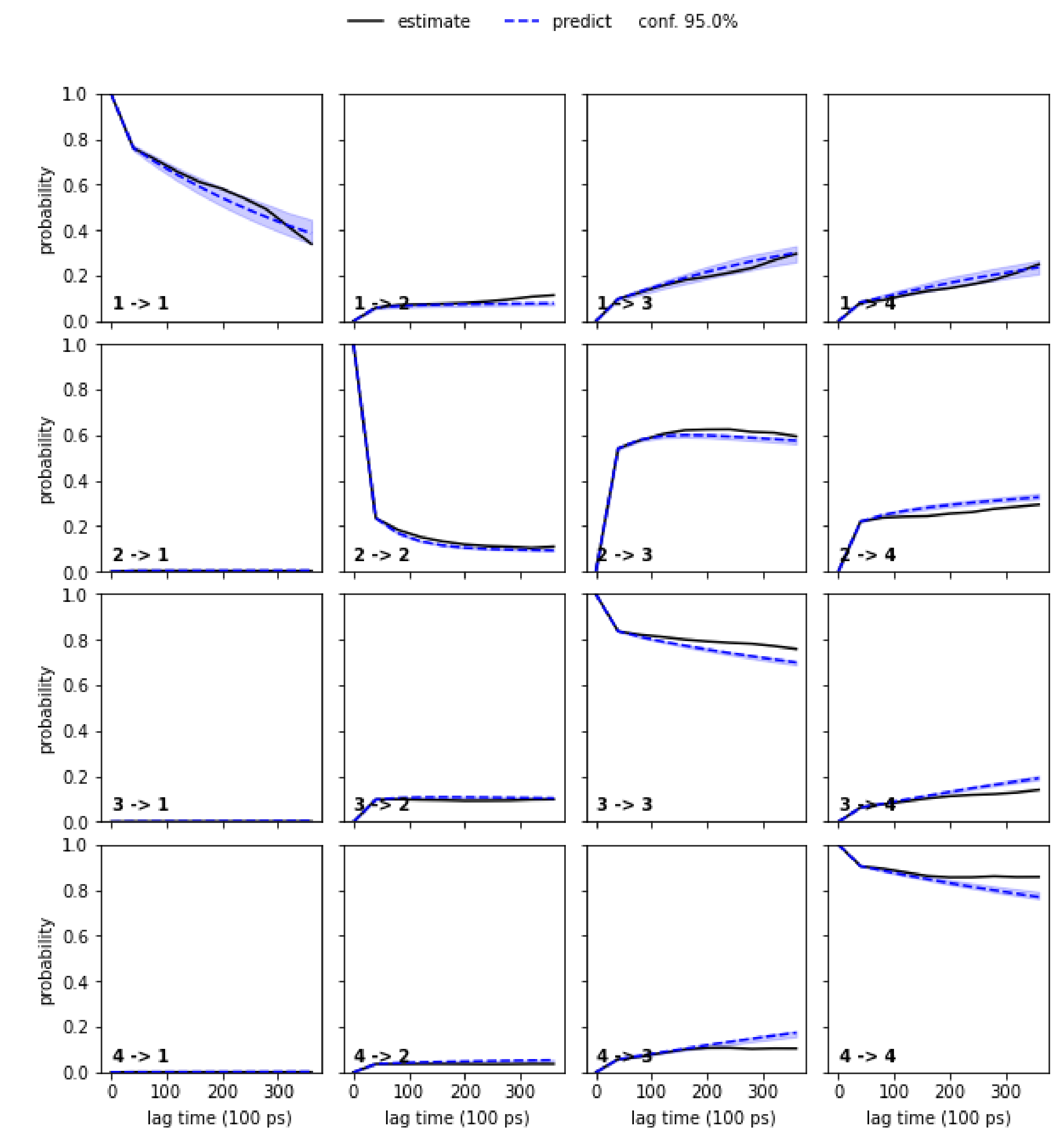}
  \caption{(Chapman-Kolmogorov (CK) test using 4 macrostates. Black lines indicate estimates while the dashed blue lines are the predictions bounded by a 95\% confidence interval)} 
\end{figure}

\textbf{CLN in distribution test set reconstructed}

\begin{figure}[H]
  \centering
  \includegraphics[width=0.8\linewidth]{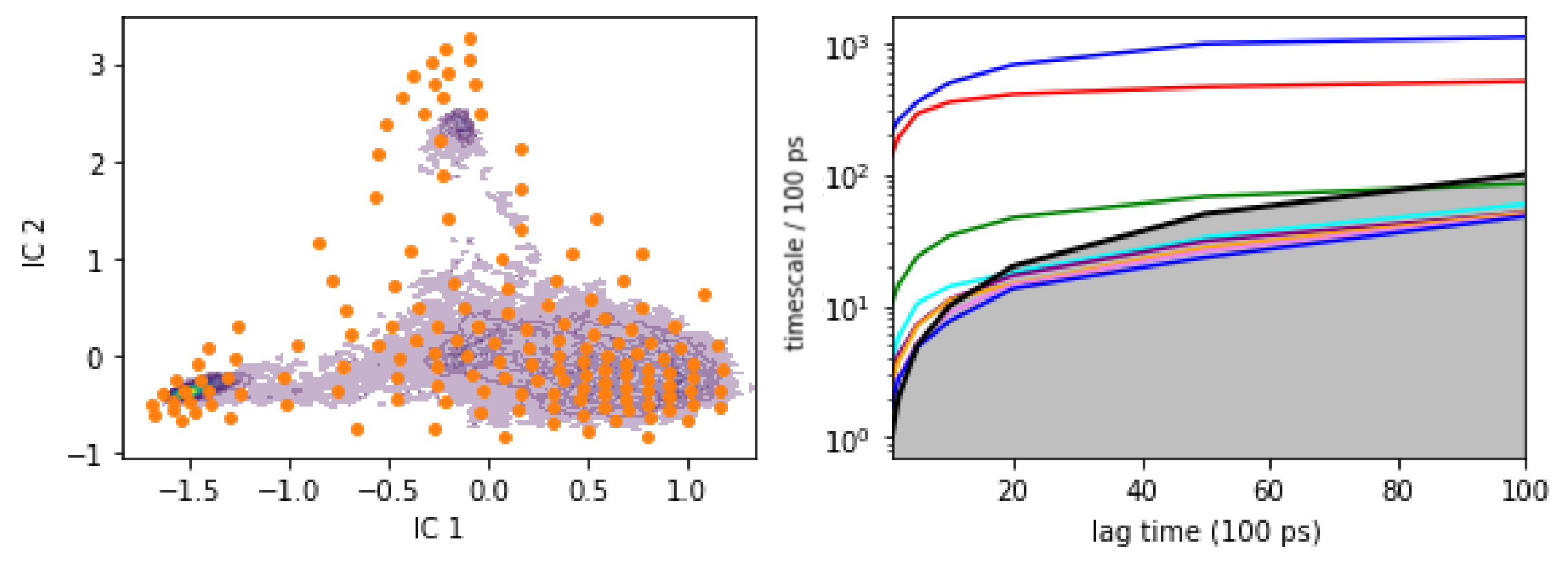}
  \caption{(left) State space clustering for the backmapped in distribution test set data using the same TICA projection and 150 k-means centers from to the reference atomistic dataset shown as orange dots over a contour of the data density. (right) Implied timescale analysis for MSM construction. Solid lines indicate the maximum likelihood MSM estimate. A lag time of 4 ns (40 frames) was selected.} 
\end{figure}

\begin{figure}[H]
  \centering
  \includegraphics[width=0.6\linewidth]{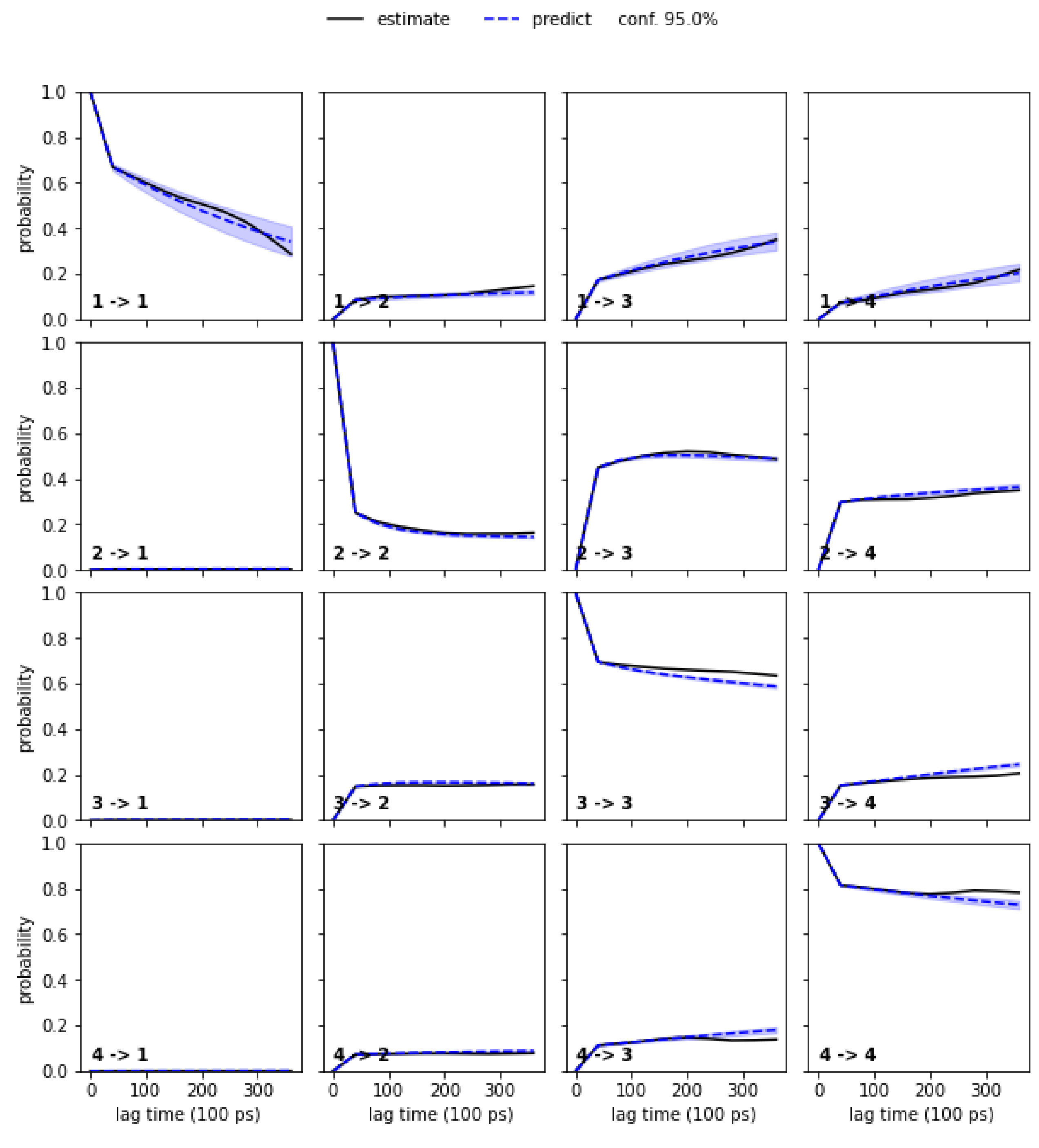}
  \caption{(Chapman-Kolmogorov (CK) test using 4 macrostates. Black lines indicate estimates while the dashed blue lines are the predictions bounded by a 95\% confidence interval)} 
\end{figure}

\textbf{CLN generalization set reconstructed}
\begin{figure}[H]
  \centering
  \includegraphics[width=0.8\linewidth]{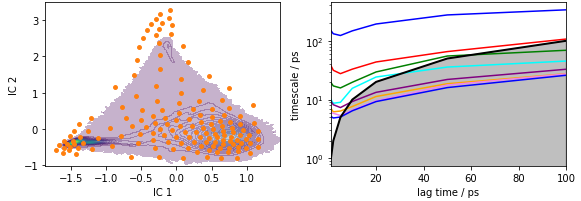}
  \caption{(left) State space clustering for the backmapped CGSchNet simulation data using the same TICA projection and 150 k-means centers from the reference atomistic dataset shown as orange dots over a contour of the data density. (right) Implied timescale analysis for MSM construction. Solid lines indicate the maximum likelihood MSM estimate. A lag time of 4 ns (40 frames) was selected.} 
\end{figure}

\begin{figure}[H]
  \centering
  \includegraphics[width=0.6\linewidth]{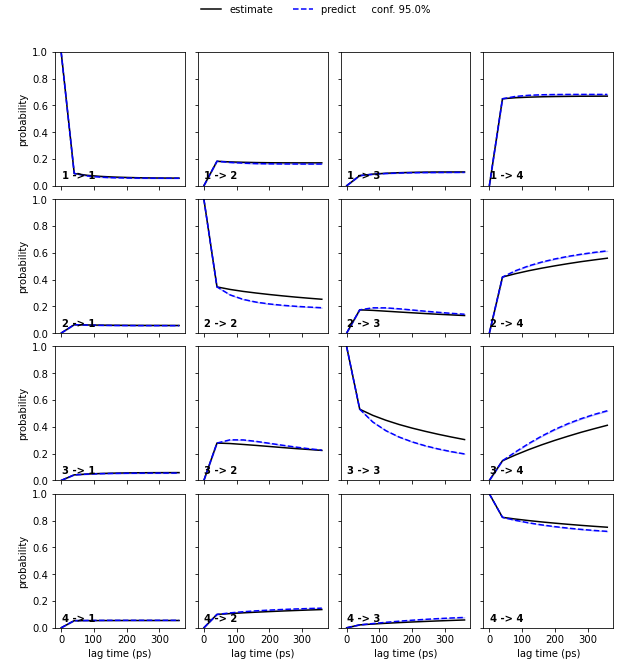}
  \caption{(Chapman-Kolmogorov (CK) test using 4 macrostates. Black lines indicate estimates while the dashed blue lines are the predictions bounded by a 95\% confidence interval)} 
\end{figure}

\textbf{CLN CGSchNet simulation}
\begin{figure}[H]
  \centering
  \includegraphics[width=0.8\linewidth]{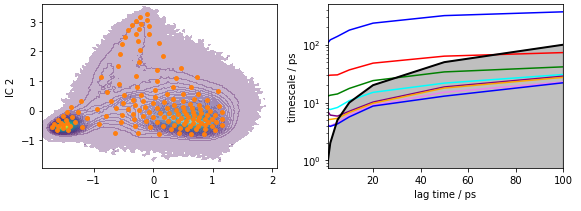}
  \caption{(left) State space clustering for the original CGSchNet simulation data using the same TICA projection and 150 k-means centers from the reference atomistic dataset shown as orange dots over a contour of the data density. (right) Implied timescale analysis for MSM construction. Solid lines indicate the maximum likelihood MSM estimate. A lag time of 4 ns (40 frames) was selected.} 
\end{figure}

\begin{figure}[H]
  \centering
  \includegraphics[width=0.6\linewidth]{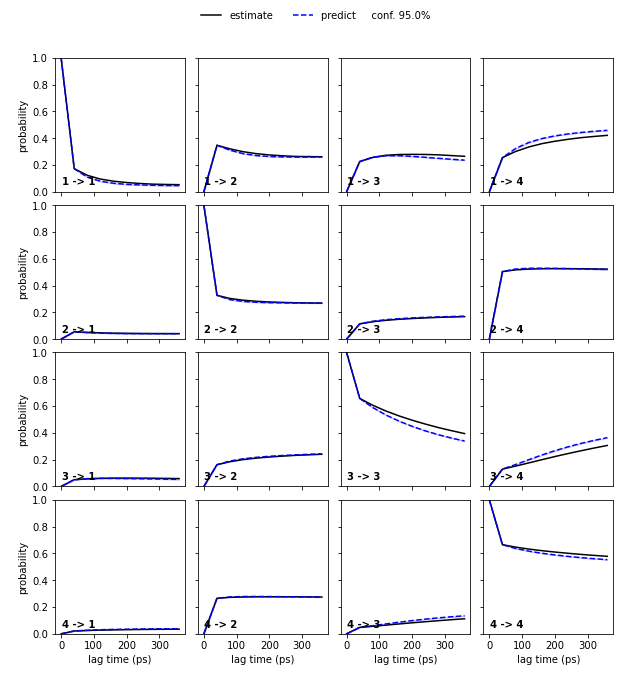}
  \caption{(Chapman-Kolmogorov (CK) test using 4 macrostates. Black lines indicate estimates while the dashed blue lines are the predictions bounded by a 95\% confidence interval)} 
\end{figure}

\clearpage
\newpage

\bibliography{manuscript}

\end{document}